\documentclass[aps,showpacs,pre,superscriptaddress,nofootinbib,notitlepage,a4paper]{revtex4}
\usepackage{dsfont}
\usepackage{amsmath}
\usepackage{amsfonts}
\usepackage{amssymb}
\usepackage{xcolor}

\usepackage{graphicx}
\usepackage{psfrag}

\usepackage{tikz}
\usetikzlibrary{patterns,decorations,arrows,shapes.geometric,arrows,mindmap,backgrounds,positioning,positioning,fit,backgrounds,positioning,arrows,matrix,calc}
\usetikzlibrary{shapes,backgrounds,mindmap,shadows,shapes.geometric,topaths,snakes,arrows,pgfplots.groupplots,fadings,calc,decorations.markings,positioning,chains,fit}
\usetikzlibrary{arrows,shapes,positioning}
\usetikzlibrary{decorations.markings}
\tikzstyle arrowstyle=[scale=1]
\tikzstyle directed=[postaction={decorate,decoration={markings,
    mark=at position .65 with {\arrow[arrowstyle]{stealth}}}}]

\newcommand{\beq}{\begin{equation}} 
\newcommand{\eeq}{\end{equation}}
\newcommand{\bem}{\begin{multline}}
\newcommand{\bes}{\begin{split}}
\newcommand{\ees}{\end{split}} 
\newcommand{\bea}{\begin{align}}
\newcommand{\eea}{\end{align}}

\newcommand{\dd}{{\rm d}}

\newcommand{\tD}{\widetilde{D}}

\newcommand{\tf}{\widetilde{f}}
\newcommand{\tb}{\widetilde{b}}

\newcommand{\hb}{\widehat{b}}

\newcommand{\hS}{\widehat{S}}
\newcommand{\hf}{\widehat{f}}
\newcommand{\hSopt}{\widehat{S}^{\rm opt}}

\newcommand{\ve}{\varepsilon}

\newcommand{\eqd}{\overset{\rm d}{=}}

\newcommand{\E}{\mathbb{E}}
\newcommand{\R}{\mathbb{R}}

\newcommand{\one}{\mathds{1}}

\newcommand{\MSE}{{\rm MSE}}
\newcommand{\MSEd}{{\rm MSE}_{\rm d}}
\newcommand{\MSEod}{{\rm MSE}_{\rm od}}
\newcommand{\MMSE}{{\rm MMSE}}
\newcommand{\MMSEB}{{\rm MSE}_{\rm BABP}}
\newcommand{\MMSEd}{{\rm MMSE}_{\rm d}}
\newcommand{\MMSEod}{{\rm MMSE}_{\rm od}}
\newcommand{\Tr}{{\rm Tr}}
\newcommand{\tr}{{\rm tr}}
\newcommand{\diag}{{\rm diag}}
\newcommand{\DB}{\mathcal{D}_{\rm BABP}}
\newcommand{\tDB}{\widetilde{\mathcal{D}}_{\rm BABP}}
\newcommand{\tDD}{\widetilde{\mathcal{D}}^{(D)}}
\newcommand{\la}{\langle}
\newcommand{\ra}{\rangle}
\newcommand{\A}{\mathcal{A}}
\newcommand{\Ad}{\mathcal{A}_{\rm d}}
\newcommand{\Aod}{\mathcal{A}_{\rm od}}
\newcommand{\M}{\mathcal{M}}
\newcommand{\cR}{\mathcal{R}}
\newcommand{\hcR}{\widehat{\mathcal{R}}}
\newcommand{\hcM}{\widehat{\mathcal{M}}}
\newcommand{\hc}{\widehat{c}}
\newcommand{\cO}{\mathcal{O}}
\newcommand{\cS}{\mathcal{S}}
\newcommand{\cI}{\mathcal{I}}
\newcommand{\cF}{\mathcal{F}}
\newcommand{\kp}{\mathfrak{p}}
\newcommand{\kq}{\mathfrak{q}}
\newcommand{\D}{\mathcal{D}}
\newcommand{\re}{{\rm Re}}
\newcommand{\im}{{\rm Im}}

\begin{document}

\bibliographystyle{myunsrt}

\title{Matrix denoising: Bayes-optimal estimators via low-degree polynomials}

\author{Guilhem Semerjian}
\affiliation{Laboratoire de physique de l'Ecole normale sup\'erieure, ENS, Universit\'e PSL, CNRS, Sorbonne Universit\'e, Universit\'e Paris Cit\'e, F-75005 Paris, France} 

\begin{abstract}

We consider the additive version of the matrix denoising problem, where a random symmetric matrix $S$ of size $n$ has to be inferred from the observation of $Y=S+Z$, with $Z$ an independent random matrix modeling a noise. For prior distributions of $S$ and $Z$ that are invariant under conjugation by orthogonal matrices we determine, using results from first and second order free probability theory, the Bayes-optimal (in terms of the mean square error) polynomial estimators of degree at most $D$, asymptotically in $n$, and show that as $D$ increases they converge towards the estimator introduced by Bun, Allez, Bouchaud and Potters in [IEEE Transactions on Information Theory {\bf 62}, 7475 (2016)]. We conjecture that this optimality holds beyond strictly orthogonally invariant priors, and provide partial evidences of this universality phenomenon when $S$ is an arbitrary Wishart matrix and $Z$ is drawn from the Gaussian Orthogonal Ensemble, a case motivated by the related extensive rank matrix factorization problem.

\end{abstract}

\maketitle

\tableofcontents

\section{Introduction}

In the matrix denoising problem considered in this paper one aims at recovering a signal $S$ from a noisy observation $Y$, both being real symmetric matrices of (large) size $n$, by computing an estimator $\hS(Y)$. In its multiplicative noise version the observation is generated as $Y=\sqrt{S} Z \sqrt{S}$ (assuming the signal $S$ is positive definite), with the noise $Z$ being a random matrix. This is a relevant model for the problem of estimating the true covariance matrix $S$ of a random vector from a sample covariance matrix $Y$, and as such has many applications in data analysis~\cite{LePe11,BuAlBoPo16,BuBoPo17}. We shall instead concentrate in this paper on the additive noise version, modeled as $Y=S+Z$, with again a noise random matrix $Z$ corrupting the signal $S$ in the observation $Y$. A motivation for this version arises from the matrix factorization problem, in which the observation matrix $Y$ is written as $Y=X X^T +Z$, with $X$ an $n \times r$ matrix and $Z$ a noise, and the goal is to infer $X$ from $Y$ (a more generic model is $Y=X F +Z$, with $X$ of size $n \times r$ and $F$ of size $r \times n$, the goal being here to infer the factors $X$ and $F$). The additive version of the matrix denoising problem can thus be seen as a simplified version of the matrix factorization one, in which one only aims at reconstructing the product of the two unknown factors (i.e. the Wishart matrix $XX^T$) instead of the factors individually. Matrix factorization appears in various problems of machine learning~\cite{OlFi96,ZiPe01,KrMuRaEnLeSe03,SpWaWr12}, and has been the subject of an intense research activity, in particular as an example of applications of methods introduced in statistical mechanics of disordered systems to inference problems~\cite{ZdKr16_review}. The low-rank regime of the matrix factorization problem, when $r$ remains finite in the large $n$ limit, is now rather well understood~\cite{RaFl12,LeKrZd17,LeMi19,BaMa19,MoWe22}, with a precise characterization of the information-theoretic optimal accuracy of estimation, as well as the performances of polynomial time algorithms for this task. By contrast the extensive rank regime, with $r$ proportional to $n$ in the large size limit, presents additional challenges that are not yet all resolved. The initial proposal of~\cite{KaKrMeSaZd16} was later unveiled to be incorrect~\cite{Sc18}, and despite a series of recent efforts~\cite{MaKrMeZd22,BaMa22,TrErKrMaZd22,PoBaMa23,BaKoRa24,PoMa23,PoMa23b,LaMeGa23,CaMe23a,CaMe23b} the current understanding of this regime is not satisfactory. For this reason it is worth coming back to the simpler matrix denoising problem.

The latter was actually studied in a landmark paper~\cite{BuAlBoPo16} by Bun, Allez, Bouchaud and Potters (BABP). Their construction of an estimator $\hS(Y)$ can be roughly summarized as follows. Imposing that $\hS(Y)$ has the same eigenvectors as $Y$, they fix its eigenvalues in order to minimize the square loss distance between $\hS(Y)$ and $S$, assuming for the sake of the argument that $S$ is known. This yields an expression for the eigenvalues of $\hS(Y)$ in terms of those of $Y$ and of the overlaps between the eigenvectors of $S$ and those of $Y$. Of course such an ``oracle'' estimator is a priori not satisfactory, since it requires the knowledge of $S$ which is precisely the unknown that has to be inferred; however in the large $n$ limit the concentration of measure phenomenon on these eigenvector overlaps has a very fortunate consequence: the optimal eigenvalues of $\hS(Y)$ become deterministic functions of those of $Y$, that can be computed from the asymptotic eigenvalue distributions of $S$ and $Z$. Since by construction $\hS(Y)$ shares the eigenbasis of $Y$ this estimator satisfies $\hS(O Y O^T) = O \hS(Y) O^T$ for all orthogonal matrices $O$, a property called rotational invariance in~\cite{BuAlBoPo16} (we shall use instead the term equivariance, that seems more precise from the point of view of invariant theory).

The derivation of~\cite{BuAlBoPo16} relies partly on the heuristic replica method to compute the eigenvector overlaps (the rigorous results of~\cite{LePe11} confirming its validity in a special case), and is not explicitly Bayesian (this point was later refined in~\cite{BuBoPo17}), in the sense that it does not postulate a prior distribution for the signal $S$. The present work revisits the denoising problem with these two aspects in mind. We introduce explicitly prior distributions on the signal $S$ and noise $Z$, and discuss the asymptotic Bayes-optimality (in the sense of the mean square error) of the BABP denoiser. This optimality is to be expected when both priors are orthogonally invariant: the Bayes-optimal estimator is then equivariant, and the BABP denoiser should be the best among the equivariant ones since it emulates the oracle one that cannot be outperformed. We shall indeed confirm this intuition to a large extent: we consider the class of estimators for which the matrix elements of $\hS(Y)$ are polynomials in the matrix elements of $Y$, of total degree at most $D$, and describe those that achieve the minimal mean square error. When the degree constraint $D$ grows to infinity (after the large size $n\to\infty$ has been taken) these optimal estimators do indeed coincide with the BABP one. This Bayesian optimality was also derived in~\cite{MaKrMeZd22} in a special case, through a relatively involved computation involving the asymptotic behavior of Harish-Chandra-Itzykson-Zuber integrals~\cite{HC57,ItZu80,Ma94,GuZe02,BuBoMaPo14}. In contrast the approach of the present manuscript is much more elementary and flexible, since it will allow us to discuss the optimal polynomial estimators of finite degree $D$, their finite $n$ corrections, and maybe more importantly the universality of this optimality. We shall indeed conjecture, and prove in a restricted setting, that the BABP denoiser is asymptotically optimal even if the prior distributions are not strictly orthogonally invariant.

From a methodological point of view this work borrows several ideas to the so-called low-degree polynomial approach~\cite{HoSt17,KuWeBa19,BrHu22,ScWe22,MoWe22}, that handles hypothesis testing and estimation problems by constructing the best polynomial approximations of complicated objective functions (the likelihood ratio in the first setting, the posterior mean in the second one). We will also discuss the connections between this estimation approach and what are called Nishimori identities in the literature inspired by the statistical mechanics of disordered systems~\cite{ZdKr16_review}.

The rest of the manuscript is organized as follows. In Section~\ref{sec_definitions} we define precisely the problem at hand and state the main results to be derived. Sec.~\ref{sec_general} discusses, in a generic setting, optimal Bayesian estimation, its approximations, and the role symmetries play in this context. This generic formalism is applied to the matrix denoising problem with orthogonally invariant priors in Sec.~\ref{sec_rot}, with in particular numerical results illustrating one specific example presented in Sec.~\ref{sec_numerics_On}. We discuss in Section~\ref{sec_universality} the universality of these results in the case of an arbitrary Wishart matrix perturbed by an arbitrary Wigner one. We present numerical simulations on finite size matrices in Sec.~\ref{sec_numerics_finiten}. Some conclusions are drawn in Sec.~\ref{sec_conclu}, more technical aspects being deferred to a series of Appendices.

\section{Main results}
\label{sec_definitions}

\subsection{Assumptions}
\label{sec_definitions_def}

We shall consider the matrix denoising problem with an additive noise modelization, where an observation $Y^{(n)}$ is generated according to $Y^{(n)}=S^{(n)} + Z^{(n)}$, with $S^{(n)}$ a signal and $Z^{(n)}$ considered as noise. $Y^{(n)}$, $S^{(n)}$ and $Z^{(n)}$ are (sequences of) random variables taking values in $M_n^{\rm sym}(\R)$, the set of real symmetric square matrices of size $n$. We are interested in the large size limit $n \to \infty$, and will sometimes drop the superscript indicating the size of the matrices to improve the readability. 

We assume that $S^{(n)}$ and $Z^{(n)}$ are independent, and that their empirical eigenvalue distribution converge in the large $n$ limit to two compactly supported probability distributions denoted $\mu_S$ and $\mu_Z$ respectively, in the sense of convergence of average moments. Since the moments of the empirical eigenvalue distribution of a matrix coincide with the (normalized) trace of its powers, we write this assumption as
\beq
\lim_{n \to \infty} \E [ \tr( (S^{(n)})^p  ) ] = \mu_{S,p} \qquad \forall \, p \in \mathbb{N} \ ,
\label{eq_convergence}
\eeq
where $\tr(\bullet) = \frac{1}{n}\Tr(\bullet)$ is the normalized trace, and where $\mu_{S,p} = \int \mu_S(\dd \lambda) \, \lambda^p$ is the $p$-th moment of the probability measure $\mu_S$. The same assumption holds for the convergence of $Z^{(n)}$, the limiting measure being denoted $\mu_Z$ and its moments $\mu_{Z,p}$.

We will actually impose additional properties on the large $n$ behavior of the matrices $S^{(n)}$ and $Z^{(n)}$, and assume that the normalized traces of their powers, $\tr((S^{(n)})^p)$, have Gaussian fluctuations around their averages, of order $1/n$. This hypothesis will be written more precisely as
\begin{align}
\lim_{n\to \infty} \E[ (\Tr(S)  & - \E[\Tr (S) ])^{q_1} (\Tr(S^2) - \E[\Tr (S^2) ])^{q_2}  \dots (\Tr(S^D) - \E[\Tr (S^D) ])^{q_D} ] 
 \nonumber \\ & = \E[(G_{S,1})^{q_1}(G_{S,2})^{q_2} \dots (G_{S,D})^{q_D}] \ ,
\label{eq_fluctuations}
\end{align}
for all $D \ge 1$ and $q_1,\dots q_D$ non-negative integers, where $\{G_{S,i}\}_{i \ge 1}$ is a centered Gaussian vector with a non-trivial covariance matrix (in the sense that all its finite-dimensional truncations are positive definite); we kept implicit the size dependency of $S=S^{(n)}$ for conciseness, and make the analogous assumption for the fluctuation moments of $Z^{(n)}$, with another Gaussian vector $\{G_{Z,i}\}$ describing their limit. Such results are known in the random matrix theory literature under the name of Central Limit Theorem (CLT) for linear statistics, and have been established for several random ensembles, including the Wigner and Wishart ones~\cite{Jo82,KhKhPa96,AnZe06,AnGuZe_book,BaSi_book}. They are also referred to as the existence of a second-order limit distribution in the context of second-order free probability~\cite{MiSp06,Re14,MiPo13,MaMiPeSp22}. Note that $\tr((S^{(n)})^p)$, which is a quantity of order 1, has fluctuations around its mean on the scale $1/n$ and not $1/\sqrt{n}$ as one could expect from the usual CLT: this difference is due to the strong interactions between the eigenvalues of random matrices, that fluctuate much less than independent random variables. Note also that (\ref{eq_fluctuations}) does not hold when $S$ is a low-rank projector with a finite number of non-zero eigenvalues of order 1, in other words the regime of low-rank matrix factorization of~\cite{RaFl12,LeKrZd17,LeMi19,BaMa19} is excluded in the following.

In the first part of the paper we assume a rotational invariance for the laws of $S$ and $Z$, namely $O S^{(n)} O^T \eqd S^{(n)} $ and $O Z^{(n)} O^T \eqd Z^{(n)} $ for any matrix $O\in \cO_n = \{O \in M_n(\R) : OO^T = O^TO=\one_n \}$ the orthogonal group, and with $\eqd$ meaning equality in distribution of random variables.

As a consequence of the hypotheses made so far $S^{(n)}$ and $Z^{(n)}$ are asymptotically free (in the sense of free probability~\cite{Vo91}, see for instance~\cite{PoBo_book,MiSp_book,NiSp_book}), hence the empirical eigenvalue distribution of $Y^{(n)}=S^{(n)} + Z^{(n)}$ converges in the large $n$ limit to a probability measure $\mu_Y$ (in the sense of equation~(\ref{eq_convergence})), the latter being given by the free additive convolution of $\mu_S$ and $\mu_Z$, denoted $\mu_Y=\mu_S \boxplus \mu_Z$ (see Appendix~\ref{app_fp} for an explicit definition of this operation). As a matter of fact the concentration hypothesis expressed in Eq.~(\ref{eq_fluctuations}) allows to extend the proof of the asymptotic freeness between Haar distributed random matrices and deterministic ones to the random case, see remark 5.14 in~\cite{MiSp_book} for more details on this point, or theorem 54 in~\cite{MiPo13}. Moreover the theory of second-order freeness~\cite{MiSp06,Re14,MiPo13,MaMiPeSp22} shows that the fluctuations of the powers of $Y^{(n)}=S^{(n)} + Z^{(n)}$ (or more generically of any non-commutative polynomials in $S^{(n)}$ and $Z^{(n)}$) have an asymptotic Gaussian distribution of the form (\ref{eq_fluctuations}) (with a covariance structure that can be in principle computed from the one of $G_S$ and $G_Z$), see in particular theorem 54 in~\cite{MiPo13}.

We shall call ``Gaussian noise'' the special case in which $Z^{(n)} = \sqrt{\Delta} B^{(n)}$, with $\Delta \ge 0$ the noise intensity, and $B^{(n)}$ drawn from the standard Gaussian Orthogonal Ensemble (GOE), which means that $B^{(n)}\in M_n^{\rm sym}(\R)$ with $\{B_{i,j}^{(n)} \}_{i \le j}$ independent centered Gaussian random variables of variance $1/n$ for $i < j$, and $2/n$ for $i=j$. It is well known that the GOE satisfies all the hypotheses spelled out above, see e.g.~\cite{AnGuZe_book}.

Our interest in this paper focuses on estimators of the signal given the observations, i.e. functions $\hS : M_n^{\rm sym}(\R) \to M_n^{\rm sym}(\R)$ such that $\hS(Y^{(n)})$ is ``close'' to the (unobserved) signal $S^{(n)}$. We measure this ``closeness'', and hence the quality of the estimator, in terms of the normalized sum of the Mean Square Error (MSE) for each matrix element, namely
\beq
\MSE(\hS) 
= \frac{1}{n} \sum_{i,j=1}^n \E[(S_{i,j} - \hS(Y)_{i,j})^2 ]  
= \E[\tr((S -\hS(Y))^2)] \ ,
\label{eq_MSE_matrix}
\eeq
where the average is over the laws of $S^{(n)}$ and $Z^{(n)}$ (the size superscript being kept implicit for clarity).
The estimator that minimizes the MSE, called Bayes-optimal in this context, is well-known to be the posterior mean, $\hS(Y)_{i,j} = \E[S_{i,j}|Y]$ for each matrix element (an explicit derivation of this fact will be given in Sec.~\ref{sec_gene_optimal}). However this conditional expectation is in general very difficult to compute, the paper is thus devoted to approximations and simplifications of this optimal estimator, in particular in the large $n$ limit.

Note that we place ourselves in the Bayesian setting in which the observer knows the probability laws of the signal and noise, and in particular their asymptotic distributions $\mu_S$ and $\mu_Z$; the different problem of reconstructing $\mu_S$ from the knowledge of $\mu_Z$ and $\mu_Y = \mu_S \boxplus \mu_Z$ is considered in~\cite{Ta20}.

\subsection{The BABP denoiser}

Let us now describe the prescription for denoising introduced in~\cite{BuAlBoPo16} in the additive case (\cite{BuAlBoPo16} also deals with multiplicative noise). For a given model specified by the asymptotic distributions $\mu_S$ and $\mu_Z$ (which sets as a consequence $\mu_Y=\mu_S \boxplus \mu_Z$), consider the function $\tDB : \R \to \R$ defined by
\beq
\qquad \tDB(\lambda) = \frac{\underset{\eta \to 0^+}{\lim} \im( g_Y(\lambda-i \, \eta) R_Z(g_Y(\lambda-i \, \eta))  ) }{\underset{\eta \to 0^+}{\lim}\im(g_Y(\lambda-i \, \eta)) } \ ,
\label{eq_def_DB}
\eeq
where we denote $z = \re(z) + i \, \im(z)$ the decomposition of a complex number into its real and imaginary part, $g_Y$ is the Cauchy transform of $\mu_Y$ and $R_Z$ the $R$-transform of $\mu_Z$ (see Appendix~\ref{app_fp} for definitions). In the special case of Gaussian noise one has $R_Z(g) = \Delta g$, this definition simplifies into
\beq
\tDB(\lambda) = 2 \Delta \underset{\eta \to 0^+}{\lim} \re( g_Y(\lambda-i \, \eta) )\ ,
\label{eq_def_DB_Gaussian}
\eeq
which corresponds to the Hilbert transform of the density of $\mu_Y$ (up to a normalization constant).

Call now $\DB : \R \to \R$ the function
\beq
\DB(\lambda) = \lambda - \tDB(\lambda) \ ,
\label{eq_def_DB_2}
\eeq
this decomposition being convenient in the following. The estimator proposed in~\cite{BuAlBoPo16} is then $\hS(Y) = \DB(Y)$, where here and in the following it is understood that a scalar function from $\R$ to $\R$ applied to a symmetric matrix acts on the eigenvalues and leaves the eigenvectors unchanged. This means that if $Y$ is diagonalized as $Y=P \, \diag(\lambda_1,\dots,\lambda_n) \, P^T$ with $P\in \cO_n$, one has by convention
\beq
\DB(Y) = P \, \diag(\DB(\lambda_1),\dots,\DB(\lambda_n) ) P^T \ ,
\label{eq_convention1}
\eeq
in agreement with the standard rules of functional calculus; note in particular that when the function is a polynomial its action on a matrix coincide with the polynomial where powers are interpreted in the matrix multiplication sense.

\subsection{Optimality of the BABP denoiser for orthogonally invariant priors}
\label{sec_definitions_res}

Our first main achievement is a partial justification of the asymptotic Bayes-optimality of the BABP denoiser, in the following sense. Let us define $\MMSE^{(n,D)}$ as the minimal MSE among polynomial estimators $\hS$ of degree at most $D$, for matrices of size $n$; we mean by this that each matrix element $\hS(Y)_{i,j}$ of the estimator is a polynomial in the matrix elements of $Y^{(n)}$, of total degree at most $D$. We denote $\MMSE^{(D)}$ the limit of $\MMSE^{(n,D)}$ as $n$ goes to infinity. Thanks to the invariance of the laws of $S^{(n)}$ and $Z^{(n)}$ under conjugation by orthogonal matrices, $\MMSE^{(n,D)}$ will be achieved by an equivariant estimator, i.e. such that $\hS(OYO^T) = O \hS(Y)O^T$ for all $O \in \cO_n$ (the role of symmetry in Bayesian estimation is discussed in a generic setting in Sec.~\ref{sec_gene_sym}). Such equivariant polynomials are shown in Appendix~\ref{app_equivariant} to be linear combinations of terms of the form $Y^p (\Tr \, Y)^{q_1} (\Tr(Y^2))^{q_2} \dots$ for non-negative integers $p,\{q_i\}$, with $p+\sum_i i q_i$ the total degree of the polynomial. We shall call scalar polynomial estimators the combination of these terms with all $q_i=0$, in other words those for which $\hS(Y) = \D(Y)$ is a polynomial in the matrix $Y$ itself, and denote $\MMSE^{(n,D,{\rm s})}$ the minimal MSE among scalar polynomial estimators of degree at most $D$, and $\MMSE^{(D,{\rm s})}$ its $n\to\infty$ limit.

We show in Section~\ref{sec_rot_largen} how to compute the coefficients of the polynomial $\D^{(D)}$ of degree at most $D$ that achieves $\MMSE^{(D,{\rm s})}$ (asymptotically in the large $n$ limit) from the definition of the model (i.e. $\mu_S$ and $\mu_Z$) by solving a linear system (of size $D+1$) explicitly given in Sec.~\ref{sec_rot_expressions}. The connection with the BABP prescription is made in Sec.~\ref{sec_rot_largeD}: we show there that $\D^{(D)}$, the optimal scalar polynomial of degree at most $D$, is the orthogonal projection of $\DB$ on the subspace of polynomials of degree $\le D$, within the $L^2(\R,\mu_Y)$ Hilbert space. This implies that, as $D$ goes to infinity after $n \to \infty$, one has $\D^{(D)} \to \DB$ (in the $L^2$ sense, hence $\mu_Y$ almost everywhere), in other words the optimal scalar polynomial estimator becomes indeed the BABP one when the constraint on its degree is released. We also obtain very simple expressions of the MMSEs in Sec.~\ref{sec_rot_MMSE}, namely
\beq
\MMSE^{(D,{\rm s})} = \mu_{S,2} - \int \mu_Y(\dd \lambda) (\D^{(D)}(\lambda))^2 = \mu_{Z,2} - \int \mu_Y(\dd \lambda) (\tD^{(D)}(\lambda))^2 \ ,
\eeq
with $\D^{(D)}(\lambda) = \lambda - \tD^{(D)}(\lambda)$. Taking $D \to \infty$ in this expression yields, thanks to the convergence of $\D^{(D)}$ in the $L^2$ sense, the expressions of the MSE of the BABP estimator,
\beq
\MMSEB = \mu_{S,2} - \int \mu_Y(\dd \lambda) (\DB(\lambda))^2 = \mu_{Z,2} - \int \mu_Y(\dd \lambda) (\tDB(\lambda))^2 \ ;
\label{eq_MMSEB_first}
\eeq
these were previously known, to the best of our knowledge, only in the Gaussian noise case~\cite{MaKrMeZd22,PoBaMa23} and under slighlty different forms, which we show in in Sec.~\ref{sec_rot_MMSE} to be equivalent to ours (we also point out that the first expression in Eq.~(\ref{eq_MMSEB_first}) can be derived rather directly from the computations of~\cite{BuAlBoPo16}, even if it is not written explicitly in that paper).

We finally show in Sec.~\ref{sec_irrelevance} that $\MMSE^{(D)} = \MMSE^{(D,{\rm s})}$, in other words that the inclusion in the estimator of the prefactors proportional to the traces of powers of $Y$ does not improve (asymptotically) the accuracy of the denoiser, by a careful control of the fluctuations of the random matrices. This concludes the justification of the asymptotic Bayes-optimality of the BABP denoiser, in the sense that it achieves $\lim_{D \to \infty} \lim_{n \to \infty}  \MMSE^{(n,D)}$ (the possible interversion of the order of these limits is further discussed in Sections \ref{sec_numerics_finiten} and \ref{sec_conclu}).

We illustrate our findings with numerical results in Sec.~\ref{sec_numerics_On} in a case motivated by the matrix factorization problem, namely a signal drawn as a (Gaussian) Wishart matrix perturbed by a Gaussian noise. We show in particular that for a large range of parameters a satisfactory accuracy can be achieved with rather small values of $D$.

\subsection{Universality conjecture}
\label{sec_conjecture}

The results presented up to now are derived under the assumption of strict orthogonal invariance for the laws of the signal and the noise. One should however expect that they are valid in a wider setting, since random matrix theory exhibits an aboundance of universality results~\cite{AnGuZe_book,BaSi_book,PoBo_book}. For instance, the asymptotic semi-circular distribution appears not only in the GOE case, but more generically for all Wigner matrices with independent entries (not necessarily Gaussian), as long as their distribution has a sufficiently fast decaying tail. Similarly the Marcenko-Pastur distribution emerges in the study of large Wishart matrices, be they Gaussian or not (under again some moment conditions). One can certainly argue that these universality results for the eigenvalue distributions are not enough in the denoising problem, which should also be sensitive to the behavior of the eigenvectors of these random matrices. But it turns out that universality also shows up in several properties more refined than eigenvalue distributions:
\begin{itemize}
\item Wigner and Wishart matrices have eigenvectors that are asymptotically delocalized and approximately orthogonally invariant in the large size limit~\cite{BaMiPa07,ErScYa09,Be11,Bo13,OrVuWa16,BaSi_book}. 
\item the fluctuations described in (\ref{eq_fluctuations}) have covariance matrices that are not universal~\cite{Jo82,KhKhPa96,AnZe06,MaMiPeSp22}, but their scaling with $n$, which is the only ingredient we use in Sec.~\ref{sec_irrelevance},  and the Gaussianity of the limit are independent of the microscopic details of the random matrix ensembles.
\item the results of~\cite{LePe11} on the overlap between the eigenvectors of the signal and observation matrices (for multiplicative noise), that underlies the approach of~\cite{BuAlBoPo16}, are once again largely universal.
\item non-trivial estimation problems and algorithmic approaches like Approximate Message Passing have been shown to exhibit rather wide universality classes, see for instance~\cite{DuSeLu22,WaZhFa22}.
\end{itemize}

One can thus surmise that the asymptotic Bayes-optimality of the BABP estimator holds beyond the strictly orthogonally invariant case, as long as the laws of the signal and noise are asymptotically ``invariant enough'', in a sense that should be specified (a minimal requirement being that these two sequences of matrices are asymptotically free, see~\cite{AnFa14} for results on asymptotic freeness under conjugation by subgroups of the orthogonal group), universality classes depending only on the asymptotic eigenvalue distribution $\mu_S$ and $\mu_Z$. We investigate this point in Sec.~\ref{sec_universality}, focusing on a specific case, namely the denoising of an arbitrary Wishart matrix corrupted by an arbitrary Wigner noise (following again the motivation of the matrix factorization). It turns out that in this setting there is no universality with respect to the noise distribution: since in a Wigner matrix the elements $Z_{i,j}$ that corrupt each entry of the signal are independent of each other and have a sizable effect on the corresponding observation $Y_{i,j}$, their complete distribution matters, and not only their first two moments. On the contrary in the signal drawn as a Wishart matrix $X X^T$ the elements of $X$ are not observed directly but only through sums containing $r$ terms; when the rank $r$ is proportional to $n$ in the large size limit this mechanism can be expected to wash out the details of the distribution of the matrix elements of $X$. In this context the conjecture is that the asymptotic Bayes-optimal estimator, and its associated MSE, does not depend on the details of the laws of the matrix elements of $X$ (provided all their moments exist and once their lowest moments are normalized in such a way to ensure the universality of $\mu_S$), whenever $r/n$ tends to a strictly positive constant. If in addition the noise matrix is extracted from the GOE a corollary of this conjecture is the optimality of the BABP denoiser (since one can replace the arbitrary law of the elements of $X$ by a Gaussian and recover orthogonally invariant priors on $S$ and $Z$). More precisely, a weak version of the conjecture would restrict its validity to estimators of finite but arbitrary degree $D$, while the strong version would posit the universality to hold for all estimators. Sec.~\ref{sec_universality} is organized as follows. In Sec.~\ref{sec_universality_def} we define precisely the signal and noise distributions; we explain in Sec.~\ref{sec_nonuniversal_noise} the absence of universality in the noise, using a scalar denoising problem as a toy model. After presenting in Sec.~\ref{sec_universality_equivariant} the generalized form of the estimators that are equivariant when the orthogonal group is replaced by the permutation one, we show in Sec.~\ref{sec_order2} that the conjecture is true for $D=2$, and compute the $1/n$ corrections to $\MMSE^{(n,2)}$ (that are not universal). We then discuss the case $D=3$ in Sec.~\ref{sec_order3}, and show in particular how the denoiser of degree 3 is to be amended when the noise is not Gaussian. We finally outline in Sec.~\ref{sec_universality_allD} a possible path towards a proof of the weak version of the conjecture for $D$ arbitrary, and the challenges it presents. 

\section{Optimal and approximate Bayesian estimation}
\label{sec_general}

This Section reviews some basic notions on Bayesian estimation and its approximation via the low-degree polynomial approach; most of these results appear in~\cite{ScWe22,MoWe22} in specific cases, for the sake of pedagogy and self-containedness we present them here in a generic perspective.

\subsection{Bayes-optimal estimation}
\label{sec_gene_optimal}

Consider a pair of random variables $(S,Y)$, with $S\in \R^N$ and $Y \in \R^M$, to be interpreted as, respectively, a signal and some observations. We are interested in estimators $\hS(Y)$, that predict the value of the signal as a function of the observations. A choice of a distance between $S$ and $\hS(Y)$ must be made in order to discuss the accuracy of the estimation; in the following we will use the standard square error, denoting $\la S,T \ra = \sum_{i=1}^N S_i T_i$ the canonical scalar product in $\R^N$, and $|| \cdot ||$ the associated norm. The accuracy of an estimator will thus be measured by its Mean Square Error (MSE),
\beq
\MSE(\hS) = \E[||S-\hS(Y)||^2] = \sum_{i=1}^N \E [ (S_i - \hS_i(Y))^2 ]\ .
\label{eq_MSE}
\eeq
With this choice the ``optimal'' estimator (in the sense that it minimizes the MSE) is $\hSopt(Y)=\E[S|Y]$, the posterior mean; to justify this well-known fact, write for an arbitrary estimator $\hS$
\begin{align}
\E[(S_i-\hS_i(Y))^2] &= \E[(S_i- \E[S_i|Y] + \E[S_i|Y] -\hS_i(Y))^2] \nonumber \\ 
& = \E[(S_i- \E[S_i|Y])^2] + 2 \E[ (S_i- \E[S_i|Y] ) (\E[S_i|Y] -\hS_i(Y))  ] + \E[ (\E[S_i|Y] -\hS_i(Y))^2] \ .
\label{eq_Si}
\end{align}
The second term in this equation vanishes: by construction of the conditional expectation $\E[AB]=\E[\E[A|Y] B ]$ if $B$ depends only on $Y$ (in technical terms, if $B$ is $\sigma(Y)$-measurable), one can thus apply this identity with $A = S_i- \E[S_i|Y]$ and $B=\E[S_i|Y] -\hS_i(Y)$, which is indeed $\sigma(Y)$-measurable, and note that $\E[A|Y]=0$ (almost surely). Since the first term of (\ref{eq_Si}) does not depend on $\hS$, the minimization of the third term yields $\hS_i(Y) = \E[S_i|Y]$, for all the indices $i$.

The Minimal MSE (MMSE) is thus reached by the posterior mean and reads
\beq
\MMSE = \E[||S-\E[S|Y]||^2] = \E[||S||^2] - \E[||\E[S|Y]||^2] \ ,
\eeq
where the second form is easily derived from the properties of the conditional expectation stated above.

Denoting $\hSopt(Y)=\E[S|Y]$ the optimal estimator one has
\beq
\E[\hSopt(Y) \varphi(Y)  ] = \E[S \varphi(Y)] \ ,
\label{eq_PMI_scalar}
\eeq
for an arbitrary test function $\varphi : \R^M \to \R$, using again the defining property of the conditional expectation. This family of identities are referred to as the orthogonality principle in the signal processing context, and can be viewed as a form of the so-called Nishimori identities~\cite{ZdKr16_review}. The latter correspond to the observation that if $S'$ is a random variable drawn conditionally on $Y$ from the posterior law $P(\cdot|Y)$ then $(S,Y) \eqd (S',Y)$, which yields (\ref{eq_PMI_scalar}) by interpreting $\E[\hSopt(Y) \varphi(Y)]$ as $\E[S' \varphi(Y)]$.  In many inference problems the methods of statistical mechanics~\cite{ZdKr16_review} (replica, cavity, Belief Propagation and Approximate Message Passing) provide asymptotically exact formulas for the Bayes-optimal estimator. In these cases the identities (\ref{eq_PMI_scalar}) are essentially self-consistency checks of the approach. On the contrary the extensive rank matrix factorization problem, and the associated denoising one, has eluded up to now such approaches; in such a case these identities can be exploited to derive systematic approximations to the optimal estimator, as we shall see in the following.

\subsection{Approximations}
\label{sec_gene_approx}

A direct computation of the optimal estimator $\hSopt(Y)=\E[S|Y]$ is often impossible, in particular in high-dimensional settings in which intricate dependencies between $S$ and $Y$ develop. In this case one can aim more modestly at an approximate estimator,
\beq
\hS(Y) = \sum_{\beta \in \A} c_\beta b_\beta(Y) \ ,
\label{eq_CL}
\eeq
where the $b_\beta$ are simple enough functions from $\R^M$ to $\R^N$ (for instance polynomials, but the reasoning here is more general), $\A$ is a finite index set, and the $c_\beta$ are free parameters in this linear combination. The latter should be chosen in such a way that $\hS$ is as close as possible to $\hSopt$; a first idea to achieve this goal would be to impose the identities (\ref{eq_PMI_scalar}) on $\hS$ with the largest possible set of test functions $\varphi$, but it is not immediately obvious how to choose the latter in such a way to get many compatible constraints. To resolve this apparent ambiguity a simple and well-founded strategy is instead to minimize the MSE with respect to the coefficients of the linear combination. Inserting the definition (\ref{eq_CL}) in the expression (\ref{eq_MSE}) of the $\MSE$ yields
\beq
\MSE(\hS) = \E[||S||^2] + \sum_{\beta,\beta' \in \A} c_\beta \M_{\beta,\beta'} c_{\beta'} - 2 \sum_{\beta \in \A} \cR_\beta c_\beta = \E[||S||^2] + c^T \M c -2 \cR^T c\ ,
\label{eq_MSE_cb}
\eeq
where $\M$ is a square matrix and $\cR$ a vector, both of size $|\A|$ , with elements given by
\beq
\M_{\beta,\beta'} = \E[\la b_\beta(Y),b_{\beta'}(Y) \ra ] \ , \qquad
\cR_\beta = \E[\la S,b_\beta(Y) \ra] \ .
\eeq
Let us write $\MMSE_\A$ for the minimal MSE achievable with an estimator built from linear combinations of basic functions indexed by $\A$, which can be expressed as
\beq
\MMSE_\A = \E[||S||^2] + \inf_{c \in \R^{|\A|}}[c^T \M c -2 \cR^T c] \ .
\eeq
Imposing the stationarity of this quadratic plus linear form with respect to $c$ yields the conditions
\beq
\sum_{\beta' \in \A} \M_{\beta,\beta'} c_{\beta'} = \cR_\beta \quad \forall \beta \in \A \ ,
\label{eq_conditions}
\eeq
or equivalently in vectorial notations $\M c = \cR$. $\M$ being a Gram matrix it is clearly positive semi-definite; suppose first that it is definite, i.e. that all its eigenvalues are strictly positive. Then $\M$ is invertible, and the function of $c$ defined in (\ref{eq_MSE_cb}) has a unique minimum, reached in  $c = \M^{-1} \cR$, yielding
\beq
\MMSE_\A = \E [||S||^2 ] - \cR^T  \M^{-1} \cR = \E [||S||^2 ] - \cR^T c \ ,
\label{eq_MMSE_A}
\eeq
where in the last form it is understood that $c$ is solution of $\M c = \cR$.

If instead $\M$ has some zero eigenvalues there exists an infinity of solutions to the system of linear equations $\M c = \cR$, forming the affine space made of the kernel of $\M$ translated by an arbitrary solution of $\M c = \cR$; all these solutions correspond to the degenerate minima of (\ref{eq_MSE_cb}), and yield the same optimal value of the MSE. To justify this claim one can diagonalize $\M$ in an orthonormal basis, $\M={\cal P}^T {\rm diag}(\lambda_1,\dots,\lambda_{|\A|}) {\cal P}$, and rotate correspondingly the vector $\cR$, defining $\cR' = {\cal P} \cR$. The minimization of the MSE can then be performed independently in each direction,
\beq
\MMSE_\A = \E[||S||^2] + \sum_{i=1}^{|\A|} \inf_{c \in \R} [\lambda_i c^2 - 2 \cR'_i c ] \ .
\eeq
It turns out that if $\lambda_i=0$ then necessarily $\cR'_i=0$, hence the MSE is constant along the directions corresponding to zero eigenvalues of $\M$. Consider indeed such an eigenvector of $\M$, whose elements are given by $\{{\cal P}_{i,\beta}\}_{\beta \in \A}$, and form the linear combination $f(Y)=\sum_\beta  {\cal P}_{i,\beta} b_\beta(Y)$. Then $\E[||f(Y)||^2]=0$, hence $f(Y)=0$ almost surely, and as a consequence $\cR'_i=\E[\la S,f(Y)\ra]=0$. Alternatively one can notice that by definition (\ref{eq_MSE_cb}) is bounded below by zero, hence along a direction with vanishing curvature the linear term must also vanish, otherwise the latter could take arbitraly large negative values, and thus violate the lower bound on $\MSE(\hS)$. A particular solution of $\M c = \cR$ can thus always be found as $c = \M^+ \cR$, where $\M^+$ denotes the Moore-Penrose pseudo-inverse of $\M$.

Note that the conditions (\ref{eq_conditions}) can be rewritten as
\beq
\E[ \la \hS(Y) , b_\beta(Y) \ra ] = \E[ \la S , b_\beta(Y) \ra ] \quad \forall \beta \in \A \ ,
\eeq
in other words the minimization of the MSE can be interpreted as imposing the identities (\ref{eq_PMI_scalar}) to $\hS$ with the test functions corresponding to the $b_\beta$ themselves (or more precisely to the scalar products with the $b_\beta$'s).

Let us make a final remark, formalized as lemma B.2 in~\cite{MoWe22}, that will be useful several times later on; one is often interested in cases in which the law of $(S,Y)$ (and possibly also $N$ and $M$) depends on some additional parameter, to be denoted $n$, while $\A$ is independent from $n$. We will denote $\M^{(n)}$ and $\cR^{(n)}$ the matrix and vector describing the optimization problem for a given value of $n$, $c^{(n)}$ an arbitrary solution of $\M^{(n)} c^{(n)} = \cR^{(n)}$, and $\MMSE_\A^{(n)}$ the corresponding optimal MSE. Suppose now that, as $n \to \infty$, $\M^{(n)}$ and $\cR^{(n)}$ converge to some finite limits denoted $\M^{(\infty)}$ and $\cR^{(\infty)}$ (since these are finite-dimensional objects this corresponds to the convergence of each entry of $\M^{(n)}$ and $\cR^{(n)}$), and assume in addition that $\M^{(\infty)}$ is invertible. Then $\MMSE_\A^{(n)}$ converges as $n\to \infty$, and its limit can be written as
\beq
\left( \lim_{n \to \infty} \E [||S||^2 ] \right)  - \left(\cR^{(\infty)}\right)^T \left(\M^{(\infty)}\right)^{-1} \cR^{(\infty)} \ .
\label{eq_MMSE_limit}
\eeq
Indeed for all $n$ beyond a finite index the matrices $\M^{(n)}$ are invertible, hence the optimum $c^{(n)}$ is uniquely defined as $\left(\M^{(n)}\right)^{-1} \cR^{(n)}$, and (\ref{eq_MMSE_limit}) directly follows from the continuity of the matrix inversion when restricted to the set of invertible matrices. Even if the final statement might seem obvious note that the justification of the exchange of order between the limit $n\to\infty$ and the minimization over the free parameters $c$ relies crucially on the hypothesis of invertibility of the limit $\M^{(\infty)}$. It would indeed be violated (even interpreting the inverse in (\ref{eq_MMSE_limit}) in the Moore-Penrose sense) if an eigenvalue $\lambda_i^{(n)}$ of $\M^{(n)}$ is strictly positive for all finite $n$ and vanishes as $n\to\infty$, with a corresponding linear coefficient ${\cR'}^{(n)}_i$ vanishing proportionally to $\sqrt{\lambda_i^{(n)}}$. Such a direction would indeed contribute to the asymptotic MMSE, and contradict the statement (\ref{eq_MMSE_limit}) .

\subsection{Symmetries}
\label{sec_gene_sym}

Both the complexity of the computation presented above and the quality of the approximation achieved increases with the number $|\A|$ of basic functions considered, a compromise should thus be found between these two aspects. We shall see now that if the problem at hand exhibit some symmetries one can reduce the set of candidate basic functions without altering the quality of the approximation, it is thus wise to exploit this phenomenon.

To formalize this notion let us first introduce some definitions and notations. We consider a group $G$ denoted multiplicatively, with neutral element written $e$, and two linear representations of $G$ on $\R^N$ and $\R^M$, denoting them as group actions. Namely for $g \in G$ and $S \in \R^N $ we write $g \cdot S \in \R^N$ the image of $S$ under the transformation $g$, with the properties $e \cdot S = S$, $(gh) \cdot S = g \cdot (h \cdot S)$, and $g \cdot (u S + v T) = u (g \cdot S) + v (g \cdot T)$ whenever $u,v \in \R$, $S,T \in \R^N$. We use the same symbol $g \cdot Y$ to denote the action of $G$ on $\R^M$, even if in general the two representations are different. We shall further assume that the action on $\R^N$ is isometric, $\la g \cdot S , g \cdot T \ra = \la S, T \ra$. A function $f : \R^M \to \R^N$  is called equivariant if $f(g\cdot Y) = g \cdot f(Y)$ for all $g \in G$ and $Y \in \R^M$, invariant if $f(g\cdot Y) = f(Y)$.

With these preliminary definitions clarified we can come back to the inference problem; we will say that the group $G$ (and its representations) is a symmetry for it if and only if $(g\cdot S , g \cdot Y) \eqd (S,Y)$ for all $g \in G$, where we recall that $\eqd$ denotes the equality in distribution of random variables. One can then check that the optimal estimator $\hSopt(Y)=\E[S|Y]$  is equivariant.

Let us now turn to the interplay between symmetries and approximate estimation. Since $\hSopt$ is equivariant it is rather intuitive that one should approximate it as a linear combination of basic functions $b_\beta$ that are themselves equivariant. One can give a formal justification of this intuition as follows. Suppose one has a probability law on $G$, with the corresponding expectation denoted $\E_g$, that is invariant in the sense that $\E_g[\phi(gh)]=\E_g[\phi(hg)]=\E_g[\phi(g)]$ for all fixed $h \in G$ and functions $\phi : G \to \R$ (this would be the Haar measure for a compact group $G$). Then from an arbitrary function $f : \R^M \to \R^N$, we define its symmetrization as
\beq
f^{\rm eq}(Y) =\E_g[g^{-1} \cdot f(g \cdot Y) ] \ . 
\label{eq_projection_equi}
\eeq
Let us check that $f^{\rm eq}$ is equivariant: for a given $h \in G$,
\beq
f^{\rm eq}(h \cdot Y) =\E_g[g^{-1} \cdot f((gh) \cdot Y) ]
= h \cdot \E_g[ (gh)^{-1} \cdot f((gh) \cdot Y) ] = h \cdot f^{\rm eq}(Y) \ , 
\eeq
where we successively used $g \cdot (h \cdot Y)  = (gh) \cdot Y$, the linearity of the group action to commute it with $\E_g$, and the invariance of the law on $G$. Note that the map $f \to f^{\rm eq}$ is a projection in the vector space of functions from $\R^M$ to $\R^N$: it is indeed linear, and if $f$ is already equivariant then $f^{\rm eq}=f$, hence it is idempotent and projects to the subspace of equivariant functions. The latter can thus all be written as the symmetrization of some function.

Coming back to the quality of approximations in presence of symmetries, we shall now show that for an arbitrary estimator $\hS$ one has $\MSE(\hS^{\rm eq}) \leq \MSE(\hS)$, in other words the symmetrization cannot deteriorate the quality of the estimator. This is a Bayesian version of the Hunt-Stein lemma, whose proof can be found in the Appendix B.3 of~\cite{MoWe22};  for completeness we reproduce it here with some adaptations to the present setting. Let us first write explicitly the two MSEs to be compared:
\begin{align}
  \MSE(\hS) &= \E[||\hS(Y)||^2 ] - 2 \E[\la \hS(Y), S \ra] + \E[||S||^2] \ , \label{eq_HS1} \\
  \MSE(\hS^{\rm eq}) &= \E[||\hS^{\rm eq}(Y)||^2 ] - 2 \E[\la \hS^{\rm eq}(Y), S \ra] + \E[||S||^2] \label{eq_HS2} \ .
\end{align}
The third terms in (\ref{eq_HS1}) and (\ref{eq_HS2}) are obviously equal; this is also the case of the second terms, thanks to the symmetry of the inference problem:
\begin{align}
\E[\la \hS^{\rm eq}(Y), S \ra] & = \E[\la \E_g[ g^{-1} \cdot  \hS(g \cdot Y)], S \ra] = \E \E_g[ \la  g^{-1} \cdot  \hS(g \cdot Y), S \ra]= \E_g \E[ \la  \hS(g \cdot Y), g \cdot S \ra]= \E_g \E[ \la  \hS(Y), S \ra] \nonumber \\
& = \E[\la \hS(Y), S \ra] \ ,
\end{align}
where we used in turns the linearity of the scalar product, the isometricity assumption for the action on $\R^N$, and the invariance assumption $(g\cdot S , g \cdot Y) \eqd (S,Y)$. We will now derive a bound between the first terms of (\ref{eq_HS1}) and (\ref{eq_HS2}); Jensen's inequality applied to the convex function $x \to x^2$ and the expectation $\E_g$ yields, for one component $i \in [N] = \{1,\dots ,N\}$,
\beq
(\hS^{\rm eq}(Y)_i)^2 = (\E_g[ (g^{-1} \cdot  \hS(g \cdot Y))_i  ])^2 \leq \E_g[( (g^{-1} \cdot  \hS(g \cdot Y))_i)^2] \ .
\eeq
Summing these inequalities on $i$ yields
\beq
|| \hS^{\rm eq}(Y) ||^2 \leq \E_g[ || g^{-1} \cdot  \hS(g \cdot Y) ||^2 ] = \E_g[ || \hS(g \cdot Y) ||^2 ] \ ,
\eeq
thanks to the isometric property of the representation. Hence
\beq
\E[|| \hS^{\rm eq}(Y) ||^2 ] \leq \E_g \E[ || \hS(g \cdot Y) ||^2 ] = \E[ || \hS(Y) ||^2 ] \ ,
\eeq
since the symmetry assumption yields $g \cdot Y \eqd Y $ for all $g$. Combining this various observations completes the justification of $\MSE(\hS^{\rm eq}) \leq \MSE(\hS)$.

We can finally conclude that it is always optimal to impose the equivariance of the basic functions $b_\beta$: suppose indeed that $\hS$ achieves $\MMSE_\A$ as a linear combination of $\{b_\beta\}_{\beta \in \A}$, where the $b_{\beta}$ are arbitrary. Then $\hS^{\rm eq}$, whose MSE is not strictly greater than $\MMSE_\A$, is a linear combination of the $\{b_\beta^{\rm eq}\}_{\beta \in \A}$; however the latter has in general a reduced dimensionality since two distinct $b_\beta$ and $b_\gamma$ can be in the same orbit of the symmetrization, and hence have $b_\beta^{\rm eq}$ proportional to $b_\gamma^{\rm eq}$.

\section{Matrix denoising, orthogonally invariant case}
\label{sec_rot}

We come back now to our main object of study, namely the matrix denoising problem defined in Sec.~\ref{sec_definitions_def}. We shall apply the results derived in Sec.~\ref{sec_general} by defining the scalar product on $M_n^{\rm sym}(\R)$ as $\la S, T \ra =\tr(ST)$, where we recall that $\tr(\bullet) = \frac{1}{n}\Tr(\bullet)$ is the normalized trace. In this way the definitions of the MSE of an estimator given in equations (\ref{eq_MSE_matrix}) and (\ref{eq_MSE}) coincide. The group of symmetry we shall exploit is here $G=\cO_n$, the orthogonal group with identity $e=\one_n$, that acts on $M_n^{\rm sym}(\R)$ via conjugation, namely $O \cdot S = O S O^T$; this is indeed a linear representation of $G$. The invariance assumptions stated in Sec.~\ref{sec_definitions_def} read in these notations $O \cdot S \eqd S$ and $O \cdot Z \eqd Z$ for all $O \in \cO_n$, hence $(O\cdot S,O\cdot Y) \eqd (S,Y)$, in such a way that $G$ is a symmetry of the inference problem, in the terminology of Sec.~\ref{sec_general}. One can also check that its representation on $S$ is isometric, $\la O \cdot S , O \cdot S' \ra = \tr(OSO^T O S' O^T) = \la S,S' \ra$, using the cyclicity of the trace.

\subsection{Equivariant estimators}
\label{sec_rot_equivariant}

The discussion of Sec.~\ref{sec_gene_sym} emphasized the important role played by equivariant functions, which in the present context are functions $f:M_n^{\rm sym}(\R) \to M_n^{\rm sym}(\R)$ such that $f(OYO^T)=Of(Y)O^T$ for all $Y \in M_n^{\rm sym}(\R)$ and $O \in \cO_n$. We will show in the appendix~\ref{app_equivariant} (see also~\cite{KuMoWe24,Pr76,GoWa_book}) that such functions are of the form 
\beq
f(Y) = \hf(Y;\Tr(Y),\Tr(Y^2),\dots,\Tr(Y^{n-1})) \ ,
\label{eq_equi_ortho}
\eeq
where $\hf$ is a function from $\R^n$ to $\R$, with the convention stated in equation (\ref{eq_convention1}) for the action of a scalar function (here the dependency in the first argument of $\hf$) on a symmetric matrix. More explicitly, if $Y$ is diagonalized as $Y=P \, \diag(\lambda_1,\dots,\lambda_n) \, P^T$ with $P\in \cO_n$, the definition of (\ref{eq_equi_ortho}) has to be interpreted as
\beq
f(Y) = P \, \diag(\{ \hf(\lambda_i;\Tr(Y),\Tr(Y^2),\dots,\Tr(Y^{n-1}))\}_{i\in \{1,\dots,n\}} ) P^T \ .
\label{eq_convention}
\eeq
It is immediate to check that the functions $f$ defined according to (\ref{eq_equi_ortho}) are equivariant for any $\hf$; what is non-trivial, and proven in appendix~\ref{app_equivariant} in two different ways, is that all equivariant functions (for the conjugation action of $\cO_n$) can be written in this way.

Turning now to the approximate estimation problem, we will consider estimators $\hS(Y)$ whose entries are multivariate polynomials in the entries of $Y$, of total degree at most $D$. Since we do not loose any accuracy by imposing their equivariance (recall the discussion of Sec.~\ref{sec_gene_sym} in the generic case), they can be written as linear combinations of the following equivariant polynomials:
\beq
Y^p (\Tr(Y))^{q_1} (\Tr(Y^2))^{q_2} \dots (\Tr(Y^D))^{q_D} \ ,
\label{eq_equiv_poly_On}
\eeq
with $\{p,q_1,q_2,\dots,q_D\}$ non-negative integers. The total degree of the polynomial is $p+\sum_{i = 1}^{D} i q_i$, and should thus be restricted to values smaller or equal to $D$. Note that, for a given $D$, the number of such polynomials is finite and independent of $n$, which confirms the interest of exploiting the invariances of the problem: a parametrization of arbitrary polynomial estimators of matrices of size $n$ would have involved a number of basic functions diverging with $n$.

In the next subsection we will further restrict our study to the case of scalar polynomial estimators, obtained with $q_1=\dots=q_D=0$; we will come back to the complete set of equivariant polynomials in Sec.~\ref{sec_irrelevance}, and show then that asymptotically in the large $n$ limit the inclusion of the multiplicative factors of traces of powers of $Y$ does not improve the MMSE.

\subsection{Scalar polynomial estimators}

\subsubsection{Equations at finite $n$}

Following the computations of Sec.~\ref{sec_gene_approx}, the optimal estimator for matrices of finite size $n$ based on scalar polynomials of degree at most $D$ is
\beq
\hS^{(n,D,{\rm s})}(Y) = \sum_{p=0}^D c_p^{(n,D,{\rm s})} Y^p \ ,
\eeq
where $c^{(n,D,{\rm s})}$ is a $D+1$ dimensional vector, solution of the equation $\M^{(n,D,{\rm s})} c^{(n,D,{\rm s})} = \cR^{(n,D,{\rm s})}$. The elements of the matrix $\M^{(n,D,{\rm s})}$ and vector $\cR^{(n,D,{\rm s})}$ are
\begin{align}
\M^{(n,D,{\rm s})}_{p,p'} & = \E[\tr ( Y^{p+p'}) ] \ , \\
\cR^{(n,D,{\rm s})}_p & = \E[\tr ( Y^p S) ]  \ ,
\end{align}
the dependency on $n$ being kept implicit in the law of the matrices, $Y=Y^{(n)}$ and $S=S^{(n)}$. We denote $\MMSE^{(n,D,{\rm s})}$ the optimal MSE achieved by polynomial estimators of degree at most $D$ for matrices of size $n$; following (\ref{eq_MMSE_A}) this quantity reads
\beq
\MMSE^{(n,D,{\rm s})} = \E[\tr((S^{(n)})^2)] - \sum_{p=0}^D c_p^{(n,D,{\rm s})} \cR^{(n,D,{\rm s})}_p \ .
\label{eq_MMSE_nD}
\eeq

\subsubsection{The large $n$ limit}
\label{sec_rot_largen}

We will now discuss the simplifications that this optimization problem undergo in the large matrix size limit. We shall denote $\hcM^{(D)}$ and  $\hcR^{(D)}$ the limits as $n\to \infty$ of $\M^{(n,D,{\rm s})}$ and $\cR^{(n,D,{\rm s})}$. Their matrix elements read
\begin{align}
\hcM^{(D)}_{p,p'} & = \mu_{Y,p+p'} = \hcM_{p,p'} \ , \label{eq_def_hcM} \\
\hcR^{(D)}_p & = \lim_{n\to \infty} \E[\tr((Y^{(n)})^p S^{(n)} )] =\hcR_p \ ,  \label{eq_def_hcR} 
\end{align}
where the first line is a consequence of the convergence in average moments of $Y^{(n)}$ towards the measure $\mu_Y$ (in the sense of Eq.~(\ref{eq_convergence})), and where we emphasized the independence of these quantities with respect to $D$.
Consider $\hc^{(D)} $ solution of $\hcM^{(D)} \hc^{(D)} = \hcR^{(D)}$; we will see shortly that $\hcM^{(D)}$ is invertible, hence this solution is unique. These vectors of coefficients hence correspond to estimators that achieve the asymptotic (as $n\to \infty$) optimal MSE among scalar polynomial functions of degree at most $D$, that we shall denote $\MMSE^{(D,{\rm s})}$. Taking the large $n$ limit in (\ref{eq_MMSE_nD}) one obtains
\begin{align}
\MMSE^{(D,{\rm s})} = \lim_{n \to \infty }\MMSE^{(n,D,{\rm s})} & = \mu_{S,2} - \sum_{p=0}^D \hcR_p  \hc_p^{(D)} \label{eq_MMSE_D} \\
& = \mu_{S,2} -  (\hcR^{(D)})^T (\M^{(D)})^{-1} \hcR^{(D)} \ .
\end{align}
One can interchange the order of the minimization and $n\to \infty$ limit since the asymptotic matrix of the quadratic form is invertible (recall the discussion at the end of Sec.~\ref{sec_gene_approx}).

The corresponding estimator will be denoted
\beq
\hS(Y) = \D^{(D)}(Y) \ , \qquad \D^{(D)}(\lambda) = \sum_{p=0}^D \hc_p^{(D)} \lambda^p \ ;
\label{eq_def_DD}
\eeq
in this last expression the coefficients of the polynomial $\D^{(D)} : \R \to \R$ are solutions of $\hcM^{(D)} \hc^{(D)} = \hcR^{(D)}$.

\subsubsection{Properties of $\hcM$ and expression of $\hcR$}
\label{sec_rot_expressions}

We will now discuss how the quantities $\hcM_{p,p'}$ and $\hcR_p$, defined in Eqs.~(\ref{eq_def_hcM},\ref{eq_def_hcR}), can be expressed in terms of the definition of the model, namely the limit eigenvalue distributions $\mu_S$ and $\mu_Z$ of the signal and the noise. We shall write $\hcM$ and $\hcR$ the infinite matrix and vector with elements $\hcM_{p,p'}$ and $\hcR_p$, in such a way that $\hcM^{(D)}$ and $\hcR^{(D)}$ are the $D+1$-dimensional truncations of these infinite objects. We will also denote $\tau(\bullet) = \lim_{n\to \infty} \E[ \tr(\bullet )]$; since the hypotheses made in Sec.~\ref{sec_definitions_def} are such that $S^{(n)}$, $Z^{(n)}$ are asymptotically free, we will also use the same symbol for the trace in a non-commutative probability space~\cite{NiSp_book,MiSp_book}, with elements denoted $S,Z$ and $Y=S+Z$, that correspond formally to the limit of the sequence of matrices. In this asymptotic framework $S$ and $Z$ are free, with distributions $\mu_S$ and $\mu_Z$, and as a consequence $\mu_Y = \mu_S \boxplus \mu_Z$, the free additive convolution of these two distributions.

Recall that (\ref{eq_def_hcM}) provided an expression of the matrix elements of $\hcM$ as $\hcM_{p,p'}=\mu_{Y,p+p'}$, in other words $\hcM$ is the Hankel matrix for the probability measure $\mu_Y = \mu_S \boxplus \mu_Z$, and can hence be in principle computed from the model definition in terms of $\mu_S$ and $\mu_Z$. It is known that $\hcM^{(D)}$ is then positive definite for all $D$ if the support of $\mu_Y$ has infinite cardinality, see for instance theorem 1 in~\cite{Si98}, which implies the uniqueness of the solution of $\hcM^{(D)} \hc^{(D)} = \hcR^{(D)}$, for all $D$. This assumption of infinite cardinality (not to be confused with unboundedness) for the support of $\mu_Y$ will be made implicitly in the following. It is in any case quite reasonable, since it will be fulfilled as soon as $\mu_Y$ has an absolutely continuous part; in particular for Gaussian noise $\mu_Y$ is the free convolution of the signal distribution with a semi-circular distribution, and~\cite{Bi97} shows that $\mu_Y$ is then absolutely continuous.

Consider now the elements of $\hcR$; from (\ref{eq_def_hcR}) we obtain
\beq
\hcR_p = \tau(S Y^p) = \tau((Y-Z) Y^p) = \tau(Y^{p+1}) - \tau(Z (S+Z)^p) = \mu_{Y,p+1} - \tau(Z (S+Z)^p)  \ .
\label{eq_hcR}
\eeq
Our goal in the following is to give an explicit expression of the second term above in terms of the distributions of $S$ and $Z$. This will be achieved thanks to the freeness of $S$ and $Z$, via some combinatorial manipulations that are standard in the free probability literature (see in particular lecture 11 in~\cite{NiSp_book}). For the convenience of the reader we present in Appendix~\ref{app_IPP} an alternative derivation that relies only on elementary techniques (Gaussian integration by parts) and bypasses the free probability manipulations, but which is only valid in the Gaussian noise case.

Let us start by recalling some combinatorial definitions. A partition $\pi$ of $[N]=\{1,\dots,N\}$ is a decomposition of this set as the disjoint union of non-empty blocks. A partition is crossing if there exists four indices $1\le i_1 < i_2 <i_3 <i_4 \le N$ such that $i_1$ and $i_3$ are in the same block, while $i_2$ and $i_4$ are both in another block; a partition is non-crossing if there are no indices with this property. The mixed free cumulants $\kappa_N$ are $N$-multilinear functions on the non-commutative probability space, defined implicitly according to
\beq
\tau(a_1 \dots a_N) = \sum_{\pi \in NC(N)} \kappa_\pi[a_1,\dots,a_N] \ ,
\eeq
where the sum is over the non-crossing partitions of $[N]$, and where $\kappa_{\pi}$ factorizes as a product over the blocks of $\pi$, the contribution to this product of a block of size $l$ containing the elements $1\le i_1 < \dots < i_l \le N$ being $\kappa_l[a_{i_1},\dots,a_{i_l}]$. The crucial property of the $\kappa_N$ is their vanishing as soon as two distinct free elements appear in their arguments, otherwise their value is the $N$-th free cumulant of the variable (see App.~\ref{app_fp} for a definition), for instance $\kappa_N[Z,\dots,Z] = \kappa_{Z,N}$.

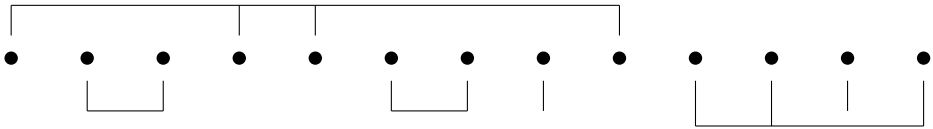
\begin{figure}
\begin{tikzpicture}
\fill[black] (0,0) circle (2.5pt);
\fill[black] (1,0) circle (2.5pt);
\fill[black] (2,0) circle (2.5pt);
\fill[black] (3,0) circle (2.5pt);
\fill[black] (4,0) circle (2.5pt);
\fill[black] (5,0) circle (2.5pt);
\fill[black] (6,0) circle (2.5pt);
\fill[black] (7,0) circle (2.5pt);
\fill[black] (8,0) circle (2.5pt);
\fill[black] (9,0) circle (2.5pt);
\fill[black] (10,0) circle (2.5pt);
\fill[black] (11,0) circle (2.5pt);
\fill[black] (12,0) circle (2.5pt);
\draw (0,.7) -- (8,.7);
\draw (0,.3) -- (0,.7);
\draw (3,.3) -- (3,.7);
\draw (4,.3) -- (4,.7);
\draw (8,.3) -- (8,.7);

\draw (1,-.7) -- (2,-.7);
\draw (1,-.3) -- (1,-.7);
\draw (2,-.3) -- (2,-.7);

\draw(5,-.7) -- (6,-.7);
\draw (5,-.3) -- (5,-.7);
\draw (6,-.3) -- (6,-.7);
\draw (7,-.3) -- (7,-.7);

\draw (11,-.3) -- (11,-.7);
\draw (9,-.3) -- (9,-.9);
\draw (10,-.3) -- (10,-.9);
\draw (12,-.3) -- (12,-.9);
\draw (9,-.9) -- (12,-.9);
\end{tikzpicture}

\caption{Illustration of the non-crossing partition of Eq~(\ref{eq_NC}); for clarity the block containing 1 has been drawn above the horizontal axis, the other blocks below. Here $p+1=13$, the block containing the first element has cardinality $m=4$ and reads $\{1,4,5,9\}$, corresponding to $j_1=2$, $j_2=0$, $j_3=3$, the length of the successive intervals it does not cover. Because of the non-crossing condition the other blocks of the partitions decompose into non-crossing partitions of the intervals not covered, of lenght $j_1=2$, $j_2=0$, $j_3=3$, $p+1-m-j_1-j_2-j_3=4$.}
\label{fig_NC}
\end{figure}

Applying this formula to the term we need to compute in (\ref{eq_hcR}) yields
\beq
\tau(Z(S+Z)^p) = \sum_{\pi \in NC(p+1)} \kappa_\pi[Z,S+Z\dots,S+Z] \ . 
\label{eq_NC}
\eeq
We shall decompose this sum by first summing over the block $B$ of $\pi$ containing $1$; we will denote $m$ its cardinality, and parametrize it as $B=\{1,2+j_1,3+j_1+j_2,\dots,m+j_1+\dots+j_{m-1}\}$, where the $j_i$'s are non-negative integers, see Fig.~\ref{fig_NC} for an illustration. The contribution of $B$ to the product in $\kappa_\pi$ is $\kappa_m[Z,S+Z,\dots,S+Z]=\kappa_{Z,m}$ since $S$ and $Z$ are free. Since $\pi$ is non crossing the rest of the summation factorizes on non-crossing partitions of size $j_1,j_2,\dots,j_{m-1},p+1-m-j_1-\dots-j_{m-1}$. Moreover
\beq
\sum_{\pi \in NC(j)} \kappa_\pi[S+Z,\dots,S+Z] = \tau((S+Z)^j) = \tau(Y^j) = \mu_{Y,j} \ ,
\eeq
one thus gets
\begin{align}
\tau(Z(S+Z)^p) &= \sum_{m=1}^{p+1} \kappa_{Z,m} \sum_{\substack{j_1,\dots j_{m-1}\ge 0 \\ j_1+\dots+j_{m-1} + m \le p+1}} \mu_{Y,j_1} \dots \mu_{Y,j_{m-1}} \mu_{Y,p+1-m-j_1-\dots-j_{m-1}} \\
& = \sum_{m=1}^{p+1} \kappa_{Z,m} \sum_{\substack{j_1,\dots j_m \ge 0 \\ j_1+\dots+j_m = p+1-m}} \mu_{Y,j_1} \dots \mu_{Y,j_m} \ .
\end{align}
Inserting this expression in (\ref{eq_hcR}) yields
\beq
\hcR_p = \mu_{Y,p+1} - \sum_{m=1}^{p+1} \kappa_{Z,m} \sum_{\substack{j_1,\dots j_m \ge 0 \\ j_1+\dots+j_m = p+1-m}} \mu_{Y,j_1} \dots \mu_{Y,j_m} \ .
\label{eq_hcR_result}
\eeq
In the Gaussian noise case the only non-zero free cumulant of $Z$ is $\kappa_{Z,2} = \Delta$, this expression then simplifies into
\beq
\hcR_p = \mu_{Y,p+1} - \Delta \sum_{j=0}^{p-1} \mu_{Y,j} \mu_{Y,p-1-j} \ ,
\label{eq_hcR_Gaussian}
\eeq
which is the result derived in Appendix~\ref{app_IPP} via more elementary means.

In the next section we will discuss the connection between this result and the one of~\cite{BuAlBoPo16}. But note that even without relying on~\cite{BuAlBoPo16} one can in practice compute the matrix $\hcM^{(D)}$ and the vector $\hcR^{(D)}$, thanks to the formulas (\ref{eq_def_hcM},\ref{eq_hcR_result}), and then with a matrix inversion find numerically the coefficients $\hc^{(D)}$ of the optimal estimator among scalar polynomials of degree at most $D$. This approach will be illustrated by numerical results in Sec.~\ref{sec_numerics_On} on a specific example. Note also that if one knows the orthogonal polynomials associated to the measure $\mu_Y$ one can use them as a basis instead of the monomials $Y^p$, thus obtaining directly a diagonal matrix $\hcM$; but in practice the determination of the orthogonal polynomials of an arbitrary measure is in general not much easier if the latter is given in terms of its moments or free cumulants (see~\cite{Le03} for a formula connecting the Jacobi parameters of the orthogonal polynomial and the free cumulants of its measure).

\subsubsection{The large $D$ limit}
\label{sec_rot_largeD}

We shall now show that in the large $D$ limit one has $\D^{(D)} \to \DB$, proving the optimality of the BABP formula among scalar estimators (with the limit $n\to\infty$ taken before $D\to\infty$). A first step in this discussion is to remark that $\D^{(D)}$ can be defined as the unique polynomial of degree at most $D$ such that
\beq
\int \mu_Y(\dd \lambda) \D^{(D)}(\lambda) \lambda^p = \hcR_p \qquad \forall p \in \{0,1,\dots,D\} \ .
\label{eq_PMI_DD}
\eeq
These equations correspond indeed to a rewriting of the optimality condition $\hcM^{(D)} \hc^{(D)} = \hcR^{(D)}$ using the expression of $\hcM_{p,p'}$ in (\ref{eq_def_hcM}). The link with the BABP formula will be established through the following fact:
\beq
\int \mu_Y(\dd \lambda) \DB(\lambda) \lambda^p = \hcR_p \qquad \forall p\ge 0 \ . 
\label{eq_PMI_DB}
\eeq
Before proving this identity let us explore its consequences. The comparison of  (\ref{eq_PMI_DD}) with (\ref{eq_PMI_DB}) reveals that $\D^{(D)}$ is the orthogonal projection of $\DB$ on the subspace of polynomials of degree at most $D$, within the $L^2(\R,\mu_Y)$ Hilbert space whose scalar product and associated norm are
\beq
(f,g)_{\mu_Y} = \int \mu_Y(\dd \lambda) f(\lambda) g(\lambda) \ , \qquad ||f||^2_{\mu_Y} = \int \mu_Y(\dd \lambda) f(\lambda)^2 \ .
\label{eq_def_Hilbert}
\eeq
Since the polynomials are dense in this space one can conclude that $||\D^{(D)} - \DB||_{\mu_Y} \to 0$ as $D \to \infty$, in other words that the convergence of $\D^{(D)}$ to $\DB$ happens in the $L^2$ sense (and as a consequence also $\mu_Y$ almost everywhere).

Let us now start the proof of (\ref{eq_PMI_DB}), where the expressions of $\DB$ and $\hcR_p$ to connect are given in the equations (\ref{eq_def_DB},\ref{eq_def_DB_2}) and (\ref{eq_hcR_result}). Since from (\ref{eq_def_DB_2}) one has $\DB(\lambda)=\lambda - \tDB(\lambda)$, and since $\int \mu_Y(\dd \lambda) \, \lambda \, \lambda^p = \mu_{Y,p+1}$ is the first term of $\hcR_p$, the proof amounts to show that:
\beq
\int \mu_Y(\dd \lambda) \tDB(\lambda) \lambda^p = \sum_{m=1}^{p+1} \kappa_{Z,m} \sum_{\substack{j_1,\dots j_m \ge 0 \\ j_1+\dots+j_m = p+1-m}} \mu_{Y,j_1} \dots \mu_{Y,j_m} \ .
\label{eq_PMI_DB_2}
\eeq

We will first treat the Gaussian noise case, which is slightly simpler, with the versions of $\tDB$ and $\hcR_p$ of (\ref{eq_def_DB_Gaussian}) and (\ref{eq_hcR_Gaussian}). For $\eta \neq 0$ real, one has 
\beq
g_Y(\lambda-i\eta) = \int \mu_Y(\dd \lambda') \frac{1}{\lambda-\lambda' - i \eta} = \int \mu_Y(\dd \lambda') \frac{\lambda-\lambda' + i \eta}{(\lambda-\lambda')^2 + \eta^2} \ .
\eeq
Taking the real part of this equation, multiplying by $2\Delta \lambda^p$ and integrating against $\mu_Y$ yields
\begin{align}
\int \mu_Y(\dd \lambda) \lambda^p 2 \Delta \re (g_Y(\lambda-i\eta)) &= 
\Delta \int \mu_Y(\dd \lambda) \mu_Y(\dd \lambda') \frac{2 \lambda^p (\lambda-\lambda')}{(\lambda-\lambda')^2 + \eta^2} \\
&= \Delta \int \mu_Y(\dd \lambda) \mu_Y(\dd \lambda') \frac{ (\lambda^p-\lambda'^p) (\lambda-\lambda')}{(\lambda-\lambda')^2 + \eta^2} \ ,
\end{align}
by symmetrizing the integrand. Since for $\lambda\neq\lambda'$ (an event of probability one in this integral since the free convolution with a semi-circle is absolutely continuous~\cite{Bi97}) one has the identity
\beq
\frac{\lambda^p-\lambda'^p}{\lambda-\lambda'} = \lambda^{p-1} + \lambda^{p-2}\lambda' + \dots + \lambda \lambda'^{p-2} + \lambda'^{p-1} = \sum_{j=0}^{p-1} \lambda^j \lambda'^{p-1-j} \ ,
\label{eq_symp}
\eeq
one can safely take the limit $\eta \to 0$ to conclude that
\beq
\int \mu_Y(\dd \lambda) \lambda^p \tDB(\lambda) = \Delta \int \mu_Y(\dd \lambda) \mu_Y(\dd \lambda') \sum_{j=0}^{p-1} \lambda^j \lambda'^{p-1-j} = \Delta \sum_{j=0}^{p-1} \mu_{Y,j} \mu_{Y,p-1-j} \ ,
\eeq
which completes the proof of  (\ref{eq_PMI_DB}) in the Gaussian noise case, comparing the last equation with (\ref{eq_hcR_Gaussian}).

We shall now generalize this computation to an arbitrary noise distribution, connecting (\ref{eq_def_DB}) and (\ref{eq_PMI_DB_2}). The link between free cumulants and $R$-transform is such that $g R_Z(g) = \sum_{m=1}^\infty \kappa_{Z,m} g^m$. Let us denote $\underset{\eta \to 0^+}{\lim}\im(g_Y(\lambda-i \eta)) = \pi \rho_Y(\lambda)$, where $\rho_Y(\lambda)$ is the density of $\mu_Y$ (that we assume to be absolutely continuous), in such a way that $\mu_Y(\dd \lambda) = \rho_Y(\lambda) \dd \lambda$. Simplifying the integration over $\mu_Y(\dd \lambda)$ with the denominator of (\ref{eq_def_DB}) thus yields
\begin{align}
\int \mu_Y(\dd \lambda) \lambda^p \tDB(\lambda) & = \sum_{m=1}^\infty \kappa_{Z,m} \underset{\eta \to 0^+}{\lim}
\int \frac{\dd \lambda}{ \pi} \lambda^p \, \im( g_Y(\lambda-i \eta)^m) \\
& = \sum_{m=1}^\infty \kappa_{Z,m} \underset{\eta \to 0^+}{\lim}
\int \frac{\dd \lambda}{ \pi} \lambda^p \mu_Y(\dd \lambda_1) \dots \mu_Y(\dd \lambda_m) \, \im \left( \prod_{k=1}^m \frac{\lambda-\lambda_k + i \eta}{(\lambda-\lambda_k)^2 + \eta^2} \right) \\
& = \sum_{m=1}^\infty \kappa_{Z,m} \underset{\eta \to 0^+}{\lim}
\int \frac{\dd \lambda}{ \pi} \lambda^p \mu_Y(\dd \lambda_1) \dots \mu_Y(\dd \lambda_m) \frac{ \eta \underset{j=1}{\overset{m}{\sum}} \underset{ \substack {j'=1 \\ j'\neq j}}{\overset{m}{\prod}} (\lambda-\lambda_{j'}) + O(\eta^3)}{\underset{k=1}{\overset{m}{\prod}}  ((\lambda-\lambda_k)^2 + \eta^2)} \ . \label{eq_proof_BABP}
\end{align}
Suppose that $\lambda_1,\dots,\lambda_m$ are all distinct (an event of probability one under $\mu_Y^{\otimes m}$ with the absolute continuity assumption), and recall that 
\beq
\lim_{\eta \to 0^+} \frac{1}{\pi}\frac{\eta}{(\lambda -x)^2 + \eta^2} = \delta(\lambda - x) \ ,
\eeq
in the sense of distributions; as a consequence,
\beq
\lim_{\eta \to 0^+} \frac{1}{\pi}\frac{\eta}{\underset{k=1}{\overset{m}{\prod}} ((\lambda-\lambda_k)^2 + \eta^2)} = \sum_{k=1}^m \delta(\lambda - \lambda_k) \frac{1}{\underset{\substack{k'=1 \\ k'\neq k}}{\overset{m}{\prod}} (\lambda_k - \lambda_{k'})^2} \ .
\eeq
Inserting this identity in (\ref{eq_proof_BABP}) and performing the integral on $\lambda$ thus yields
\begin{align}
\int \mu_Y(\dd \lambda) \lambda^p \tDB(\lambda) & =
\sum_{m=1}^\infty \kappa_{Z,m} \int \mu_Y(\dd \lambda_1) \dots \mu_Y(\dd \lambda_m) \sum_{k=1}^m \lambda_k^p \underset{j=1}{\overset{m}{\sum}} \frac{ \underset{\substack{j'=1 \\ j'\neq j}}{\overset{m}{\prod}}(\lambda_k-\lambda_{j'})}{\underset{\substack{k'=1 \\ k'\neq k}}{\overset{m}{\prod}} (\lambda_k - \lambda_{k'})^2} \ .
\end{align}
The product over $j'$ vanishes unless $j=k$, hence
\beq
\int \mu_Y(\dd \lambda) \lambda^p \tDB(\lambda)  =
\sum_{m=1}^\infty \kappa_{Z,m} \int \mu_Y(\dd \lambda_1) \dots \mu_Y(\dd \lambda_m) 
\sum_{k=1}^m \frac{\lambda_k^p}{\underset{\substack{k'=1 \\ k'\neq k}}{\overset{m}{\prod}} (\lambda_k - \lambda_{k'})} \ .
\label{eq_int_generic}
\eeq
There is a generalization of (\ref{eq_symp}), which corresponds to the case $m=2$, to an arbitrary $m$, that reads
\beq
\sum_{k=1}^m \frac{\lambda_k^p}{\underset{\substack{k'=1 \\ k'\neq k}}{\overset{m}{\prod}} (\lambda_k - \lambda_{k'})} = \sum_{\substack{j_1,\dots,j_m \ge 0 \\ j_1 + \dots + j_m = p-m+1 }} \lambda_1^{j_1} \dots \lambda_m^{j_m} \ ;
\label{eq_symp2}
\eeq
this identity is well-known in the literature on symmetric polynomials, see for instance exercise 7.4 in~\cite{Stanley_vol2} and~\cite{GuMi83} for a proof and an history of this identity. Simplifying (\ref{eq_int_generic}) with (\ref{eq_symp2}) and performing the integrals over the $\lambda_i$'s yields (\ref{eq_PMI_DB_2}), which concludes the justification of (\ref{eq_PMI_DB}) for generic noises.

\subsubsection{Expressions of the MMSE}
\label{sec_rot_MMSE}

We will now present various forms of $\MMSE^{(D,{\rm s})}$, the large $n$ limit of the MMSE among polynomial scalar estimators of degree at most $D$, and of $\MMSEB$, its large $D$ limit.

From (\ref{eq_MMSE_D}) we obtain
\beq
\MMSE^{(D,{\rm s})} = \mu_{S,2} - \sum_{p,p'=0}^D \hc_p^{(D)} \hcM_{p,p'} \hc_{p'}^{(D)} = \mu_{S,2} - \int \mu_Y(\dd \lambda) (\D^{(D)}(\lambda))^2 = \mu_{S,2} - ||\D^{(D)} ||_{\mu_Y}^2 \ ,
\label{eq_MMSE_D_2}
\eeq
where in the first step we used the equation $\hcM^{(D)} \hc^{(D)} = \hcR^{(D)}$ obeyed by the optimal coefficients, and in the second one the expression (\ref{eq_def_hcM}) of $\hcM$ and the definition (\ref{eq_def_DD}); the last step is a mere rewriting using the definition (\ref{eq_def_Hilbert}) of the norm on $L^2(\R,\mu_Y)$. Since the convergence of $\D^{(D)}$ to $\DB$ holds in the $L^2$ sense one obtains
\beq
\lim_{D \to \infty} \MMSE^{(D,{\rm s})} = \MMSEB = \mu_{S,2} - ||\DB ||_{\mu_Y}^2 \ .
\label{eq_MSEB}
\eeq
Note that this quantity has been obtained by taking first the $n\to\infty$ limit, and then the $D\to\infty$ limit; it is not obvious a priori that it coincides with the large $n$ limit of the MSE reached by applying the BABP estimator to matrices of size $n$, that would somehow correspond to the limits taken in the reverse order. We will check numerically in Sec.~\ref{sec_numerics_finiten} that this interversion seems to be valid. Moreover we will see now that this formula derives from the computations of~\cite{BuAlBoPo16}. Consider indeed a given realization of $S$ and $Y$, and diagonalize these matrices of size $n$ in orthonormal bases as
\beq
S = \sum_{i=1}^n \zeta_i \, v_i v_i^T \ , \qquad Y = \sum_{i=1}^n \lambda_i \, u_i u_i^T \ ,
\eeq
where the $u_i$ (resp. $v_i$) are the eigenvectors of $Y$ (resp. $S$) associated to the eigenvalues $\lambda_i$ (resp. $\zeta_i$). The square error between $S$ and the estimator $\hS(Y) = \DB(Y)$ thus reads
\begin{align}
\frac{1}{n} \Tr((S - \hS(Y)^2) & = \frac{1}{n} \Tr(S^2) + \frac{1}{n} \Tr(\hS(Y)^2) - 2 \frac{1}{n} \Tr(S \hS(Y)) \\
& = \frac{1}{n} \Tr(S^2) + \frac{1}{n} \sum_{i=1}^n \DB(\lambda_i)^2 - 2 \frac{1}{n} \sum_{i,j=1}^n \DB(\lambda_i) \zeta_j (u_i^T v_j)^2 \ .
\label{eq_BABP_SE}
\end{align}
The reasoning of~\cite{BuAlBoPo16} is based on the computation of the overlaps between the eigenvectors $v$ and $u$, and the function $\DB$ arises from the asymptotic computation in the large size limit:
\beq
\sum_{j=1}^n \zeta_j (u_i^T v_j)^2 \approx \DB(\lambda_i) \ .
\eeq 
Inserting this last identity in (\ref{eq_BABP_SE}), and noting that the $\lambda_i$ are asymptotically distributed according to $\mu_Y$, one recovers (\ref{eq_MSEB}); this is the reasoning performed in~\cite{PoBaMa23}.

To facilitate the comparison with other expressions already present in the literature one can derive the following equivalent forms of the MMSEs:
\beq
\MMSE^{(D,{\rm s})} = \mu_{Z,2} - ||\tDD ||_{\mu_Y}^2 \ , \qquad \MMSEB = \mu_{Z,2} - ||\tDB ||_{\mu_Y}^2 \ .
\label{eq_MMSE_D_3}
\eeq
Indeed, writing $\D^{(D)}(\lambda)=\lambda - \tDD(\lambda)$ in (\ref{eq_MMSE_D_2}) one obtains
\beq
\MMSE^{(D,{\rm s})} = \mu_{S,2} - \mu_{Y,2} + 2 \int \mu_Y(\dd \lambda) \lambda \, \tDD(\lambda) - ||\tDD ||_{\mu_Y}^2 \ .
\eeq
From the expression (\ref{eq_hcR_result}) of $\hcR_1$ one deduces that $\int \mu_Y(\dd \lambda) \lambda \, \tDD(\lambda) = \kappa_{Z,1} \mu_{Y,1} + \kappa_{Z,2}$. Since $\mu_Y$ is the free convolution of $\mu_S$ and $\mu_Z$ the corresponding free cumulants are additive, namely $\kappa_{Y,p} = \kappa_{S,p} + \kappa_{Z,p}$; moreover the relation between the first two moments and the first two free cumulants coincide with the one of classical cumulants, namely $\mu_{Y,1}=\kappa_{Y,1}$ and $\mu_{Y,2}=\kappa_{Y,2}+(\kappa_{Y,1})^2$. Combining these various observations yields after simplification the form (\ref{eq_MMSE_D_3}). 

These expressions of $\MMSEB$ for generic distributions of the signal and the noise did not appear previously in the literature, to the best of our knowledge; however some equivalent forms were presented in the case of Gaussian noise, for which (\ref{eq_MMSE_D_3}) becomes
\beq
\MMSEB = \Delta - 4 \Delta^2 \int \mu_Y(\dd \lambda) h_Y(\lambda)^2 \ ,
\label{eq_MMSEB_Gaussian}
\eeq
where we denote $h_Y(\lambda) = \underset{\eta \to 0^+}{\lim} \re \, ( g_Y(\lambda-i \eta) )$, the Hilbert transform of the density of $\mu_Y$ (up to some multiplicative factors). The other forms of this quantity that appeared respectively in Eq.~(58) of~\cite{MaKrMeZd22} and Eq.~(6) of~\cite{PoBaMa23} are (after some change of notations to comply with the present ones):
\begin{align}
\MMSEB & = \Delta - 2 \Delta^2 \frac{\partial}{\partial \Delta} \int \mu_Y(\dd \lambda) \mu_Y(\dd \lambda') \ln |\lambda -\lambda'| \label{eq_MMSEB_Gaussian2} \\
& = \Delta - \frac{4 \pi^2}{3} \Delta^2 \int \dd \lambda \ \rho_Y(\lambda)^3 \ , \label{eq_MMSEB_Gaussian3}
\end{align}
where $\rho_Y$ is the density of the measure $\mu_Y$. The equivalence between (\ref{eq_MMSEB_Gaussian}) and (\ref{eq_MMSEB_Gaussian3}) follows from a lemma of harmonic analysis on the Hilbert transform (see lemma C.1 in~\cite{MaKrMeZd22} and lemma B.1 in~\cite{PoBaMa23}). The equivalence with (\ref{eq_MMSEB_Gaussian2}) derives from~\cite{Vo93}, in the context of the free probability equivalents of entropy and Fisher information, see~\cite{PoBaMa23} for more details. The advantage of (\ref{eq_MMSEB_Gaussian}) and (\ref{eq_MMSEB_Gaussian3}) with respect to (\ref{eq_MMSEB_Gaussian2}) is that it can be evaluated in practice more easily, since it only requires the numerical integration of a scalar function, instead of the computation of a bidimensional integral (with a logarithmic singularity on the diagonal) followed by a numerical differentiation.

Note that the previous works~\cite{MaKrMeZd22,PoBaMa23} connected, in the Gaussian noise setting, the MMSE and the mutual information between $S$ and $Y$: in~\cite{MaKrMeZd22} this mutual information was rephrased in terms of a spherical (Harish-Chandra-Itzykson-Zuber~\cite{HC57,ItZu80}) integral, whose asymptotic behavior was computed thanks to the results of~\cite{Ma94,GuZe02,BuBoMaPo14}. The connection to the MMSE was then established thanks to the I-MMSE theorem of~\cite{GuShVe05}. In~\cite{PoBaMa23} a somehow reverse direction was followed: assuming the Bayes-optimality of the BABP estimator the asymptotic mutual information, and hence the asymptotics of the spherical integral, was extracted from the I-MMSE relation. There seems however to be several difficulties in the generalization of these connections to arbitrary noise (one could hope to exploit them to find more explicit solutions to Matytsin's equations~\cite{Ma94}); in particular if $Z$ is rotationally invariant but not extracted from the GOE then its matrix elements are not independent and the mutual information between $S$ and $Y$ takes a more complicated form. Moreover the generalization of the I-MMSE relation to non-Gaussian channels discussed for instance in~\cite{GuShVe05b} relates the derivative of the mutual information with correlation functions that in general do not coincide with the MMSE.

\subsection{Irrelevance of non-scalar terms}
\label{sec_irrelevance}

The goal of this section is to prove that $\MMSE^{(D)}=\MMSE^{(D,{\rm s})}$, in other words that using the full set of equivariant polynomials written in Eq.~(\ref{eq_equiv_poly_On}) does not improve the accuracy of estimation with respect to the scalar polynomials, asymptotically in the large $n$ limit.

A first reasoning suggests that this should be obviously true: a simple consequence of (\ref{eq_fluctuations}) (for the matrix $Y$ instead of $S$) is the concentration of $\tr(Y^i)$ around $\mu_{Y,i}$, hence the function in Eq.~(\ref{eq_equiv_poly_On}) reduces, apart from a multiplicative constant, to the simpler scalar monomial $Y^p$. A second thought reveals however that this intuition could be too naive: the full statement expressed in (\ref{eq_fluctuations}), i.e. the CLT for linear statistics, is that once centered and properly normalized the traces of powers of $Y$ are non-trivial random variables, and as such might contain an information on the signal that could be exploited to improve the accuracy of its estimation. The correct justification of the identity $\MMSE^{(D)}=\MMSE^{(D,{\rm s})}$ will in fact rely on the asymptotic vanishing of this amount of information.

Let us now formalize this reasoning. We define the set of basic functions as
\beq
b_\beta(Y) = Y^p (\Tr(Y) - \E[\Tr(Y)])^{q_1} (\Tr(Y^2)-\E[\Tr(Y^2)] )^{q_2} \dots (\Tr(Y^D) - \E[\Tr(Y^D)])^{q_D} \ ,
\label{eq_b_rot}
\eeq
where the indices $\beta=(p,q) = (p,q_1,q_2,\dots,q_D)$ made of $D+1$ non-negative integers lie in the index set 
\beq
\A_D=\left\{(p,q_1,\dots,q_D) : p+\sum_{i = 1}^{D} i q_i \le D \right\} \ .
\eeq
The linear combinations of the $\{b_\beta\}_{\beta \in \A_D}$ span the full space of equivariant polynomials of degree at most $D$. Following the generic formalism of Sec.~\ref{sec_gene_approx} we write the corresponding matrix elements of $\M$ as
\begin{align}
  \M^{(n,D)}_{\beta,\beta'} & = \E\left[\tr(Y^{p+p'}) \prod_{i=1}^D (\Tr(Y^i)-\E[\Tr(Y^i)] )^{q_i+q'_i}   \right] \\
  & = \E[\tr(Y^{p+p'})] \, \E\left[\prod_{i=1}^D (\Tr(Y^i)-\E[\Tr(Y^i)] )^{q_i+q'_i}   \right] \\ 
& + \frac{1}{n} \E\left[(\Tr(Y^{p+p'})-\E[\Tr(Y^{p+p'})] ) \prod_{i=1}^D (\Tr(Y^i)-\E[\Tr(Y^i)] )^{q_i+q'_i}   \right] \ .
\end{align}
According to the theory of second-order freeness~\cite{MiSp06,Re14,MiPo13,MaMiPeSp22}, and in particular theorem 54 in~\cite{MiPo13}, $Y^{(n)}=S^{(n)}+Z^{(n)}$ satisfies a CLT for its linear statistics, hence the last expectation in this equation admits a limit as $n\to\infty$. The second line is thus of order $1/n$, and we obtain the matrix elements of $\M^{(\infty,D)} = \lim \M^{(n,D)}$:
\beq
\M^{(\infty,D)}_{\beta,\beta'} = \hcM_{p,p'} \alpha_{q+q'} \ , 
\qquad \text{with} \qquad
\alpha_q = \E\left[\prod_{i \ge 1}(G_{Y,i})^{q_i} \right] \ ,
\label{eq_def_alpha}
\eeq
where $\hcM$ is the matrix introduced above in (\ref{eq_def_hcM}) in the scalar case, and $\{G_{Y,i}\}$ is a centered Gaussian vector describing the asymptotic fluctuations of the traces of the powers of $Y$, as in (\ref{eq_fluctuations}).

We treat similarly the vector $\cR$ describing the linear terms in the variational expression of the $\MSE$:
\begin{align}
  \cR^{(n,D)}_\beta & = \E\left[\tr(Y^p S) \prod_{i=1}^D (\Tr(Y^i)-\E[\Tr(Y^i)] )^{q_i}   \right] \\
  & = \E[\tr(Y^p S)] \, \E\left[\prod_{i=1}^D (\Tr(Y^i)-\E[\Tr(Y^i)] )^{q_i}   \right] \\ 
& + \frac{1}{n} \E\left[(\Tr(Y^p S)-\E[\Tr(Y^p S)] ) \prod_{i=1}^D (\Tr(Y^i)-\E[\Tr(Y^i)] )^{q_i}   \right] \ .
\end{align}
The last expectation is again a product of centered traces of polynomials in $S^{(n)}$ and $Z^{(n)}$, the theory of second-order freeness allows thus to conclude that it admits a limit as $n \to \infty$. As a consequence the elements of $\cR^{(\infty,D)} = \lim \cR^{(n,D)}$ read
\beq
\cR^{(\infty,D)}_{\beta} = \hcR_p \alpha_q \ ,
\eeq
where $\alpha_q$ was defined in (\ref{eq_def_alpha}) and $\hcR$ in (\ref{eq_def_hcR}).

At this point we can check that the matrix $\M^{(\infty,D)}$ is invertible: for a vector $c$ indexed by $\A_D$ we have
\beq
\sum_{\beta,\beta' \in \A_D} c_\beta \M^{(\infty,D)}_{\beta,\beta'} c_{\beta'} = \E\left[\left(\sum_{\beta \in \A_D} c_\beta \lambda^p \prod_{i \ge 1}(G_{Y,i})^{q_i} \right)^2 \right] \ ,
\eeq
where $\lambda$ is here a random variable with the law $\mu_Y$, independently from the $\{G_{Y,i}\}$. This quantity is strictly positive unless $c=0$ (thanks to the invertibility of $\hcM^{(D)}$ justified previously, and the assumption that the covariance of the $\{G_{Y,i}\}$ has no zero eigenvalues), which shows that all the eigenvalues of $\M^{(\infty,D)}$ are strictly positive, this matrix is thus invertible. According to the discussion at the end of Sec.~\ref{sec_gene_approx} this property allows to exchange the limit $n\to\infty$ and the minimization, hence
\beq
\MMSE^{(D)} = \mu_{S,2} - \sum_{\beta \in \A_D} \cR^{(\infty,D)}_{\beta} c_\beta^{(\infty,D)} \ ,
\label{eq_sum_MMSE}
\eeq
where $c^{(\infty,D)}$ is the (unique) solution of $\M^{(\infty,D)} c^{(\infty,D)} = \cR^{(\infty,D)}$. The latter is finally recognized to be $c_\beta^{(\infty,D)}=\hc_p^{(D)}\delta_{q,0}$, with $\hc_p^{(D)}$ the solution of $\hcM^{(D)} \hc^{(D)} = \hcR^{(D)}$ studied above. Indeed,
\beq
\sum_{\beta' \in \A_D} \M^{(\infty,D)}_{\beta,\beta'} c_{\beta'}^{(\infty,D)}
= \alpha_q \hcR_p^{(D)} = \cR^{(\infty,D)}_\beta \ .
\eeq
Since in the sum of (\ref{eq_sum_MMSE}) the fluctuation exponents $q_i$ all vanish the expression of the MMSE reduces to the one of the scalar case in (\ref{eq_MMSE_D}), concluding the justification of $\MMSE^{(D)}=\MMSE^{(D,{\rm s})}$.

\subsection{Example: Wishart signal corrupted by Gaussian noise}
\label{sec_numerics_On}

We will now illustrate our findings on a specific example, providing numerical results on the accuracy of the finite $D$ optimal denoisers and their convergence towards the result of~\cite{BuAlBoPo16} in the large $D$ limit. Motivated by the matrix factorization problem we take as signal a Wishart matrix, defined as
\beq
S^{(n)} = \frac{1}{\sqrt{n r}} X^{(n)} (X^{(n)})^T - \frac{1}{\sqrt{\alpha}} \one_n \ .
\label{eq_Wishart_Gaussian}
\eeq
where $X^{(n)}$ is an $n \times r$ matrix whose entries are independent Gaussian random variables of mean zero and variance 1, with $\alpha=n/r$ the aspect ratio of this rectangular matrix. The law of $S^{(n)}$ is invariant under the orthogonal group thanks to the transformation rules of Gaussian vectors under linear combinations. The normalization and substraction of the diagonal terms in (\ref{eq_Wishart_Gaussian}) agrees with the convention used in~\cite{MaKrMeZd22}, in order to facilitate the comparison with their results and to have a well-behaved limit when $\alpha \to 0$. The large $n$ limit will be taken with a finite value of $\alpha$, hence $r$ also diverges proportionally to $n$; it is well-known (see e.g.~\cite{PoBo_book,BaSi_book}) that the limit $\mu_S$ of the empirical spectral distribution is then the Marcenko-Pastur law~\cite{MaPa67}, which reads with these conventions:
\beq
\mu_S(\dd \lambda) = \left(1-\frac{1}{\alpha} \right)_+ \delta_{-\frac{1}{\sqrt{\alpha}}} + \frac{\sqrt{(\sqrt{\alpha}+2-\lambda)(\lambda-(\sqrt{\alpha}-2)) }}{2 \pi (1+\sqrt{\alpha} \, \lambda)} \, \dd \lambda \, \one(\lambda \in [\sqrt{\alpha}-2,\sqrt{\alpha}+2]) \ ,
\eeq
with $x_+=\max(x,0)$.
The corresponding $R$-transform is
\beq
R_S(g) = \frac{1}{\sqrt{\alpha}} \frac{1}{1-\sqrt{\alpha} g} - \frac{1}{\sqrt{\alpha}} = \frac{g}{1-\sqrt{\alpha} g} \ ,
\label{eq_RS_example}
\eeq
whose expansion in powers of $g$ provides the free cumulants, $\kappa_{S,1}=0$, $\kappa_{S,p} = \alpha^{\frac{p}{2} -1}$ for $p \ge 2$.

We consider a corruption of the signal by a Gaussian noise, which as defined previously means that $Z^{(n)}$ is $\sqrt{\Delta}$ times a standard GOE matrix, i.e. $Z^{(n)}$ is an $n \times n$ symmetric matrix whose entries $\{Z^{(n)}_{i,j}\}_{i \le j}$ are independent centered Gaussian random variables, of variance $\frac{\Delta}{n}$ for $i < j$, and $\frac{2 \Delta}{n}$ for $i=j$. In the large $n$ limit its eigenvalue distribution is known~\cite{AnGuZe_book,PoBo_book,BaSi_book} to converge to Wigner's semi-circle distribution (rescaled by $\Delta$),
\beq
\mu_Z(\dd \lambda) = \frac{1}{2 \pi \Delta} \sqrt{4\Delta - \lambda^2} \, \dd \lambda \, \one (\lambda \in [-2\sqrt{\Delta},2\sqrt{\Delta}]) \ .
\eeq
The only non-zero free cumulant of the semi-circle distribution is the second one (which makes this distribution the free probability analogue of the Gaussian), namely $R_Z(g) = \Delta g$ and $\kappa_{Z,p} = \Delta \delta_{p,2}$.

To compute explicitly the denoisers and their associated MSE we shall need to manipulate the limit distribution $\mu_Y$ of the observation, which is the free convolution of those of the signal and of the noise, $\mu_Y = \mu_S \boxplus \mu_Z$. This operation is more easily described in terms of the $R$ transform and of the free cumulants, which are additive under this convolution, hence
\beq
R_Y(g) = R_S(g) + R_Z(g) = \frac{g}{1-\sqrt{\alpha} g}  + \Delta g \ , \quad
\kappa_{Y,1}=0 \ , \ \ \kappa_{Y,2}=1+\Delta \ , \  \ \kappa_{Y,p} = \alpha^{\frac{p}{2} -1} \ \text{for} \ p \ge 3 \ .
\label{eq_RY_example}
\eeq
The Cauchy transform $g_Y(z)$ of $\mu_Y$ is then obtained, for $z \in \mathbb{C} \setminus \R$, by inverting the $R$ transform; it is indeed the solution $g$ of the equation
\beq
z = \frac{g}{1-\sqrt{\alpha} g} + \Delta g + \frac{1}{g} \ ,
\eeq
which can be rewritten as a cubic equation,
\beq
g^3 - \left(\frac{1}{\Delta} z + \frac{1}{\sqrt{\alpha} \Delta} + \frac{1}{\sqrt{\alpha} } \right) g^2 + \frac{1}{\sqrt{\alpha} \Delta} (z + \sqrt{\alpha}) g - \frac{1}{\sqrt{\alpha} \Delta} = 0 \ .
\label{eq_cubic}
\eeq
What is needed to compute numerically the various quantities of interest is actually the density of $\mu_Y$, namely $\rho_Y(\lambda) = \frac{1}{\pi} \im ( g_Y(\lambda - i \eta))$ in the limit $\eta \to 0^+$, as well as $h_Y(\lambda)=\re ( g_Y(\lambda - i \eta))$, again for infinitesimal $\eta$, that appears in the expression (\ref{eq_def_DB_Gaussian}) of $\tDB$. These two quantities can be easily and accurately determined numerically on the support of $\mu_Y$: Cardan's formulas provide explicit expressions for the root of cubic equations like (\ref{eq_cubic}). Moreover these can be solved for $z=\lambda \in \R$, i.e. setting directly $\eta=0$. As a matter of fact cubic equations with real coefficients have either three real roots, and then $\lambda$ is outside the support of $\mu_Y$, or one real root and two complex conjugate ones. In the latter case the relevant solution is the one with strictly positive imaginary part, corresponding to $\rho_Y(\lambda) > 0$, whose real part yields $h_Y(\lambda)$. This completes the description of the numerical procedure followed to determine $\DB(\lambda)$; the computation of $\MMSEB$ from (\ref{eq_MMSEB_Gaussian}) requires in addition an integration over the support of $\mu_Y$ of $\rho_Y h_Y^2$, which we performed with the trapeze method.

In order to compute the finite $D$ optimal estimators $\D^{(D)}$ one needs to determine first the matrix elements of $\hcM$ and $\hcR$, which from (\ref{eq_def_hcM}) and (\ref{eq_hcR_Gaussian}) amounts to compute the moments $\mu_{Y,p}$ of the observation matrix. For small values of $p$ this can be done manually thanks to the known formulas between moments and free cumulants, the latter being very simple for $\mu_Y$ and given above in (\ref{eq_RY_example}), yielding
\beq
\mu_{Y,1} = 0 \ , \ \ \mu_{Y,2} = 1+ \Delta \ , \ \ \mu_{Y,3} = \sqrt{\alpha} \ , \ \ \mu_{Y,4} = \alpha + 2 (1+\Delta)^2 \ .
\label{eq_moments_example_small}
\eeq
To reach numerically arbitrary large values of $D$ we derived a formula that allows an efficient computation of all the moments $\mu_{Y,p}$,
\beq
\mu_{Y,p} = \one(p \ \text{is even} ) \Delta^{\frac{p}{2}} \frac{1}{p+1} \binom{p+1}{\frac{p}{2}} + \sum_{\substack{q \ge 0, r \ge 1 \\ 2(q+r) \le p}} \Delta^q \alpha^{\frac{p}{2}-q-r} \frac{1}{p+1} \binom{p+1}{q,r} \binom{p-2q-r-1}{r-1} \ ,
\label{eq_moments_example}
\eeq
whose proof is deferred to Appendix~\ref{app_fp}.

\begin{figure}
\includegraphics[width=8cm]{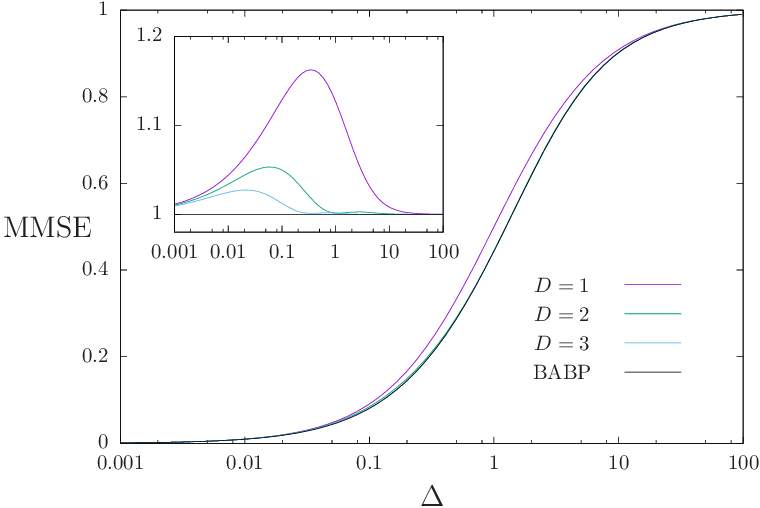}
\hspace{1cm}
\includegraphics[width=8cm]{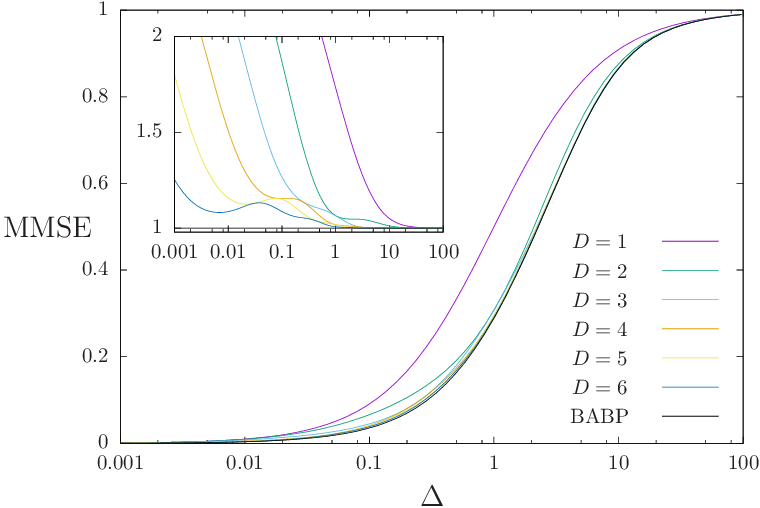}
\caption{The curves of $\MMSE^{(D)}$ as a function of $\Delta$, for $\alpha=1$ (left panel) and $\alpha=5$ (right panel), different colors corresponding to different values of $D$; the black curve labeled BABP is $\MMSEB$, the large $D$ limit of $\MMSE^{(D)}$. The insets present the ratios $\MMSE^{(D)}/\MMSEB$, with the same color code than in the main plots.}
\label{fig_MMSE}
\end{figure}

We present in Fig.~\ref{fig_MMSE} the numerical results obtained in this way for $\MMSE^{(D)}$, the optimal MSE among degree $D$ polynomial estimators, as a function of the noise intensity $\Delta$, for a few values of $D$ along with its large $D$ limit $\MMSEB$ (we checked that our numerical results for $\MMSEB$ coincide with the ones of~\cite{MaKrMeZd22}). The two panels correspond to two values of the aspect ratio $\alpha$, on the left panel $\alpha=1$ and on the right $\alpha=5$. In both cases one sees, as expected, that for a given value of $(\alpha,\Delta)$ $\MMSE^{(D)}$ decreases monotonically towards its limit $\MMSEB$ as $D$ increases (the insets represent the ratio $\MMSE^{(D)}/\MMSEB$ to better appreciate this convergence). A detail (particularly visible in the inset of the right panel) might puzzle the reader: there are some points where the curves corresponding to two consecutive values of $D$ touch each other. The explanation of this phenomenon is the following: a closer scrutiny of the numerically obtained coefficients $\hc^{(D)}$ of the optimal denoisers reveals that for $D\ge 3$ their highest order terms $\hc_D^{(D)}$ change sign as $\Delta$ varies, and hence vanish for some specific values of $\Delta$ (see Eq.~(\ref{eq_coefs_order3}) below for the analytical expression when $D=3$). When this occurs the polynomial $\D^{(D)}$ is effectively of degree $D-1$, and hence coincides with $\D^{(D-1)}$.

\begin{figure}
\includegraphics[width=8cm]{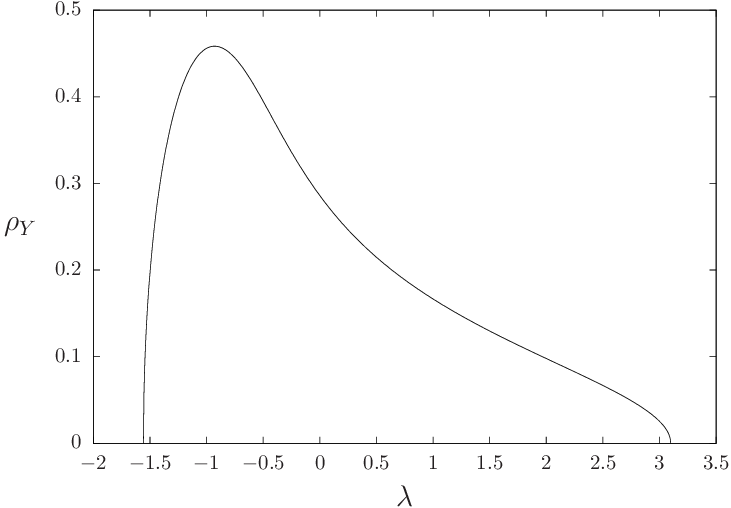}
\hspace{1cm}
\includegraphics[width=8cm]{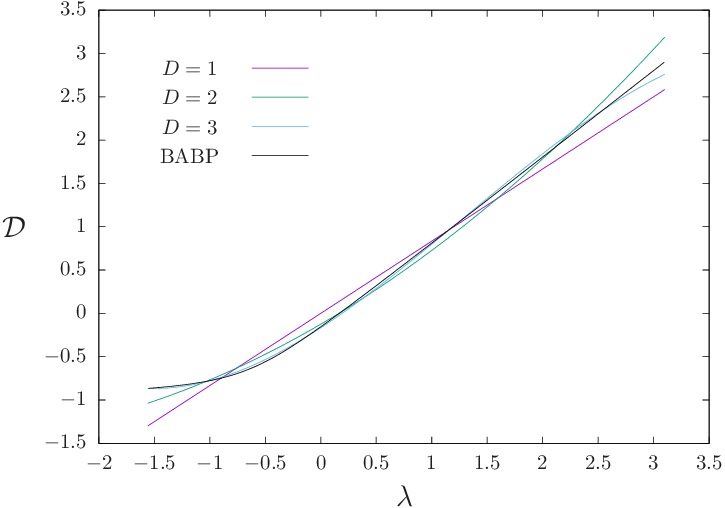}
\caption{The density $\rho_Y$ of the observation matrix (left panel), and the optimal denoising function $\DB$ along with its low degree approximations $\D^{(D)}$ (right panel), for $\alpha=1$ and $\Delta=0.2$.}
\label{fig_a1}
\end{figure}

\begin{figure}
\includegraphics[width=8cm]{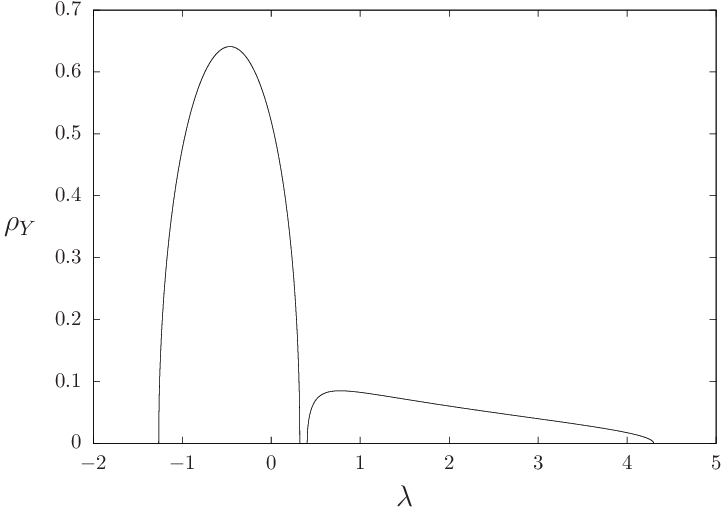}
\hspace{1cm}
\includegraphics[width=8cm]{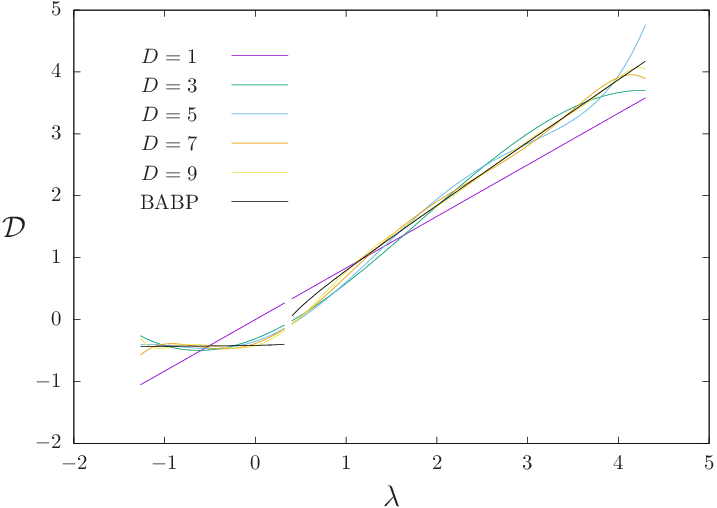}
\caption{The density $\rho_Y$ of the observation matrix (left panel), and the optimal denoising function $\DB$ along with its low degree approximations $\D^{(D)}$ (right panel), for $\alpha=5$ and $\Delta=0.2$.}
\label{fig_a5}
\end{figure}

The comparison of the two panels of Fig.~\ref{fig_MMSE} reveals that the convergence of the degree $D$ approximations towards their large $D$ limit is much faster for $\alpha=1$, where $D=2$ is already a very good approximation to the limit, than for $\alpha=5$, where notable discrepancies are still visible for $D=6$. The additional numerical results presented in Figs.~\ref{fig_a1} and \ref{fig_a5} help to understand this behavior. The left panel of these two figures display the density $\rho_Y(\lambda)$ for the same value of $\Delta=0.2$ and $\alpha=1$ (Fig.~\ref{fig_a1}) and $\alpha=5$ (Fig.~\ref{fig_a5}), while the right panel show the denoising functions $\DB$ and their sequence of approximations $\D^{(D)}$. When $\alpha$ is relatively small (see Fig.~\ref{fig_a1}) the support of $\rho_Y$ is a single interval, on which $\DB$ varies smoothly and is rather close to a linear function, hence can be accurately approximated by a low degree polynomial. In fact when $\alpha \to 0$ the matrix elements of $S^{(n)}$ become approximately independent, the signal matrix becomes a Wigner one (this is particularly obvious on the expression of the $R$-transform given in Eq.~(\ref{eq_RS_example})), hence the optimal estimator becomes a scalar Gaussian denoiser for each matrix element, which in the present notation corresponds to $\D(\lambda) = \frac{1}{1+\Delta} \lambda$. On the contrary when $\alpha$ gets large (as e.g. $\alpha=5$ in Fig.~\ref{fig_a5}) the support of $\mu_Y$ is made of two intervals (the one on the right drifting to infinity as $\alpha \to \infty$), and the function $\DB$ is roughly constant on the left interval, and almost affine on the right interval. An accurate approximation of this type of function clearly requires polynomials of higher degree, as illustrated in the right panel of Fig.~\ref{fig_a5}.

Besides these numerical results let us also write down the analytical formulas for the optimal polynomial denoisers of degree $D=1$ and $D=2$. Their coefficients are the solutions of $\hcM^{(D)} \hc^{(D)} = \hcR^{(D)}$, where the matrices $\hcM$ and vector $\hcR$ can be explicitly computed thanks to (\ref{eq_def_hcM},\ref{eq_hcR_Gaussian}) and the expressions of the small moments in (\ref{eq_moments_example_small}):
\beq
\hcM^{(2)} = \begin{pmatrix}
1 & 0 & 1+\Delta \\
0 & 1+\Delta & \sqrt{\alpha} \\
1+\Delta & \sqrt{\alpha}  & \alpha + 2 (1+\Delta)^2
\end{pmatrix} \ , \qquad
\hcR^{(2)} = \begin{pmatrix} 0 \\ 1 \\ \sqrt{\alpha} \end{pmatrix} \ ,
\eeq
$\hcM^{(1)}$ and $\hcR^{(1)}$ corresponding to the first two rows and columns. We let the reader check that
\beq
\hc^{(1)} = \begin{pmatrix} 0 \\ \frac{1}{1+\Delta} \end{pmatrix} \ , \qquad
\hc^{(2)} = \frac{1}{(1+\Delta)^3 + \alpha \Delta} \begin{pmatrix} -\sqrt{\alpha} \Delta (1+\Delta) \\ (1+\Delta)^2 \\ \sqrt{\alpha} \Delta \end{pmatrix} \ , 
\label{eq_example_coefs}
\eeq
are the solutions of $\hcM^{(D)} \hc^{(D)} = \hcR^{(D)}$ for $D=1,2$. This gives for the denoiser functions:
\beq
\D^{(1)}(\lambda) = \frac{1}{1+\Delta} \lambda \ , \qquad
\D^{(2)}(\lambda) = \frac{\sqrt{\alpha} \Delta}{(1+\Delta)^3 + \alpha \Delta} \lambda^2 +
\frac{(1+\Delta)^2}{(1+\Delta)^3 + \alpha \Delta} \lambda
- \frac{\sqrt{\alpha} \Delta (1+\Delta)}{(1+\Delta)^3 + \alpha \Delta} \ .
\eeq
The minimal MSE among estimators of degree 1 and 2 can be easily computed from (\ref{eq_MMSE_D_2}) and the coefficients $\hc^{(D)}$ we just obtained, one finds after some simplifications:
\beq
  \MMSE^{(1)} = \frac{\Delta}{1+\Delta} \ , \qquad
  \MMSE^{(2)}  = \frac{\Delta (1+\Delta)^2}{(1+\Delta)^3 + \alpha \Delta} \ .
\label{eq_example_MMSE}
\eeq

Let us remark that $\D^{(1)}$ coincides with a scalar Gaussian denoiser applied to each matrix element independently, that as explained above corresponds to the optimal denoising strategy in the limit $\alpha \to 0$. Moreover, the expansion of $\D^{(2)}$ for small $\alpha$ yields
\beq
\D^{(2)}(\lambda) =\frac{1}{1+\Delta} \lambda + \sqrt{\alpha} \frac{\Delta}{(1+\Delta)^3}(\lambda^2 - (1+\Delta)) + O(\alpha) \ ;
\eeq
this coincides, as it should, with the equation (73) of~\cite{MaKrMeZd22} that studied the expansion of $\DB$ in powers of $\sqrt{\alpha}$. We should nevertheless underline that the sequence $\D^{(D)}$ of approximations of $\DB$ introduced here is not an expansion in powers of $\sqrt{\alpha}$, or of $\lambda$, but rather the sequence of best polynomial approximations of $\DB$ in $L^2(\R,\mu_Y)$. Note in particular that power series expansion have domain of convergence which are disks in the complex plane, hence intervals symmetric around 0 when restricted to the real axis, which cannot coincide with the support of $\mu_Y$, in particular when the latter is made of two disjoint intervals.

The analytic expressions of the optimal denoisers can in principle be obtained similarly for higher values of $D$, but they quickly become rather long and not particularly illuminating. For future reference we shall only write down three of the elements of the solution of $\hcM^{(3)} \hc^{(3)} = \hcR^{(3)}$:
\beq
\begin{pmatrix}
  \hc^{(3)}_1 \\ \hc^{(3)}_2 \\ \hc^{(3)}_3
\end{pmatrix} =\frac{1}{(1+\Delta)^6+3 \alpha \Delta (1+\Delta)^3+\alpha^2\Delta(1+\Delta^2)}
\begin{pmatrix}
  (1+\Delta)^2((1+\Delta)^3-2\alpha\Delta(\Delta-2)) \\ \sqrt{\alpha}\Delta((1+\Delta)^3+2\alpha) \\ \alpha \Delta (\Delta^2-1) \end{pmatrix} \ .
\label{eq_coefs_order3}
\eeq

\section{Universality beyond orthogonal invariance}
\label{sec_universality}

\subsection{Definitions}
\label{sec_universality_def}

As argued in Sec.~\ref{sec_conjecture} there are good reasons to expect that the asymptotic optimality of the BABP estimator holds even if the priors on the signal and noise are not strictly orthogonally invariant for finite size matrices, as long as their asymptotic behavior comply approximately with this invariance. To investigate this phenomenon in a concrete case we shall consider here a generalization of the model studied in Sec.~\ref{sec_numerics_On}, where the signal was a Gaussian Wishart matrix corrupted by a noise extracted from the GOE ensemble, to an arbitrary Wishart signal and a Wigner matrix noise. More explicitly, concerning the signal we extend (\ref{eq_Wishart_Gaussian}) to
\beq
(S^{(n)})_{i,j} = \frac{1}{\sqrt{n r}} \sum_{\mu=1}^r X_{i,\mu} X_{j,\mu} - \frac{1}{\sqrt{\alpha}} \delta_{i,j} \ ,
\label{eq_def_Wishart_gene}
\eeq
where the $\{ X_{i,\mu}\}_{i\in[n]}^{\mu \in [r]}$ are i.i.d. random variables admitting all moments, with $\E[X_{1,1}]=0$ and $\E[(X_{1,1})^2]=1$, and as before $\alpha$ denotes the aspect ratio $\alpha=n/r$. We assume that the large size limit is taken with a finite value of $\alpha$ (the different sub-extensive regime where $\alpha$ diverges as a power of $n$ is considered in~\cite{PoBaMa23,BaKoRa24}). The situation considered in Sec.~\ref{sec_numerics_On} is recovered when $X_{1,1}$ has a standard Gausian distribution. The matrix elements of the noise are taken as $(Z^{(n)})_{i,j} = \sqrt{\frac{\Delta}{n}} B_{i,j}$, with the $\{B_{i,j}\}_{i \le j}$ independent, the matrix being completed by the symmetry $B_{i,j}=B_{j,i}$. All the off-diagonal elements have the law of $B_{1,2}$, that admits all moments, with $\E[B_{1,2}]=0$ and $\E[(B_{1,2})^2]=1$, the diagonal elements have the law of $B_{1,1}$, that also admits all moments, with $\E[B_{1,1}]=0$, and an arbitrary variance. The laws of $X_{1,1}$, $B_{1,1}$ and $B_{1,2}$ are independent of $n$.

To simplify some of the following computations we shall sometimes furthermore assume that the laws of $X_{1,1}$, $B_{1,1}$ and $B_{1,2}$  are symmetric, in the sense that $X_{1,1} \eqd -X_{1,1}$,  $B_{1,1} \eqd - B_{1,1} $ and $B_{1,2} \eqd - B_{1,2} $; this implies in particular that all their odd moments vanish. We shall call inversion invariance this property.

Our objective in the following is to discuss the universality of the results derived in the orthogonally invariant case, namely the dependency (or not) of the large $n$ limit of $\MMSE^{(n,D)}$ on the details of the laws $X_{1,1}$, $B_{1,1}$ and $B_{1,2}$ (provided the first two moments of $X_{1,1}$ and $B_{1,2}$ and the first moment of $B_{1,1}$  are fixed as above).

\subsection{Absence of universality in the noise}
\label{sec_nonuniversal_noise}

We already mentioned in Sec.~\ref{sec_conjecture} that one should not expect universality with respect to the distribution of the noise, we shall now make this point more precise and quantitative. Consider indeed one given element of the (rescaled) observation matrix, $\sqrt{n} Y_{i,j} = \sqrt{n} S_{i,j} + \sqrt{\Delta} B_{i,j}$. It turns out that both terms of the right hand side are of order 1 in the large $n$ limit, hence all the details of the law of the noise $B$ impact the distribution of a matrix element of $Y$, the optimal denoising strategy can thus exploit the knowledge of all the moments of $B$ and will not coincide with the orthogonally invariant one when the noise is not Gaussian.

It will be useful in the following to expand quantitatively on this toy model of a scalar denoising problem. Consider indeed two independent real random variables $S$ and $B$, and the problem of estimating $S$ from the observation of $Y=S+\sqrt{\Delta} B \in \R$. For simplicity we assume that the laws of $S$ and $B$ are even, in the sense that $S\eqd -S$ and $B \eqd -B$, and normalized in such a way that $\E[S^2]=\E[B^2]=1$. If $S$ and $B$ are both Gaussian, the estimator $\hS(Y)$ that minimizes the MSE $\E[(S-\hS(Y))^2]$, i.e. the posterior mean, can be computed exactly: one finds then $\E[S|Y] = \frac{1}{1+\Delta} Y$. As soon as $S$ or $B$ are not Gaussian such a simple and exact formula for the posterior mean is generically not available anymore. One can nevertheless consider an approximate estimation via a polynomial $\hS(Y)$, with coefficients adjusted to minimize the MSE. Since $S$ and $B$ are even it is sufficient to include only the monomials of odd degree in $\hS(Y)$, the simplest non-trivial deviations from the Gaussian denoiser is thus observable for a polynomial of degree 3, that we shall denote $\hS(Y)=u_1 Y + u_3 Y^3$. It is a simple exercise to show that the choice of $(u_1,u_3)$ that minimizes the MSE is given by
\beq
\begin{pmatrix} u_1 \\ u_3 \end{pmatrix} = \begin{pmatrix} \E[Y^2] & \E[Y^4] \\  \E[Y^4] & \E[Y^6] \end{pmatrix}^{-1}  \begin{pmatrix} 1 \\ \E[S^4] + 3 \Delta \end{pmatrix} \ ,
\eeq
where the moments of $Y$ can be expressed in terms of those of $S$ and $B$. This reveals that generically $u_3 \neq 0$ as soon as $S$ or $B$ is not Gaussian. For future convenience let us consider the case where $S$ is Gaussian (hence $\E[S^4]=3$ and $\E[S^6]=15$) while $B$ is arbitrary. The coefficients of the optimal third order polynomial are then found to be
\beq
\begin{pmatrix} u_1 \\ u_3 \end{pmatrix} = \begin{pmatrix} \frac{6 (1+\Delta)^3 + 3 \Delta^2(4-\Delta) (\E[B^4]-3) + \Delta^3 (\E[B^6]-15)}{6 (1+\Delta)^4 + 3 \Delta^2 (1+\Delta)(3-2\Delta) (\E[B^4]-3) + \Delta^3 (1+\Delta) (\E[B^6]-15) - \Delta^4 (\E[B^4]-3)^2} \\
\frac{\Delta^2 (3-\E[B^4])}{6 (1+\Delta)^4 + 3 \Delta^2 (1+\Delta)(3-2\Delta) (\E[B^4]-3) + \Delta^3 (1+\Delta) (\E[B^6]-15) - \Delta^4 (\E[B^4]-3)^2} \end{pmatrix} \ ,
\label{eq_scalar_order3}
\eeq
which only reduce to $(u_1,u_3)=\left(\frac{1}{1+\Delta},0 \right)$ when $B$ is also Gaussian (or more precisely when its first sixth moments match those of a Gaussian).

\subsection{Equivariant estimators}
\label{sec_universality_equivariant}

The invariance under conjugation by orthogonal matrices of the Wishart ensemble is in general broken when the law of $X_{1,1}$ is arbitrary, and only restored when the latter is Gaussian; similarly the GOE is the only member of the family of Wigner ensembles that is in addition invariant under $\cO_n$. As a consequence we cannot exploit this symmetry to restrict the space of polynomial estimators as we did in Sec.~\ref{sec_rot_equivariant}. Fortunately we can nevertheless rely on a weaker symmetry of the arbitrary Wishart and Wigner ensembles, namely their invariance under permutations of their indices, that we formalize now.

We denote $\cS_n$ the symmetric group of permutations of $[n]$, whose neutral element is the identity and where the multiplication corresponds to the composition of functions. We let one of its element $\sigma \in \cS_n$ act on a symmetric matrix $Y \in M_n^{\rm sym}(\R)$ according to $(\sigma \cdot Y)_{i,j} = Y_{\sigma^{-1}(i),\sigma^{-1}(j)}$. One can check that $\tau \cdot (\sigma \cdot Y) = (\tau \circ \sigma) \cdot Y$, hence this definition provides a linear representation of $\cS_n$ on $M_n^{\rm sym}(\R)$. One can view $\cS_n$ as a subgroup of $\cO_n$: for $\sigma \in \cS_n$ let us denote $O(\sigma) \in \cO_n$ the permutation matrix associated to $\sigma$, with matrix elements $O(\sigma)_{i,j} = \delta_{i,\sigma(j)}$. One has indeed $O(\tau) O(\sigma) = O(\tau \circ \sigma)$, and the action of $\cS_n$ defined above can be rewritten as $\sigma \cdot Y = O(\sigma) Y O(\sigma)^T$, where one recognizes the conjugation by an orthogonal matrix. As a consequence of this inclusion the representation is isometric in the sense of Sec.~\ref{sec_gene_sym}.

The i.i.d. hypothesis on the random variables $X_{i,\mu}$ imply that $S^{(n)}$ is invariant under $\cS_n$, $\sigma \cdot S^{(n)} \eqd S^{(n)}$ for all $\sigma$ (this would still be true under the weaker hypothesis of $X_i = \{ X_{i,\mu}\}$ i.i.d. vectors in $\R^r$); the same conclusion holds for $Z^{(n)}$ by the i.i.d. character of its diagonal and off-diagonal matrix elements (note indeed that the action of $\cS_n$ on matrices does not mix these two families of elements), hence $(\sigma \cdot S,\sigma \cdot Y) \eqd (S,Y)$. The group of permutations being a symmetry of the inference problem in the sense of Sec.~\ref{sec_general}, we can therefore restrict the variational space of polynomial estimators to the equivariant ones, i.e. the (polynomial) functions $b_\beta : M_n^{\rm sym}(\R) \to M_n^{\rm sym}(\R)$ such that $b_\beta(\sigma \cdot Y) = \sigma \cdot b_\beta(Y)$ for all $\sigma \in \cS_n$. We show in Appendix~\ref{app_equivariant_permu} (see also~\cite{PuLiKiMaLi23}) that these functions are indexed by unlabelled marked multigraphs, in the following sense. Consider a finite set $V$ of vertices, and a multiset $E$ of edges, that are unordered pairs of (not necessarily distinct) vertices (at variance with a simple graph, in $(V,E)$ we allow multiple edges and self-loops, i.e. edges from one vertex to itself); distinguish in addition two vertices $v,w \in V$, called the marked vertices, that are not necessarily distinct. For a given $G=(V,E,v,w)$ one defines the function $\hb_G : M_n^{\rm sym}(\R) \to M_n^{\rm sym}(\R)$ according to
\begin{align}
  (\hb'_G(Y))_{i,j} & = 
\sum_{\substack{\phi \in \cI(V,[n]) \\ \phi(v)=i,\phi(w)=j}} \prod_{e=\{a,b\} \in E} Y_{\phi(a),\phi(b)} \ , \label{eq_bpG} \\
  (\hb_G(Y))_{i,j} & = \frac{1}{2} ( (\hb'_G(Y))_{i,j} + (\hb'_G(Y))_{j,i}) \ , \label{eq_bG}
\end{align}
with $\cI(E,F)$ denoting the set of injective functions from $E$ to $F$.

One can easily check that such functions are equivariant, the reasoning presented in App.~\ref{app_equivariant_permu} shows that all equivariant polynomials can be written as linear combinations of such $\hb_G$'s. We present in Fig.~\ref{fig_bG} a few examples of these graphs, their algebraic translations being written in Eq.~(\ref{eq_babcd}). Let us make a series of elementary remarks on this definition:
\begin{itemize}

\item similar parametrizations of permutation invariant functions are quite frequent in the litterature, see for instance~\cite{Ma20} in the context of traffic distribution, \cite{PuLiKiMaLi23} for their use in graph neural networks, and~\cite{MoWe22} for the low-degree polynomial approach to rank-one matrix factorization.

\item the matrix elements of $\hb_G(Y)$ are polynomial functions of the matrix elements of $Y$, of total degree equal to $|E|$, the number of edges of the graph (counting the possible edge multiplicities).

\item an isolated unmarked vertex only contribute by a multiplicative constant to $\hb_G$, one can therefore assume that $G$ does not contain any such vertex.

\item if $G$ and $H$ are isomorphic, i.e. if there exists a bijection between the vertex sets of $G$ and $H$ that preserves the edges (with their multiplicities) and the marked vertices, then $\hb_G=\hb_H$. As a consequence one can keep a single representant of each of these equivalence classes in the set $\A$ of basic functions, in other words consider unlabelled graphs. We will denote $\A^{(D)}$ the set of such unlabelled graphs with at most $D$ edges.

\item if the two marks are on the same vertex, $v=w$, then $\hb_G(Y) = \hb'_G(Y)$ is a diagonal matrix; if on the contrary $v \neq w$ then the diagonal of $\hb_G(Y)$ vanishes. This property reflects the fact that the orbits of diagonal and off-diagonal elements are disjoint under the action of $\cS_n$, as already mentioned.

\item the injectivity condition on $\phi$ imply that the indices $\{\phi(v)\}_{v \in V}$ are all distinct. One can actually build an equivalent set of basic functions $\tb_G$ by dropping this condition and summing instead on all functions $\phi:V \to [n]$. Indeed, $\tb_G$ can then be written as a linear combinations of some $\hb_H$, where $H$ runs over the graphs obtained from $G$ by the identification of some vertices of $G$ (namely those that share a common value of $\phi$). Using the M\"obius function of the lattice of partitions (see e.g. exercise 10.33 in~\cite{NiSp_book} and lemma 2.6 in~\cite{Ma20}) one can invert this relation and write the $\hb_G$ as linear combinations of the $\tb_H$. 

\item one can give an intuitive justification of the form (\ref{eq_bpG}) (or more precisely of $\tb_G$ discussed in the previous point) according to the following reasoning: a matrix element $f(Y)_{i,j}$ of an equivariant polynomial of degree $m$ is a linear combination of terms of the form $Y_{i_1,i_2} \dots Y_{i_{2m-1},i_{2m}}$. Among these indices some will be equal to $i$ in all the terms, some to $j$, and others should be summed over. For the linear combination to be equivariant under the permutations no index in $[n]$ can play a special role, hence the indices that are summed upon should be summed uniformly over $[n]$. The graph structure allows precisely to encode which among the $2m$ indices $i_1,\dots,i_{2m}$ are fixed to $i$ or $j$, and which are summed upon; since the $m$ matrices $Y$ are identical all the edges play the same role, and since $Y$ is symmetric the edges are undirected.

\end{itemize}

If the distributions of the random variables $X_{1,1}$ and $B_{1,2}$ have the additional inversion invariance one can further restrict the candidate basic functions $\hb_G$. As shown in App.~\ref{app_equivariant_inversion} the only non-trivial contributions arise from the $G$ that satisfy the following conditions:
\begin{itemize}
\item unmarked vertices must have an even degree, where the degree of a vertex is the number of edges in which it appears, counting their multiplicities, and with each self-loop counting for 2.
\item if $v=w$ the marked vertex must have an even degree (in the above sense).
\item if $v \neq w$ the marked vertices must have odd degrees.
\end{itemize}

Since the orbits of the action of the permutation group on matrices do not mix the diagonal and off-diagonal elements, one can consider separately the optimization of the MSE for the diagonal and off-diagonal parts. We shall thus decompose the MSE of an estimator as
\begin{align}
\MSE(\hS) &= \MSEd(\hS) + \MSEod(\hS) \ , \\
\MSEd(\hS) &= \frac{1}{n} \sum_{i=1}^n \E[(S_{i,i} - \hS(Y)_{i,i})^2 ]  \ , \\
\MSEod(\hS) &= \frac{1}{n} \sum_{\substack{i,j=1 \\ i \neq j}}^n \E[(S_{i,j} - \hS(Y)_{i,j})^2 ]  \ .
\end{align}
We define as before $\MMSE^{(n,D)}$ as the minimal MSE among estimators that are polynomial of degree at most $D$; one has $\MMSE^{(n,D)} = \MMSEd^{(n,D)} + \MMSEod^{(n,D)}$, the minimization can be performed independently on the two parts, with basis functions indexed by $\Ad^{(D)}$ and $\Aod^{(D)}$ respectively, where $\Ad^{(D)}$ is the set of unlabelled multigraphs with at most $D$ edges and a vertex marked twice, while in $\Aod^{(D)}$ the two marked vertices are necessarily distinct. Since the number $n$ of diagonal elements is negligible compared to the $n(n-1)$ off-diagonal elements, the contribution $\MMSEd^{(n,D)}$ is of order $1/n$; in fact the trivial estimator $\hS(Y)_{i,i}=0$, which corresponds to the prior mean with the centering of the signal matrix, achieves a diagonal MSE of $\MSEd(0) =\E[S_{1,1}^2]= \frac{1}{n}(\E[X_{1,1}^4] - 1)$ (see the Appendix \ref{app_order2}  for details on these moment computations), and the minimal MSE can only be smaller.

\subsection{The $D=2$ case}
\label{sec_order2}

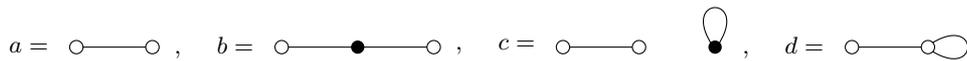
\begin{figure}
\begin{tikzpicture}
\draw (0,0) -- (1,0);
\fill[white] (0,0) circle (2.5pt);
\draw (0,0) circle (2.5pt);
\fill[white] (1,0) circle (2.5pt);
\draw (1,0) circle (2.5pt);
\draw (0,0) node [left] {$a=$\phantom{=}  };
\begin{scope}[xshift=2.7cm]
\draw (0,0) -- (2,0);
\fill[white] (0,0) circle (2.5pt);
\draw (0,0) circle (2.5pt);
\fill[white] (2,0) circle (2.5pt);
\draw (2,0) circle (2.5pt);
\fill[black] (1,0) circle (2.5pt);
\draw (0,0) node [left] {, \phantom{=} $b=$\phantom{=}  };
\end{scope}
\begin{scope}[xshift=6.4cm]
\draw (0,0) -- (1,0);
\fill[white] (0,0) circle (2.5pt);
\draw (0,0) circle (2.5pt);
\fill[white] (1,0) circle (2.5pt);
\draw (1,0) circle (2.5pt);
\fill[black] (2,0) circle (2.5pt);
\draw (1.98,0) ..controls (1.5,.7) and (2.5,.7).. (2.02,0);
\draw (0,0) node [left] {, \phantom{=} $c=$\phantom{=}  };
\end{scope}
\begin{scope}[xshift=10.2cm]
\draw (0,0) -- (1,0);
\draw (1,0.02) ..controls (1.7,.5) and (1.7,-.5).. (1,-0.02);
\fill[white] (0,0) circle (2.5pt);
\draw (0,0) circle (2.5pt);
\fill[white] (1,0) circle (2.5pt);
\draw (1,0) circle (2.5pt);
\draw (0,0) node [left] {, \phantom{=} $d=$\phantom{=}  };
\end{scope}
\end{tikzpicture}
\caption{The graphs in $\Aod^{(2)}$ that contribute to the off-diagonal estimators at order 2, in presence of the inversion symmetry.
White (resp. black) circles represent the marked (resp. unmarked) vertices. The corresponding formulas for these four estimators can be found in equation (\ref{eq_babcd}). }
\label{fig_bG}
\end{figure}

In this section we present the results of a detailed study of the case $D=2$, including the (non-universal) corrections of order at most $1/n$ to $\MMSE^{(n,2)}$. As explained before the dominant contribution comes from the off-diagonal terms; we present in Fig.~\ref{fig_bG} the four graphs that contribute to it up to degree 2, if one assumes the inversion invariance. These are indeed the only graphs with two distinct marked vertices and at most two edges respecting the degree constraints stated above (without the inversion symmetry one would have 28 additional graphs to consider, that would contribute to the corrections of order $1/n$). The corresponding four estimators, denoted $\hb_a,\dots,\hb_d$, vanish on their diagonal and have for $i\neq j$:
\beq
(\hb_a(Y))_{i,j} = Y_{i,j} \ , \quad
(\hb_b(Y))_{i,j} = \sum_{k \neq i,j} Y_{i,k} Y_{k,j} \ , \quad
(\hb_c(Y))_{i,j} = Y_{i,j} \sum_{k \neq i,j} Y_{k,k} \ , \quad
(\hb_d(Y))_{i,j} = \frac{1}{2} Y_{i,j} (Y_{i,i} + Y_{j,j}) \ .
\label{eq_babcd}
\eeq
It is equivalent, and will turn out to simplify the following discussion, to consider the four functions $b_1,\dots,b_4$, whose linear combinations span the same space:
\beq
(b_1(Y))_{i,j} = Y_{i,j} \ , \quad
(b_2(Y))_{i,j} =  (Y^2)_{i,j} \ , \quad
(b_3(Y))_{i,j} = Y_{i,j} \Tr(Y) \ , \quad
(b_4(Y))_{i,j} = \sqrt{n} \frac{1}{2} Y_{i,j} (Y_{i,i} + Y_{j,j}) \ .
\label{eq_order2_b1b4}
\eeq
Note that the first three functions are actually invariant under $\cO_n$ and coincide with the polynomials studied in the previous section, while the fourth one is invariant under the permutations but not under the orthogonal group.

Following the strategy already illustrated before, we write
\beq
\MMSEod^{(n,2)}= (n-1) \E[S_{1,2}^2] + \inf_{c \in \R^4}[c^T \M^{(n,2)} c - 2 c^T \cR^{(n,2)} ] \ ,
\label{eq_order2_MMSE}
\eeq
with the 4-dimensional matrix $\M^{(n,2)}$ and vector $\cR^{(n,2)}$ defined as
\beq
\M^{(n,2)}_{\beta,\beta'} = (n-1) \E[(b_\beta(Y))_{1,2} (b_{\beta'}(Y))_{1,2}  ] \ , \qquad
\cR^{(n,2)}= (n-1) \E[S_{1,2}  (b_\beta(Y))_{1,2}  ] \ .
\label{eq_order2_MR}
\eeq
We exploited the permutation invariance to single out the contribution of an arbitrary off-diagonal element in the sums. By elementary, even if slightly cumbersome, computations we obtained the exact expressions of $\M^{(n,2)}$ and $\cR^{(n,2)}$, that are reported in Appendix \ref{app_order2}. They depend on the universal parameters $\alpha$ and $\Delta$, on $n$, as well as on the non-universal quantities $\E[X_{1,1}^4]$, $\E[X_{1,1}^6]$ and $\E[B_{1,1}^2]$. A reassuring sanity check is provided by setting these quantities to the values they take when $X_{1,1}$ is Gaussian and when the noise is distributed according to the GOE (namely $\E[X_{1,1}^4]=3$, $\E[X_{1,1}^6]=15$ and $\E[B_{1,1}^2]=2$): one finds indeed in this case that the minimizer of the MSE has $c_4=0$ exactly for all finite $n$, restoring the orthogonal invariance. Coming back to generic distributions and taking the $n\to \infty$ limit in the formulas of the Appendix one finds
\beq
\M^{(\infty,2)} = \begin{pmatrix}
1+\Delta & \sqrt{\alpha} & 0 & 0 \\
\sqrt{\alpha} & (1+\Delta)^2+\alpha & 0 & 0 \\
0 & 0 & (1+\Delta)(\E[X_{1,1}^4] - 1 + \Delta  \E[B_{1,1}^2]) & 0 \\
0 & 0 & 0 & \frac{1}{2} (1+\Delta)(\E[X_{1,1}^4] - 1 + \Delta  \E[B_{1,1}^2])
\end{pmatrix} \ ,
\label{eq_order2_Minfty}
\eeq
and $\cR^{(\infty,2)} = (1,\sqrt{\alpha},0,0)^T$. The matrix $\M^{(\infty,2)}$ being invertible, we can perform the minimization with these limit coefficients, which yields for the optimal parameters $c^{(\infty,2)} = \frac{1}{(1+\Delta)^3 + \alpha \Delta}((1+\Delta)^2,\sqrt{\alpha} \Delta,0,0)^T$ and for the MMSE (using $\E[S_{1,2}^2]=1/n$):
\beq
\lim_{n \to \infty} \MMSE^{(n,2)} = \lim_{n \to \infty} \MMSEod^{(n,2)} = 1 - (\cR^{(\infty,2)})^T  c^{(\infty,2)} = \frac{\Delta (1+\Delta)^2}{(1+\Delta)^3 + \alpha \Delta} \ .
\eeq
These quantities are indeed universal, and coincide with the values given in (\ref{eq_example_MMSE}) for the MMSE and in (\ref{eq_example_coefs}) for the optimal scalar denoiser in the orthogonally invariant case (the first row of $\hc^{(2)}$ in (\ref{eq_example_coefs}) does not appear here since it only contributes to the diagonal of the denoiser).

To be precise the invertibility of $\M^{(\infty,2)}$ is only ensured if $\E[X_{1,1}^4] - 1 + \Delta  \E[B_{1,1}^2]>0$, and fails to be true in the very special case where $X_{1,1}=\pm 1$ with equal probability and where the noise vanishes on the diagonal. However the conclusion above on the large $n$ limit of $\MMSE^{(n,2)}$ remains true in this case: an inspection of the coefficients detailed in the Appendix \ref{app_order2} reveals indeed that the MSE is then strictly independent of $c_3$ and $c_4$, for all $n$.

With some additional work, and the help of a symbolic computation software (see Appendix \ref{app_order2} for more details) we obtained the first correction to the off-diagonal MMSE in the large $n$ limit,
\begin{align}
\MMSEod^{(n,2)} & =\frac{\Delta (1+\Delta)^2}{(1+\Delta)^3 + \alpha \Delta}   \\
& +\frac{1}{n}
\frac{
\alpha \Delta^2 (\alpha - 2 (1+\Delta)^3) (\E[X_{1,1}^4]-1)^2
+\Delta ( \alpha \Delta (1+\Delta)^2 - (1+\Delta)^5 - \alpha^2 \Delta  ) (\E[X_{1,1}^4]-1)
}
{((1+\Delta)^3 + \alpha \Delta)^2 (\E[X_{1,1}^4]-1+ \Delta \E[B_{1,1}^2])} \\
& + \frac{1}{n}
\frac{\Delta^2 ( \alpha \Delta (1+\Delta)^2 - (1+\Delta)^5 +\alpha^2 \Delta (\E[X_{1,1}^4]-2)  ) \E[B_{1,1}^2]
}
{((1+\Delta)^3 + \alpha \Delta)^2 (\E[X_{1,1}^4]-1+ \Delta \E[B_{1,1}^2])} + O\left(\frac{1}{n^2}\right) \ ,
\label{eq_order2_odcorrections}
\end{align}
and for completeness we also give the dominant term of the diagonal contributions, that is of order $1/n$ (see again Appendix \ref{app_order2} for additional explanations):
\begin{align}
& \MMSEd^{(n,2)} = \frac{1}{n} (\E[X_{1,1}^4] - 1)  \times \label{eq_order2_d} \\ & \times \left[1 - \frac{(\E[X_{1,1}^4] - 1) ( 2 (1+\Delta)^2 + \Delta^2 (\E[B_{1,1}^4]-3) + \alpha \Delta \E[B_{1,1}^2]) }{ (\E[X_{1,1}^4]-1+ \Delta \E[B_{1,1}^2]) ( 2 (1+\Delta)^2 + \Delta^2 (\E[B_{1,1}^4]-3) + \alpha (\E[X_{1,1}^4]-1)) - \alpha (\E[X_{1,1}^4]-1)^2 }  \right] + O\left(\frac{1}{n^2}\right) \ .
\nonumber
\end{align}

These expressions are not particularly enlightening, but the important point that we want to emphasize is the universality of the large $n$ limit of $\MMSE^{(n,2)}$, the non-universal parameters $\E[X_{1,1}^4]$, $\E[X_{1,1}^6]$ and $\E[B_{1,1}^2]$ appearing only in the terms of order $1/n$.

\subsection{The $D=3$ case}
\label{sec_order3}

\begin{figure}
\begin{tikzpicture}
\fill[black] (1,0) circle (2.5pt);
\fill[black] (2,0) circle (2.5pt);
\draw (0,0) -- (3,0);
\fill[white] (0,0) circle (2.5pt);
\fill[white] (3,0) circle (2.5pt);
\draw (0,0) circle (2.5pt);
\draw (3,0) circle (2.5pt);
\draw (3,0) node [right] {\phantom{=} ,};

\begin{scope}[xshift=4.3cm]
\fill[black] (1,0) circle (2.5pt);
\fill[black] (3,0) circle (2.5pt);
\draw (0,0) -- (2,0);
\fill[white] (0,0) circle (2.5pt);
\fill[white] (2,0) circle (2.5pt);
\draw (0,0) circle (2.5pt);
\draw (2,0) circle (2.5pt);
\draw (2.98,0) ..controls (2.5,.7) and (3.5,.7).. (3.02,0);
\draw (3,0) node [right] {\phantom{=} ,};
\end{scope}

\begin{scope}[xshift=8.5cm]
\draw (2,0.02) to[in=160,out=20] (3,0.02);
\draw (2,-0.02) to[in=200,out=-20] (3,-0.02);
\fill[black] (2,0) circle (2.5pt);
\fill[black] (3,0) circle (2.5pt);
\draw (0,0) -- (1,0);
\fill[white] (0,0) circle (2.5pt);
\fill[white] (1,0) circle (2.5pt);
\draw (0,0) circle (2.5pt);
\draw (1,0) circle (2.5pt);
\draw (3,0) node [right] {\phantom{=} ,};
\end{scope}

\begin{scope}[xshift=12.7cm]
\fill[black] (2,0) circle (2.5pt);
\draw (1.98,0) ..controls (1.5,.7) and (2.5,.7).. (2.02,0);
\fill[black] (3,0) circle (2.5pt);
\draw (2.98,0) ..controls (2.5,.7) and (3.5,.7).. (3.02,0);
\draw (0,0) -- (1,0);
\fill[white] (0,0) circle (2.5pt);
\fill[white] (1,0) circle (2.5pt);
\draw (0,0) circle (2.5pt);
\draw (1,0) circle (2.5pt);
\draw (3,0) node [right] {\phantom{=} ,};
\end{scope}

\begin{scope}[yshift=-1.3cm]
\draw (0,0) -- (2,0);
\fill[black] (1,0) circle (2.5pt);
\draw (-.02,0) ..controls (-.5,.7) and (.5,.7).. (.02,0);
\fill[white] (0,0) circle (2.5pt);
\draw (0,0) circle (2.5pt);
\fill[white] (2,0) circle (2.5pt);
\draw (2,0) circle (2.5pt);
\draw (2,0) node [right] {\phantom{=} ,};
\end{scope}

\begin{scope}[yshift=-1.3cm,xshift=3.3cm]
\draw (0,0) -- (2,0);
\fill[black] (1,0) circle (2.5pt);
\draw (.98,0) ..controls (.5,.7) and (1.5,.7).. (1.02,0);
\fill[white] (0,0) circle (2.5pt);
\draw (0,0) circle (2.5pt);
\fill[white] (2,0) circle (2.5pt);
\draw (2,0) circle (2.5pt);
\draw (2,0) node [right] {\phantom{=} ,};
\end{scope}

\begin{scope}[yshift=-1.3cm,xshift=6.5cm]
\draw (1,0.02) to[in=160,out=20] (2,0.02);
\draw (1,-0.02) to[in=200,out=-20] (2,-0.02);
\draw (0,0) -- (1,0);
\fill[white] (0,0) circle (2.5pt);
\draw (0,0) circle (2.5pt);
\fill[white] (1,0) circle (2.5pt);
\draw (1,0) circle (2.5pt);
\fill[black] (2,0) circle (2.5pt);
\draw (2,0) node [right] {\phantom{=} ,};
\end{scope}

\begin{scope}[yshift=-1.3cm,xshift=9.7cm]
\draw (1.98,0) ..controls (1.5,.7) and (2.5,.7).. (2.02,0);
\draw (1.98,0) ..controls (1.5,-.7) and (2.5,-.7).. (2.02,0);
\fill[white] (0,0) circle (2.5pt);
\draw (0,0) circle (2.5pt);
\draw (0,0) -- (1,0);
\fill[white] (0,0) circle (2.5pt);
\draw (0,0) circle (2.5pt);
\fill[white] (1,0) circle (2.5pt);
\draw (1,0) circle (2.5pt);
\fill[black] (2,0) circle (2.5pt);
\draw (2,0) node [right] {\phantom{=} ,};
\end{scope}

\begin{scope}[yshift=-1.3cm,xshift=13cm]
\draw (1.98,0) ..controls (1.5,.7) and (2.5,.7).. (2.02,0);
\draw (-.02,0) ..controls (-.5,.7) and (.5,.7).. (.02,0);
\fill[white] (0,0) circle (2.5pt);
\draw (0,0) circle (2.5pt);
\draw (0,0) -- (1,0);
\fill[white] (0,0) circle (2.5pt);
\draw (0,0) circle (2.5pt);
\fill[white] (1,0) circle (2.5pt);
\draw (1,0) circle (2.5pt);
\fill[black] (2,0) circle (2.5pt);
\draw (2,0) node [right] {\phantom{=} ,};
\end{scope}

\begin{scope}[yshift=-2.6cm]
\draw (-.02,0) ..controls (-.5,.7) and (.5,.7).. (.02,0);
\draw (0.98,0) ..controls (.5,.7) and (1.5,.7).. (1.02,0);
\draw (0,0) -- (1,0);
\fill[white] (0,0) circle (2.5pt);
\draw (0,0) circle (2.5pt);
\fill[white] (1,0) circle (2.5pt);
\draw (1,0) circle (2.5pt);
\draw (1,0) node [right] {\phantom{=} ,};
\end{scope}

\begin{scope}[yshift=-2.6cm,xshift=2.3cm]
\draw (0,0) -- (1,0);
\draw (-.02,0) ..controls (-.5,.7) and (.5,.7).. (.02,0);
\draw (-.02,0) ..controls (-.5,-.7) and (.5,-.7).. (.02,0);
\fill[white] (0,0) circle (2.5pt);
\draw (0,0) circle (2.5pt);
\fill[white] (1,0) circle (2.5pt);
\draw (1,0) circle (2.5pt);
\draw (1,0) node [right] {\phantom{=} ,};
\end{scope}

\begin{scope}[yshift=-2.6cm,xshift=4.6cm]
\draw (0,0) -- (1,0);
\draw (0,0.02) to[in=160,out=20] (1,0.02);
\draw (0,-0.02) to[in=200,out=-20] (1,-0.02);
\fill[white] (0,0) circle (2.5pt);
\draw (0,0) circle (2.5pt);
\fill[white] (1,0) circle (2.5pt);
\draw (1,0) circle (2.5pt);
\draw (1,0) node [right] {\phantom{=} .};
\end{scope}

\end{tikzpicture}
\caption{The 12 graphs in $\Aod^{(3)} \setminus \Aod^{(2)}$ that complement those of Fig.~\ref{fig_bG} to describe the off-diagonal estimators of degree at most 3, in presence of the inversion symmetry. White (resp. black) circles represent the marked (resp. unmarked) vertices. }
\label{fig_order3}
\end{figure}
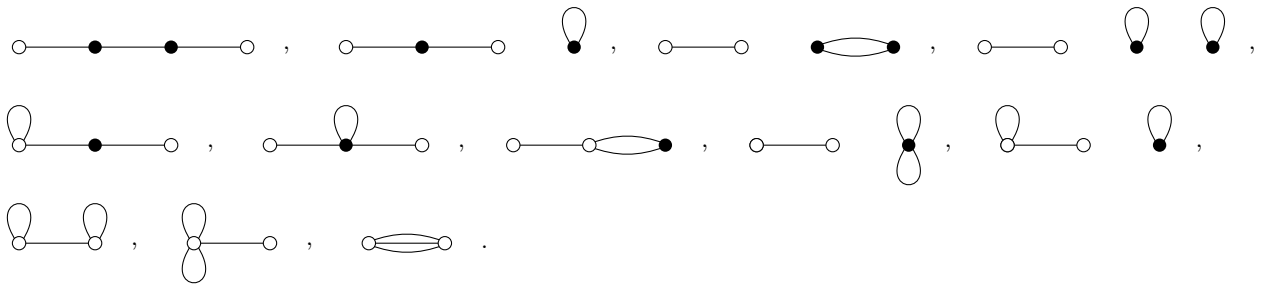

Let us now consider the estimators of degree at most 3, and discuss the large $n$ limit of $\MMSE^{(n,3)}$ (we will skip most of the details of the computations, that are generalizations of those presented in Appendix \ref{app_order2}). As previously explained we can concentrate on the off-diagonal part of the denoiser, which, in presence of the inversion invariance, is a linear combination of 16 terms: the 4 ones of degree 1 and 2 studied above, and 12 additional ones of degree 3 defined in terms of their underlying graphs in Fig.~\ref{fig_order3}. We did not attempt a full study of this 16-dimensional variational problem and settled for a simpler strategy that we believe captured nevertheless the correct asymptotics of $\MMSE^{(n,3)}$: we considered the minimization of the off-diagonal MSE in the 4-dimensional space of estimators spanned (for $i \neq j$) by $(b_1(Y))_{i,j} = Y_{i,j}$, $(b_2(Y))_{i,j} = (Y^2)_{i,j}$, $(b_3(Y))_{i,j} = (Y^3)_{i,j}$, which are the three orthogonally invariant relevant terms, complemented by $(b_4(Y))_{i,j}$ which was successively chosen among the last 8 graphs of Fig.~\ref{fig_order3} (properly centered and normalized). For 7 of these cases the asymptotic MMSE reached coincided with the orthogonally invariant result, independently of the signal and noise distributions; the coefficients $(c_1,c_2,c_3,c_4)$ of the linear combination of $b_1, \dots ,b_4$ that achieved the minimal MSE had $c_4=0$, and the first three coefficients took their orthogonally invariant values (i.e. $(c_1,c_2,c_3)=(\hc_1^{(3)},\hc_2^{(3)},\hc_3^{(3)})$, the latter having been given in Eq.~(\ref{eq_coefs_order3})). The only non-trivial case corresponds to the last graph of Fig.~\ref{fig_order3}, which amounts to take $(b_4(Y))_{i,j}=n Y_{i,j}^3$ (the factor $n$ is the proper normalization to ensure a non-trivial limit). This led us to 
\beq
\MMSE^{(3)} = \lim_{n \to \infty} \MMSE^{(n,3)} = 1 - (\cR^{(\infty,3)})^T  c^{(\infty,3)} \ , 
\eeq
where $c^{(\infty,3)}$ is the solution of $\M^{(\infty,3)} c = \cR^{(\infty,3)}$, these 4-dimensional matrices and vectors being given by:
\begin{align}
\M^{(\infty,3)}& = \begin{pmatrix}
1+\Delta & \sqrt{\alpha} & \alpha + 2 (1+\Delta)^2 & \nu_1 \\
\sqrt{\alpha} & \alpha +  (1+\Delta)^2 & \sqrt{\alpha} (\alpha + 4 (1+\Delta)) & 3 \sqrt{\alpha} (1+\Delta) \\
\alpha + 2 (1+\Delta)^2 & \sqrt{\alpha} (\alpha + 4 (1+\Delta)) & \alpha^2 + 2 \alpha (4+3 \Delta) + 5 (1+\Delta)^3 & (1+\Delta)  (3 \alpha + 2 \nu_1 ) \\
\nu_1 & 3 \sqrt{\alpha} (1+\Delta) & (1+\Delta)  (3 \alpha + 2 \nu_1 ) & \nu_2
\end{pmatrix} \ , \\
\cR^{(\infty,3)} & = ( 1 , \sqrt{\alpha} , \alpha + 2 (1+\Delta) , 3 (1+\Delta) )^T \ ,
\end{align}
where we introduced the two non-universal parameters:
\beq
\nu_1 = 3 + 6 \Delta + \Delta^2 \E[B_{1,2}^4] \ , \qquad \nu_2=15 + 45 \Delta + 15 \Delta^2 \E[B_{1,2}^4]  + \Delta^3 \E[B_{1,2}^6] \ . 
\eeq
Two special cases are worth investigating:
\begin{itemize}
\item when $\alpha=0$ one can check that the optimal vector of coefficients $c^{(\infty,3)}$ is $c_2=c_3=0$, while $(c_1,c_4)=(u_1,u_3)$, the coefficients defined in Eq.~(\ref{eq_scalar_order3}): indeed in this limit of ``very extensive rank'' with $r \gg n$ the matrix elements of $S$ behave as independent standard Gaussians, the matrix denoising thus reduces to independent scalar problems on each entry, which was precisely the toy model solved in Eq.~(\ref{eq_scalar_order3}) for polynomials of degree 3.
\item when $B_{1,2}$ is a standard Gaussian, in such a way that $\nu_1 = 3 (1+\Delta)^2$ and $\nu_2 = 15 (1+\Delta)^3$, one finds that $c_4=0$ in $c^{(\infty,3)}$, while $(c_1,c_2,c_3)=(\hc_1^{(3)},\hc_2^{(3)},\hc_3^{(3)})$, the values computed in the orthogonally invariant setting and defined in Eq.~(\ref{eq_coefs_order3}).
\end{itemize}
It seems unlikely to us that the study of the full 16-dimensional variational problem would yield a different value of $\MMSE^{(3)}$ (this would mean that linear combinations of estimators that are asymptotically irrelevant individually become relevant); if one is ready to admit this assertion the results we just presented confirm, for polynomial estimators of degree at most 3, the validity of the universality conjecture: the details of the law of $X_{1,1}$ do not appear in $\MMSE^{(3)}$, only those of the noise distribution $B_{1,2}$, and when the latter is Gaussian the asymptotic optimal estimator coincide with the one determined in the orthogonally invariant case, for any distribution of $X_{1,1}$.

\subsection{Arbitrary values of $D$}
\label{sec_universality_allD}

We did not manage to prove the universality conjecture, i.e. the independence of the MMSE on the law of $X$, and its coincidence with the orthogonally invariant result for a Gaussian noise, in its full generality; in this section and in the Appendix~\ref{app_universality} we shall content ourselves with a sketch of a possible roadmap towards a proof of the weak form of the conjecture for arbitrary $D$ (assuming of course it is true), in the Gaussian noise case; the global structure of the reasoning mirrors the proposition 3.3 in~\cite{MoWe22}, while the details are rather different.

At a first level of description the main idea is to define a set of basic functions $\{b_G \}_{G \in \A^{(D)}}$, indexed by the marked graphs with at most $D$ edges introduced in Sec.~\ref{sec_universality_equivariant}, whose linear combinations span all possible polynomial estimators that are equivariant under the permutation group. They are divided into the ``good'' and the ``bad'' ones, with the index set decomposed as the disjoint union $\A^{(D)} = \A^{(D,{\rm g})} \cup \A^{(D,{\rm b})} $. This distinction is made in such a way that the linear combinations of $\{b_G \}_{G \in \A^{(D,{\rm g})}}$ correspond to the estimators that are equivariant under the orthogonal group. The computation of $\MMSE^{(n,D)}$ goes as usual through the minimization of the quadratic plus linear form with the matrix $\M^{(n,D)}$ and vector $\cR^{(n,D)}$, indexed by $\A^{(D)}$, with elements $\M^{(n,D)}_{G,G'} = \E[\tr (b_G(Y) b_{G'}(Y) ) ]$ and $\cR^{(n,D)}_G = \E[\tr ( S \, b_G(Y))]$. These can be decomposed in blocks of good and bad components,
\beq
\M^{(n,D)} = \begin{pmatrix} \M^{(n,D,{\rm gg})} & \M^{(n,D,{\rm gb})} \\
(\M^{(n,D,{\rm gb})})^T & \M^{(n,D,{\rm bb})}
\end{pmatrix} \ , \qquad
\cR^{(n,D)} = \begin{pmatrix} \cR^{(n,D,{\rm g})} \\ \cR^{(n,D,{\rm b})} 
  \end{pmatrix} \ .
\eeq
Let us denote $\M^{(\infty,D)}$ and $\cR^{(\infty,D)}$ their large $n$ limit. The proof would be complete if one managed to show that 
\beq
\M^{(\infty,D)}\ \ \text{is invertible,}  \qquad  \M^{(\infty,D,{\rm gb})}=0\ , \qquad  \text{and} \ \cR^{(\infty,D,{\rm b})}=0 \ .
\label{eq_objective}
\eeq
As a matter of fact one could then exchange the large $n$ limit and the minimization (recall the discussion of Sec.~\ref{sec_gene_approx}) to write
\beq
\lim_{n \to \infty} \MMSE^{(n,D)} = \mu_{S,2} - (\cR^{(\infty,D)})^T (\M^{(\infty,D)})^{-1} \cR^{(\infty,D)} = \mu_{S,2} - (\cR^{(\infty,D,{\rm g})})^T (\M^{(\infty,D,{\rm gg})})^{-1} \cR^{(\infty,D,{\rm g})} \ . 
\eeq
The last expression only involves the good components, i.e. the orthogonally invariant ones, and as such reduces precisely to the computation performed in Sec.~\ref{sec_rot}. The latter involved the limit law $\mu_Y$, which is universal for the addition of an arbitrary Wishart matrix and a GOE distributed one. The fluctuation moments of the centered traces introduced in Eq.~(\ref{eq_fluctuations}) have instead a non-universal limit covariance, but the reasoning of Sec.~\ref{sec_irrelevance} only relied on the existence of a Gaussian limit, not on the explicit form of the covariance. One could thus conclude at this point in favor of the universality of $\lim_{n \to \infty} \MMSE^{(n,D)} $. Note that for $D=2$ this is precisely what we obtained above: the limit matrix in Eq.~(\ref{eq_order2_Minfty}) has the correct invertible block diagonal structure, the good components being the first three ones, and the bad row vanishes in $\cR^{(\infty,2)} $. We defer to Appendix~\ref{app_universality} some additional details on a possible implementation of this strategy, and the difficulties it involves.

\section{Numerical results on finite-size matrices}
\label{sec_numerics_finiten}

We present in this Section the results of numerical simulations on the denoising of matrices of finite size $n$. We used as a signal distribution the Wishart ensemble defined in Eq.~(\ref{eq_def_Wishart_gene}), with matrix elements $X_{i,\mu}$ drawn either from the Gaussian law with zero mean and variance 1, or from the Rademacher-Bernouilli law where $X_{i,\mu}=0$ with probability $\rho$, $1/\sqrt{\rho}$ with probability $\rho/2$ and $-1/\sqrt{\rho}$ with the same probability, in such a way that the first two moments of $X_{i,\mu}$ coincide in these two ensembles. The noise was extracted from the GOE, with a width parametrized by $\Delta$. The BABP denoiser $\hS(Y)$ was implemented by a numerical diagonalization of the matrix $Y$, the function $\DB$ of Eqs.~(\ref{eq_def_DB_Gaussian},\ref{eq_def_DB_2}) being applied on each of its eigenvalues $\lambda$; the computation of this scalar function was performed with Cardan's formula to solve the cubic equation (\ref{eq_cubic}) with $z=\lambda$ (i.e. from the asymptotic $n\to\infty$ distribution $\mu_Y$). As explained in Sec.~\ref{sec_numerics_On} the relevant root can be unambiguously selected on the support of $\mu_Y$ by choosing the one with a positive imaginary part; when $\lambda$ is outside the support (which can happen at finite $n$) we took the unique branch that ensures both the continuity of $h_Y(\lambda)$ on $\R$ and the asymptotic behavior $h_Y(\lambda) \sim 1/\lambda$ as $\lambda \to \pm \infty$.

\begin{figure}
\includegraphics[width=8cm]{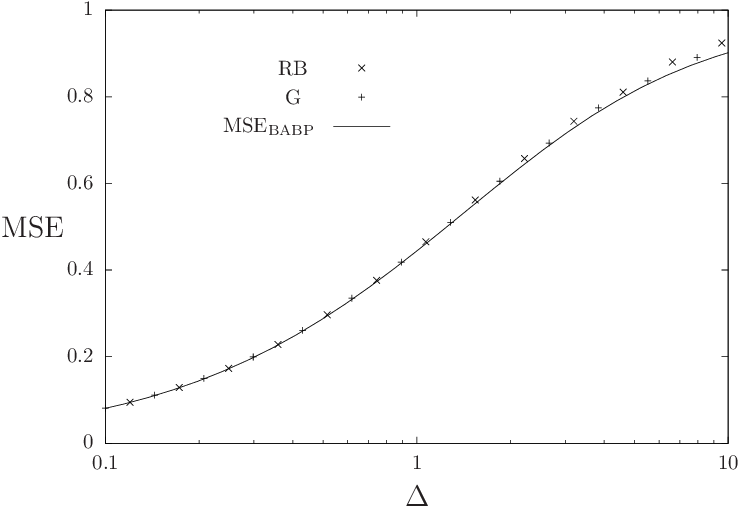}
\hspace{1cm}
\includegraphics[width=8cm]{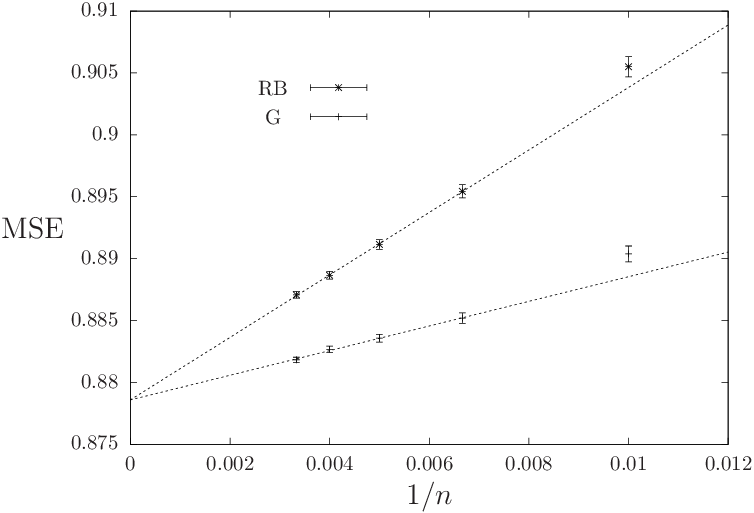}
\caption{The MSE reached by the BABP denoiser on Wishart matrices with aspect ratio $\alpha=1$ and $X_{i,\mu}$ drawn as a standard Gaussian (points denoted G) or as a Rademacher-Bernouilli random variable with $\rho=0.2$ (RB). Left panel: the MSE is plotted as a function of $\Delta$, the solid line is the analytical prediction from Eq.~(\ref{eq_MMSEB_Gaussian}), the symbols correspond to matrices of size $n=100$, each point is an average over 200 repetitions, error bars are of the order of the symbol size. Right panel: study of the finite-size effects for $\Delta=8$. The symbols have been obtained from simulations with matrices of size $n$ between 100 and 300, each point corresponding to an average over 2000 repetitions. The dashed lines are fits of the data of the form $a+b/n$, with the asymptotic value $a$ fixed to the analytical prediction $\MMSEB$.}
\label{fig_fs_BABPMMSE}
\end{figure}

In the left panel of Fig.~\ref{fig_fs_BABPMMSE} we plot the MSE achieved by the BABP denoiser as a function of the noise intensity $\Delta$ for a given aspect ratio $\alpha$, and compare it to the analytical prediction $\MMSEB$ from Eq.~(\ref{eq_MMSEB_Gaussian}). The agreement, which can be expected to hold exactly only in the large $n$ limit, is very good already for matrices of size $n=100$, except at large noise $\Delta$ where some deviations are visible. These finite-size effects are further studied in the right panel  of Fig.~\ref{fig_fs_BABPMMSE}, where the MSE is now plotted as a function of $1/n$ for one choice of $(\alpha,\Delta)$. The data extrapolates in a very accurate way to the analytical prediction in the $n\to\infty$ limit, with dominant corrections of order $1/n$. Recall that the analytical formula (\ref{eq_MMSEB_Gaussian}) was obtained by taking the $n\to\infty$ limit after the $D\to \infty$ limit, hence this numerical result supports the hypothesis that the same result is obtained when the limits are taken in the reverse order. Furthermore, the two ensembles of Wishart matrices (Gaussian and Rademacher-Bernoulli) exhibit different finite-size corrections, but the extrapolation of the MSE in the large $n$ limit coincide with $\MMSEB$ for both of them. Of course this is only an evidence for the universality of the MSE of the BABP estimator, which nevertheless shows that the properties of the eigenvectors of the matrices $S$ in both ensembles are sufficiently similar for the computations recalled in (\ref{eq_BABP_SE}) to give the same asymptotic result. This numerical result has no direct bearing on the conjecture of universality of the Bayes Optimal estimator, which could be distinct from the BABP one in the non-orthogonally invariant Rademacher-Bernoulli ensemble.

\begin{figure}
\includegraphics[width=8cm]{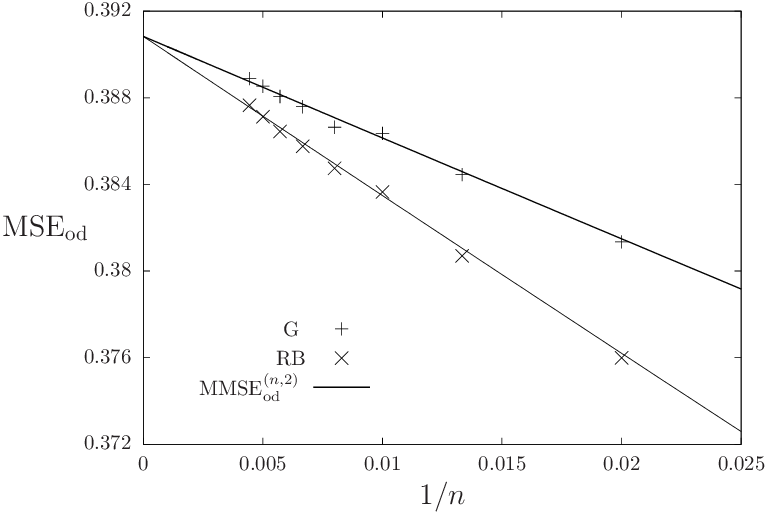}
\hspace{1cm}
\includegraphics[width=8cm]{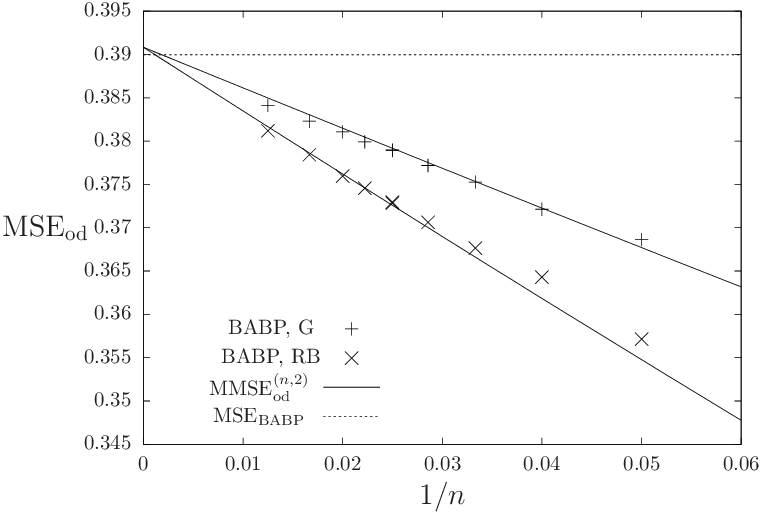}
\caption{A numerical study of the second order denoiser of Sec.~\ref{sec_order2}, for $\alpha=1$ and $\Delta=0.8$. Left panel: the two solid lines are the predictions MMSE$_{\rm od}^{(n,2)}$ of Eq.~(\ref{eq_order2_MMSE}), with the non-universal parameters corresponding to $X_{i,\mu}$ of laws either Gaussian (G) or Rademacher-Bernoulli (RN) with $\rho=0.2$. The symbols represent the results of numerical simulations with matrices of size between $n=50$ and $n=225$, averaged over $10^4$ repetitions for the smallest sizes, $10^3$ for the largest ones (error bars are of the order of the symbol size). Right panel: the solid lines are the same as in the left panel, the symbols are obtained with the BABP estimator, each point is an average over $10^4$ samples. }
\label{fig_fs_order2}
\end{figure}

In figure~\ref{fig_fs_order2} we report some results obtained with the second order permutation invariant denoiser studied in Sec.~\ref{sec_order2}, that estimates the off-diagonal elements of $S$ with the linear combinations of the four functions of Eq.~(\ref{eq_order2_b1b4}). For each finite size $n$ the optimal coefficients have been computed by solving numerically the linear system of dimension 4 $\M^{(n,2)} c = \cR^{(n,2)}$, with the exact expressions at finite $n$ of $\M^{(n,2)}$ and $\cR^{(n,2)}$ given in Appendix \ref{app_order2}. In the left panel we check, for one choice of $(\alpha,\Delta)$, that the expression of $\MMSEod^{(n,2)}$ computed from (\ref{eq_order2_MMSE}) coincides, as it should, with the MSE obtained by the optimal second order denoiser. This plot also illustrates the universality of the large $n$ limit that converges to $\MMSE^{(2)}$, while the finite $n$ corrections depend on the details of the law of $X$. The right panel of Fig.~\ref{fig_fs_order2} presents a comparison between this denoiser and the BABP one (to allow for a fair comparison we computed the off-diagonal part of the MSE for both denoisers). What is to be noticed on this plot is the fact that for small sizes the optimal denoiser of degree 2 reaches a smaller MSE than the BABP one. As a matter of fact the latter is expected to be optimal only in the infinite size limit, finite degree estimators with $n$-dependent coefficients (and possibly terms breaking the orthogonal invariance as $b_4$ for RB matrices) may well be more accurate at finite $n$. This happens here only for rather small matrices ($n \lesssim 40$ in the RB case, $n \lesssim 25$ in the Gaussian one), and with an ad-hoc choice of the parameters $(\alpha,\Delta)$ such that $\MMSE^{(2)}$ is only slightly larger than $\MMSEB$. One can however expect that including polynomial terms of degree higher than 2 in the exact finite $n$ treatment would widen the range of $(n,\alpha,\Delta)$ for which this phenomenon happens.

\section{Conclusions}
\label{sec_conclu}

Let us conclude this manuscript with a few remarks and perspectives for possible future work. We focused our attention on the additive noise model $Y=S+Z$, but the multiplicative noise one $Y =\sqrt{S} Z \sqrt{S}$ could be handled along the same lines. In particular the coefficients of the optimal scalar polynomial estimators of degree at most $D$ will again be given as the solution of the matrix equation $\hcM^{(D)} c = \hcR^{(D)}$, where $\hcM$ will be the Hankel matrix of $\mu_Y = \mu_S \boxtimes \mu_Z$, the multiplicative free convolution~\cite{NiSp_book,PoBo_book} replacing the additive one, and where the coefficients of the right hand side will read $\hcR_p=\tau(S(SZ)^p)$. Free-probability manipulations analogous to those performed in Sec.~\ref{sec_rot_expressions} and \ref{sec_rot_largeD} should allow to recover, in the large $D$ limit, the expression of the denoiser given in~\cite{BuAlBoPo16}.

Another direction that could be investigated concerns the case of non-symmetric or rectangular matrices, for which~\cite{TrErKrMaZd22,PoMa23b,LaMeGa23} derived an estimator that acts on the singular values of the observed matrix, keeping its singular vectors untouched. The low-degree polynomial approach presented here could most probably be adapted to this case; this rectangular setting is also relevant for the estimation of cross-correlations in datasets~\cite{BeBoPo23}.

One could also  investigate further the case of an orthogonally invariant signal perturbed by a non-Gaussian Wigner noise: the matrix elements $(\hS(Y))_{i,j}$ of the asymptotically optimal polynomial estimators of degree at most $D$ will certainly contain a linear combination of $(Y^p)_{i,j}$ and $Y_{i,j}^r$, with $p,r \le D$ and $r$ odd, as was demonstrated for $D=3$ in Sec.~\ref{sec_order3}. Whether these are the only relevant terms for generic $D$ is not clear to us at the moment.

At variance with the derivation of~\cite{BuAlBoPo16} we did not directly deal with the overlaps between the eigenvectors of $S$ and those of $Y=S+Z$; however the free-probability computation of $\hcR_p = \tau(S (S+Z)^p)$ performed in Sec.~\ref{sec_rot_expressions} can actually be interpreted in terms of these overlaps, as explained in Section 6.2 of~\cite{Bi03} (see also Section 19.4 of~\cite{PoBo_book} for a related re-interpretation of this result).

A drawback of our approach is the order in which the limits $n\to\infty$ and $D\to \infty$ are taken, one could indeed worry that estimators that are not polynomials, or polynomials of growing order $D(n)$, reach a better accuracy. For orthogonally invariant priors there are however some reasons (besides the coincidence with the result of~\cite{BuAlBoPo16}) to believe that this is not an issue: recall indeed that, at any finite $n$, the optimal estimator is the posterior mean, and is thus equivariant. The generic form of such functions was given in Eq.~(\ref{eq_equi_ortho}); assuming some regularity for its dependency on the first argument of $\hf$ the latter can be approximated within an arbitrary precision by a polynomial of fixed order. Moreover the traces of growing powers of $Y$ are more and more concentrated on the right edge of the support of $\mu_Y$, and hence should not spoil the optimality result obtained by taking $D\to\infty$ after $n \to \infty$.

The status of this interversion of limit is less clear when the priors are only approximately orthogonally invariant. As a matter of fact the universality we conjecture for the denoising of Wishart matrices perturbed by GOE noise seems to contradict some of the findings of~\cite{CaMe23a,CaMe23b}. These papers present indeed numerical results on the denoising of Wishart matrices with a Rademacher-Bernouilli law for the entries of $X$ by an estimator that reaches a smaller MSE than the BABP one, in a certain range of the parameters $(\alpha,\Delta)$. Two possible explanations of this apparent contradiction are that either the universality conjecture is plainly wrong, or the numerical results from~\cite{CaMe23a,CaMe23b} suffer from finite-size effects and are not representative of the asymptotic behavior as $n \to \infty$. There is however a third, more far-fetched but richer scenario: it would be that the weak form of the conjecture is true but the strong one is wrong, i.e. that for all finite $D$ the asymptotic MMSE among polynomial estimators of degree $D$ is universal, but this is not the case for unbounded degree estimators. Since low-degree polynomials are expected to emulate all polynomial-time algorithms, and since the procedure of~\cite{CaMe23a,CaMe23b} has an exponential computational cost, this third explanation would suggest that matrix denoising with some priors that break the orthogonal invariance can have a hard phase in which the information-theoretic optimal accuracy cannot be reached in polynomial time. If this scenario is the correct one it would be worth investigating the large size limit $n\to \infty$ with a growing degree constraint $D=D(n)$, and in particular trying to determine the scaling of $D(n)$ that separates the finite $D$ universal regime from the unbounded degrees non-universal one.

\appendix

\section{Some free probability formulas}
\label{app_fp}

We collect in this Appendix some definitions on transforms of probability measures, and a computation for the moments of the free convolution of a Marcenko-Pastur distribution with a semi-circular law that was stated without proof in Eq.~(\ref{eq_moments_example}).

We consider a probability measure $\mu$ on the reals, with bounded support. Its moments are denoted $\mu_p = \int \mu(\dd \lambda) \lambda^p$. One defines its Cauchy transform $g(z)$, for $z \in \mathbb{C}\setminus \R$, by $g(z) = \int \mu(\dd \lambda) \frac{1}{z-\lambda}$. The $R$-transform of $\mu$ is the function $R(g) = z(g) - \frac{1}{g}$, where $z(g)$ is the functional inverse of $g(z)$. This transformation is invertible, i.e. the knowledge of $R$ caracterizes the probability distribution $\mu$. The free cumulants $\kappa_p$ of $\mu$ are the coefficients of the series expansion of $R(g)$ around $g=0$, namely $R(g) = \sum_{p=0}^\infty \kappa_{p+1} g^p$. When manipulating different probability measures $\mu_S$, $\mu_Z$ and $\mu_Y$ we add the corresponding index to the various transforms, and denote for instance $g_Y$, $R_Y$, $\mu_{Y,p}$ and $\kappa_{Y,p}$ the various quantities associated to $\mu_Y$.

The free convolution is an operation taking two probability measures and forming a third one, in such a way that the $R$-transform (and hence the free cumulants) is additive. Explicitly, $\mu_Y = \mu_S \boxplus \mu_Z$ is defined by $R_Y(g) = R_S(g) + R_Z(g)$, or $\kappa_{Y,p} = \kappa_{S,p} + \kappa_{Z,p}$ for all $p$.

The free cumulant of order $p$ of a measure $\mu_Y$, $\kappa_{Y,p}$, can be expressed in terms of the moments $\{\mu_{Y,p'}\}_{p' \le p}$, and reciprocally $\mu_{Y,p}$ is a function of $\{\kappa_{Y,p'}\}_{p' \le p}$. Explicit forms of these relationships can be obtained for small values of $p$ by truncating the formal expansions of the power series $R_Y(g)$ and $g_Y(z)$ (around $g=0$ and $z=\infty$), imposing the functional inverse relationship between $g_Y$ and $R_Y$ order by order. A more systematic approach is based on the Lagrange inversion theorem, that allows to compute the coefficients of the functional inverse of a formal power series in terms of the coefficients of the series itself. Its application in the present context yields a simple expression of the moments in terms of the $R$-transform (see Prop. 16.20 in~\cite{NiSp_book}):
\beq
\mu_{Y,p} = \frac{1}{p+1} [g^p] (1+g R_Y(g))^{p+1} \ ,
\eeq
where $[g^p]f(g)$ denotes the coefficient of the $p$-th term in the formal power series $f(g)$.

We will now apply this formula to the case considered in Sec.~\ref{sec_numerics_On}, where $\mu_Y$ is the free convolution of a Marcenko-Pastur distribution with a semi-circular law (its $R$-transform was given in (\ref{eq_RY_example})). This yields
\begin{align}
\mu_{Y,p} &= \frac{1}{p+1} [g^p] \left(1+\Delta g^2+ g^2 \frac{1}{1-\sqrt{\alpha} g} \right)^{p+1} \\
&= \frac{1}{p+1} \sum_{\substack{q,r \ge 0 \\ q+r \le p+1 } } \binom{p+1}{q,r} \Delta^q [g^p] g^{2q+2r} (1-\sqrt{\alpha} g)^{-r} \\
&= \frac{1}{p+1} \sum_{\substack{q,r \ge 0 \\ q+r \le p+1  \\ 2 (q+r) \le p}  } \binom{p+1}{q,r} \Delta^q [g^{p-2q-2r}] (1-\sqrt{\alpha} g)^{-r} \ ,
\label{eq_moments_example_proof}
\end{align}
with a multinomial expansion to go from the first to the second line. For $r=0$, $[g^k] (1-\sqrt{\alpha} g)^{-r} = \delta_{k,0}$, while for $r \ge 1$,
\begin{align}
[g^k] (1-\sqrt{\alpha} g)^{-r} & = (-1)^k \alpha^\frac{k}{2} \frac{1}{k!} (-r)(-r-1) \dots (-r-k+1) \\
& = \alpha^\frac{k}{2} \frac{1}{k!} r (r+1) \dots (r+k-1) \\
& = \alpha^\frac{k}{2} \binom{r+k-1}{r-1} \ .
\end{align}
Separating the contributions of $r=0$ and $r\ge 1$ in (\ref{eq_moments_example_proof}), and noting than in the latter case the condition $2 (q+r) \le p$ implies $q+r \le p+1$, one obtains (\ref{eq_moments_example}).

\section{Equivariant functions}
\label{app_equivariant}

\subsection{Orthogonal group, direct reasoning}
\label{app_equivariant_ortho_direct}

In this appendix we shall characterize the functions $f:M_n^{\rm sym}(\R) \to M_n^{\rm sym}(\R)$, where $M_n^{\rm sym}(\R)$ is the set of $n \times n$ real symmetric matrices, that are equivariant under conjugation by orthogonal matrices, i.e. such that $f(OYO^T)=O f(Y) O^T$ for all $O \in \cO_n$ and $Y \in M_n^{\rm sym}(\R) $, where $\cO_n=\{O \in M_n(\R) : OO^T = O^TO=\one_n \}$ is the orthogonal group.

In a first step we shall use the fact that any real symmetric matrix $Y$ is diagonalizable as $Y=P \, \diag(\lambda_1,\dots,\lambda_n) P^T$, with $P \in \cO_n$; for an equivariant function this yields $f(Y)=P \, f(\diag(\lambda_1,\dots,\lambda_n)) \, P^T$, it is thus enough to characterize $f$ when its argument is a diagonal matrix. It turns out that the image of a diagonal matrix by an equivariant $f$ is also diagonal (we follow here the proof of this fact given in Appendix A.1 of~\cite{PoMa23}): let us define, for $k \in [n]$, the matrix $I(k)=\diag(1,\dots,1,-1,1,\dots 1)$, with the negative element in the $k$-th row. Since $I(k)=I(k)^T$ and $I(k)^2=\one_n$, this matrix is orthogonal, and for a diagonal matrix $D$ one has $I(k)D I(k) = D$. As a consequence for an equivariant function $f$ one obtains $f(D)=f(I(k)D I(k))=I(k)f(D)I(k)$. Taking the matrix elements of this equation gives $f(D)_{i,j}=(-1)^{\one(i=k)+\one(j=k)}f(D)_{i,j}$. In particular for $i \neq j=k$, $f(D)_{i,k}=-f(D)_{i,k}=0$. Since this is valid for all $k$, one concludes that an equivariant $f$ sends diagonal matrices to diagonal matrices.

We can thus write $f(\diag(\lambda_1,\dots,\lambda_n))=\diag(f_1(\lambda_1,\dots,\lambda_n),\dots,f_n(\lambda_1,\dots,\lambda_n))$, and it remains to characterize the $n$ functions $f_i:\R^n \to \R$. For $\pi \in \cS_n$ a permutation of $[n]$, consider the matrix $\Pi$ with matrix elements $\Pi_{i,j}=\one(j=\pi(i))$, which is easily seen to belong to $\cO_n$. One has $\Pi  \, \diag(\lambda_1,\dots,\lambda_n) \, \Pi^T = \diag(\lambda_{\pi(1)},\dots,\lambda_{\pi(n)})$, hence the functions $f_i$ obey the equations
\beq
f_i(\lambda_{\pi(1)},\dots,\lambda_{\pi(n)})=f_{\pi(i)}(\lambda_1,\dots,\lambda_n) \quad \forall \pi \in \cS_n \ .
\label{eq_permutation}
\eeq
Consider first the $(n-1)!$ permutations that fixes the point $1$, i.e. such that $\pi(1)=1$. Equation (\ref{eq_permutation}) with $i=1$ becomes
\beq
f_1(\lambda_1,\lambda_2\dots,\lambda_n) = f_1(\lambda_1,\lambda_{\pi(2)}\dots,\lambda_{\pi(n)}) \ ,
\eeq
which shows that the dependency of $f_1$ on its last $n-1$ arguments is symmetric under the permutations. The theory of symmetric functions allows then to conclude on the existence of a function $\tf$ such that
\beq
f_1(\lambda_1,\lambda_2,\dots,\lambda_n) = \tf\left(\lambda_1;\sum_{j=2}^n \lambda_j , \sum_{j=2}^n \lambda_j^2,\dots,\sum_{j=2}^n \lambda_j^{n-1}\right) \ .
\label{eq_tf}
\eeq
Indeed, such a symmetric dependency on $(\lambda_2,\dots,\lambda_n)$ can be interpreted as a dependency on the multiset $\{\lambda_2,\dots ,\lambda_n\}$, with multiplicities, or equivalently on the roots of the univariate polynomial $P(t)=(t-\lambda_2)(t-\lambda_3)\dots (t-\lambda_n)$. There are several equivalent ways to encode the coefficients of this polynomial, see for instance~\cite{Stanley_vol2} for more details, one being in terms of the powersums, which justifies the expression (\ref{eq_tf}). Defining a new function $\hf$ as $\hf(\lambda;x_1,\dots,x_{n-1})=\tf(\lambda;x_1-\lambda,x_2-\lambda^2,\dots,x_{n-1}-\lambda^{n-1})$, we arrive at
\beq
f_1(\lambda_1,\lambda_2,\dots,\lambda_n) = \hf\left(\lambda_1;\sum_{j=1}^n \lambda_j , \sum_{j=1}^n \lambda_j^2,\dots,\sum_{j=1}^n \lambda_j^{n-1}\right) \ .
\eeq
Consider now an index $k\neq 1$, and a permutation $\pi$ such that $\pi(1)=k$; from (\ref{eq_permutation}) with $i=1$ we get
\begin{align}
f_k(\lambda_1,\dots,\lambda_n) &= f_1(\lambda_k,\lambda_{\pi(2)},\dots,\lambda_{\pi(n)}) \\
& = \tf\left(\lambda_k;\sum_{j\neq k} \lambda_j , \sum_{j\neq k} \lambda_j^2,\dots,\sum_{j\neq k} \lambda_j^{n-1}\right)  \\
&= \hf\left(\lambda_k;\sum_{j=1}^n \lambda_j , \sum_{j=1}^n \lambda_j^2,\dots,\sum_{j=1}^n \lambda_j^{n-1}\right) \ .
\end{align}
We can conclude at this point that an arbitrary equivariant function $f$ acts on diagonal matrices as $f(D)=\hf(D;\Tr(D),\dots,\Tr(D^{n-1}))$, with the convention that the function $\hf$ acts componentwise through its first argument. Coming back to arbitrary matrices we write $Y=PDP^T$ with $P \in \cO_n$ and $D$ diagonal, use the equivariance of $f$ to deduce $f(Y)=Pf(D)P^T$, and the invariance of the trace under orthogonal conjugation to write $\Tr(Y^j)=\Tr(D^j)$. This yields $f(Y)=\hf(Y;\Tr(Y),\dots,\Tr(Y^{n-1}))$, under the convention that a scalar function acts on $Y$ by being applied componentwise on its eigenvalues without altering its eigenvectors, and concludes the justification of the statement (\ref{eq_equi_ortho}) of the main text, namely that all equivariant functions $f$ are of the form $f(Y)=\hf(Y;\Tr(Y),\dots,\Tr(Y^{n-1}))$, for an arbitrary function $\hf:\R^n\to \R$.

As a side remark let us mention that a similar reasoning can be made to characterize the invariant functions $f:M_n^{\rm sym}(\R) \to \R$, such that $f(OYO^T)=f(Y)$ for all $O \in \cO_n$ and $Y \in M_n^{\rm sym}(\R) $: these are of the form $f(Y)=F(\Tr(Y),\Tr(Y^2),\dots,\Tr(Y^n))$, for an arbitrary function $F:\R^n \to \R$.

\subsection{Orthogonal group, via Weingarten calculus}
\label{app_ortho_weingarten}

We present here an alternative justification of a weaker version of the statement made above on the structure of the equivariant functions, based on the symmetrization defined in Eq.~(\ref{eq_projection_equi}), which in the present context amounts to take an arbitrary function $f:M_n^{\rm sym}(\R) \to M_n^{\rm sym}(\R)$, and symmetrize it with the action considered here, defining
\beq
f^{\rm eq}(Y) = \E_O[O^T f(OYO^T) O ] \ ,
\label{eq_feq_weingarten}
\eeq
where the expectation is over the Haar measure of the orthogonal group, that is invariant under multiplication by a fixed element, as required in Sec.~\ref{sec_gene_sym}. One can check indeed that $f^{\rm eq}$ is equivariant for arbitrary $f$ (provided the expectation above exists), and since $f^{\rm eq}=f$ if $f$ is equivariant, the map $f \to f^{\rm eq}$ is a linear projection in the space of functions from $M_n^{\rm sym}(\R)$ to itself, hence all equivariant functions can be obtained by such a symmetrization.

It might be difficult to compute this expectation for generic $f$, or even to ensure that the expectation is well-defined; we shall bypass this difficulty by treating a special case, assuming that each matrix element of $f(Y)$ is a polynomial of total degree $m$ in the matrix elements of $Y$. The computation of $f^{\rm eq}$ amounts then to compute the expectation of a product of $2m+2$ matrix elements of an Haar distributed orthogonal matrix $O$, a problem which has been solved in~\cite{CoSn06}, with the following result:
\beq
\E_O[O_{i_1,j_1} \dots O_{i_{2m+2},j_{2m+2}}] = \sum_{\kp,\kq} {\rm Wg}(n,\kp,\kq) \prod_{\{a,b\}\in \kp} \delta_{i_a,i_b} \prod_{\{a,b\}\in \kq} \delta_{j_a,j_b} \ , 
\label{eq_weingarten}
\eeq
where the sum is over two independent pairings $\kp,\kq$ of $[2m+2]$, a pairing being a partition of $[2m+2]$ in which all blocks denoted $\{a,b\}$ are of size 2, the Kronecker deltas imposing the equality of the row (resp. column) indices $i$ (resp. $j$) in the blocks of the pairing $\kp$ (resp. $\kq)$. We shall not need here an explicit expression of the so-called Weingarten function ${\rm Wg}(n,\kp,\kq)$, the rest of the computation only relies on the fact that the non-zero terms in the expectation are necessarily of this form.

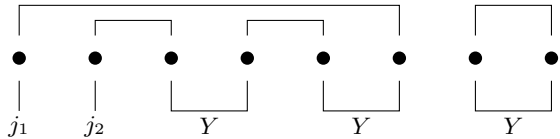
\begin{figure}
\begin{tikzpicture}
\fill[black] (0,0) circle (2.5pt);
\fill[black] (1,0) circle (2.5pt);
\fill[black] (2,0) circle (2.5pt);
\fill[black] (3,0) circle (2.5pt);
\fill[black] (4,0) circle (2.5pt);
\fill[black] (5,0) circle (2.5pt);
\fill[black] (6,0) circle (2.5pt);
\fill[black] (7,0) circle (2.5pt);

\draw (0,.3) -- (0,.7) -- (5,.7) -- (5,.3);
\draw (1,.3) -- (1,.5) -- (2,.5) -- (2,.3);
\draw (3,.3) -- (3,.5) -- (4,.5) -- (4,.3);
\draw (6,.3) -- (6,.7) -- (7,.7) -- (7,.3);

\draw (0,-.3) -- (0,-.7);
\draw (1,-.3) -- (1,-.7);
\draw (2,-.3) -- (2,-.7) -- (3,-.7) -- (3,-.3);
\draw (4,-.3) -- (4,-.7) -- (5,-.7) -- (5,-.3);
\draw (6,-.3) -- (6,-.7) -- (7,-.7) -- (7,-.3);

\draw (0,-.9) node {$j_1$};
\draw (1,-.9) node {$j_2$};
\draw (2.5,-.9) node {$Y$};
\draw (4.5,-.9) node {$Y$};
\draw (6.5,-.9) node {$Y$};

\end{tikzpicture}

\caption{An illustration of the contribution of Eq.~(\ref{eq_pairing}). In this figure $m=3$, the pairing $\kq=\{ \{1 , 6 \} , \{2 ,3 \} , \{4 ,5 \} , \{7 ,8 \}\}$ is represented above the horizontal axis, while below are the representation of the matrices $Y$ as the blocks of the canonical pairing $\{ \{3 ,4 \} , \{5 ,6 \} , \{7 ,8 \}\}$ and two ``external legs''. This yields here $(Y^2)_{j_1,j_2} \Tr (Y)$.}
\label{fig_pairing}
\end{figure}

Let us now write a polynomial $f$ of degree $m$ as
\beq
f(Y)_{i_1,i_2} = \sum_{i_3,\dots,i_{2m+2}} F_{i_1,i_2,i_3,\dots,i_{2m+2}} Y_{i_3,i_4} \dots Y_{i_{2m+1},i_{2m+2}} \ ,
\label{eq_f_arbitrary}
\eeq
where here and in the following we keep implicit the range $[n]$ of the summations over indices $i_a,j_a$, and where $F$ is an arbitrary array, invariant under the exchange of its first two arguments for the image of $f$ to be a symmetric matrix. Inserting this definition in (\ref{eq_feq_weingarten}) and computing the expectation thanks to (\ref{eq_weingarten}) yields:
 \begin{align}
 f^{\rm eq}(Y)_{j_1,j_2} &= \E_O[(O^T f(OYO^T) O)_{j_1,j_2} ] \label{eq_sym1} \\ 
 & = \sum_{i_1,i_2} \E_O[O_{i_1,j_1} O_{i_2,j_2} f(OYO^T)_{i_1,i_2} ] \\
 & = \sum_{i_1,\dots,i_{2m+2}} F_{i_1,\dots,i_{2m+2}} \E_O[O_{i_1,j_1} O_{i_2,j_2} (OYO^T)_{i_3,i_4} \dots (OYO^T)_{i_{2m+1},i_{2m+2}}] \\
 & = \sum_{\substack{i_1,\dots,i_{2m+2} \\ j_3,\dots,j_{2m+2}}} F_{i_1,\dots,i_{2m+2}} \E_O[O_{i_1,j_1} \dots O_{i_{2m+2},j_{2m+2}}] Y_{j_3,j_4} \dots Y_{j_{2m+1},j_{2m+2}} \label{eq_sym4} \\
& = \sum_{\kp,\kq} {\rm Wg}(n,\kp,\kq) \left(\sum_{i_1,\dots,i_{2m+2}} F_{i_1,\dots,i_{2m+2}}  \prod_{\{a,b\}\in \kp} \delta_{i_a,i_b} \right) \\
& \hspace{4cm}
\left(\sum_{j_3,\dots,j_{2p+2}} \left( \prod_{\{a,b\}\in \kq} \delta_{j_a,j_b} \right) Y_{j_3,j_4} \dots Y_{j_{2m+1},j_{2m+2}}  \right) \ .
\label{eq_pairing}
\end{align}
A moment of thought reveals that the last parenthesis is equal to $(Y^p)_{j_1,j_2}(\Tr(Y))^{q_1} \dots (\Tr(Y^m))^{q_m}$, where the non-negative integers $p,q_1,\dots,q_m$ depend on the pairing $\kq$, and add up to $m$, hence the dependency of $f^{\rm eq}$ on $Y$ is a linear combination of such terms. This is most easily seen through a graphical reasoning illustrated on one example in Fig.~\ref{fig_pairing}, by representing the pairing $\kq$ along with the canonical pairing $\{3,4\},\dots,\{2p+1,2p+2\}$ complemented by two ``external legs'' corresponding to the indices $1$ and $2$. The combination of these two combinatorial objects yields a single path between the two external lengths of length $p$, where length is counted in terms of the number of blocks crossed in the canonical pairing, and an union of cycles, $q_l$ corresponding to the number of cycles of length $l$. In this reasoning one uses crucially the symmetry of $Y$, in its absence one would get arbitrary words in $Y$ and $Y^T$ instead of simple matrix powers of $Y$. To fully recover the result of Appendix~\ref{app_equivariant_ortho_direct} one could use the theory of symmetric polynomials to express, for $l > n$, $\Tr(Y^l)$ as a polynomial in the $\Tr(Y),\dots,\Tr(Y^n)$, we skip these details which shall not be useful for us.

Note that a similar computation allows to characterize the polynomials $f:M_n^{\rm sym}(\R) \to \R$ that are invariant in the sense that $f(OYO^T)=f(Y)$ for all $O \in \cO_n$ and $Y \in M_n^{\rm sym}(\R) $: starting from an arbitrary polynomial $f$ one computes its symmetrized version $f^{\rm inv}(Y)=\E_O[ f(OYO^T) ] $, that is indeed invariant. Following the same steps as above shows that $f^{\rm inv}$ is a polynomial in the traces of the powers of $Y$ (one obtains the same graphical representation without the external legs).

Let us finally mention that~\cite{BlVi22} presents another method, attributed to Malgrange, to obtain the equivariant functions $f:V \to W$ for generic group actions on $V$ and $W$, via the determination of the invariant functions on $V \times W^*$, with $W^*$ the dual of $W$. It should be possible to use it to rederive the results presented here for the conjugation action of the orthogonal group.

\subsection{Symmetric group}
\label{app_equivariant_permu}

We justify in this Appendix the claims made in Sec.~\ref{sec_universality_equivariant} on the form of the polynomials equivariant under the symmetric group of permutations, adapting the proof of Appendix \ref{app_ortho_weingarten} (a similar reasoning can be found in~\cite{PuLiKiMaLi23}). Consider indeed an arbitrary polynomial function as in (\ref{eq_f_arbitrary}), and its symmetrization
\beq
f^{\rm eq}(Y) = \E_\sigma[O(\sigma)^T f(O(\sigma)YO(\sigma)^T) O(\sigma) ] \ ,
\eeq
where the average is under the uniform measure on the $n!$ permutations of $\cS_n$, and $O(\sigma)$ is the associated matrix with elements $O(\sigma)_{i,j} = \delta_{i,\sigma(j)}$. One can perform exactly the same steps as in Eqs.~(\ref{eq_sym1}-\ref{eq_sym4}), with the replacement of the Haar measure on $\cO_n$ by the one on $\cS_n$. The computation thus boils down to find the equivalent of (\ref{eq_weingarten}), i.e. the Weingarten function of the symmetric group instead of the orthogonal one, an exercise performed in Section 3 of \cite{CoMaNo22} and that we reproduce here. This function turns out to be much simpler, since
\beq
\E_\sigma[O(\sigma)_{i_1,j_1} \dots O(\sigma)_{i_{2m+2},j_{2m+2}}] = \mathbb{P}_\sigma[ i_a = \sigma(j_a) \ \forall a \in [2m+2] ] = \mathbb{P}_\sigma[i = \sigma \circ j ] \ ,
\eeq
where in the second step one interprets the multiplet of indices $(i_1,\dots,i_{2m+2})$ as a function $i:[2m+2]\to [n]$, and similarly for $j$. This expectation is thus the ratio between the number of permutations $\sigma$ such that $i = \sigma \circ j$ and the total number $n!$ of permutations. To complete the computation let us define the kernel $\ker(i)$ (resp. $\ker(j)$) as the partition of $[2m+2]$ induced by the equivalence relation $a\sim b$ iff $i_a=i_b$ (resp. $j_a=j_b$), in other words the blocks of $\ker(i)$ encode the patterns of equalities in the indices $(i_1,\dots,i_{2m+2})$. A moment of thought reveals that one can find a permutation $\sigma$ with $i = \sigma \circ j$ if and only if $\ker(i) = \ker(j)$, and that in this case there are $(n-|\ker(i)|)!$ satisfying permutations, with $|\ker(i)|$ the number of blocks in the partition: the value of $\sigma$ on the image of each block of $j$ is fixed to match the corresponding values in $i$, and $\sigma$ is a bijection from the indices absent from $j$ to those absent from $i$. In formula this reads
\beq
\E_\sigma[O(\sigma)_{i_1,j_1} \dots O(\sigma)_{i_{2m+2},j_{2m+2}}] = \one(\ker(i) = \ker(j)) \frac{(n-|\ker(i)|)!}{n!} \ ,
\eeq
which inserted in (\ref{eq_sym4}) yields
\beq
 f^{\rm eq}(Y)_{j_1,j_2} = \sum_{i_1,\dots,i_{2m+2}} F_{i_1,\dots,i_{2m+2}} \frac{(n-|\ker(i)|)!}{n!} \sum_{\substack{j_3,\dots,j_{2p+2} \\ \ker(j) = \ker(i)}} Y_{j_3,j_4} \dots Y_{j_{2m+1},j_{2m+2}} \ .
\label{eq_feq_permu}
\eeq
One can conclude from here that such a function is indeed a linear combination of the $\hb_G$ defined in (\ref{eq_bpG}): for each choice of $i$ consider the graph $G$ with $|\ker(i)|$ vertices associated to the blocks of $\ker(i)$, $m$ edges between the vertices $B(i_{2p+1})$ and $B(i_{2p+2})$, for $p=1,\dots,m$, where $B(i_a)$ denotes the block containing $i_a$, and mark the vertices $B(i_1)$ and $B(i_2)$ (which can be equal). The summation over $j$ in (\ref{eq_feq_permu}) is then equal to $(\hb'_G)_{j_1,j_2}$. Moreover the invariance of the array $F$ under the exchange of its first two indices imply the symmetrization of (\ref{eq_bG}).

\subsection{Inversion symmetry}
\label{app_equivariant_inversion}

We turn now to the consequences of the inversion symmetry that appears when the laws of $X_{1,1}$ and $B_{1,2}$ are invariant under a change of sign. Consider the group $\{-1,1\}^n$, with componentwise multiplication, and let it act on $M_n^{\rm sym}(\R)$ as $(\ve \cdot Y)_{i,j} = \ve_i \ve_j Y_{i,j}$ for $\ve=(\ve_1,\dots,\ve_n) \in \{-1,1\}^n$, which can be seen as the conjugation with the orthogonal matrix ${\rm diag}(\ve_1,\dots,\ve_n)$. The assumptions on the laws of $X_{1,1}$ and $B_{1,2}$ imply that this action is a symmetry for the inference problem. We can therefore symmetrize the permutation invariant polynomials in order to further reduce the set of relevant basic functions. Since $\ve^{-1} = \ve$, the symmetrization of $\hb'_G$ of (\ref{eq_bpG}) yields
\beq
\E_{\ve}\left[ (\ve \cdot \hb'_G( \ve \cdot Y))_{i,j} \right]= 
\sum_{\substack{\phi \in \cI(V,[n]) \\ \phi(v)=i,\phi(w)=j}} \left( \prod_{e=\{a,b\} \in E} Y_{\phi(a),\phi(b)} \right) \E_{\ve} \left[ \ve_i \ve_j \prod_{e=\{a,b\} \in E} \ve_{\phi(a)} \ve_{\phi(b)} \right] \ ,
\eeq
where $\E_{\ve}$ denotes the uniform average over the $2^n$ elements of $\{-1,1\}^n$. The expectation in the right hand side vanishes unless for all $k\in[n]$, the power of $\ve_k$ is even. If $k$ is not in the image of $\phi$ this power is zero; if it is, the power arising from the product over the edges is the degree of the vertex $a=\phi^{-1}(k)$ (here the injectivity of $\phi$ is crucial), while each of the marked vertices (possibly equal) add 1 to the corresponding powers. Hence the average over $\ve$ is independent of $\phi$, and is equal to 1 if the degree constraints stated at the end of Sec.~\ref{sec_universality_equivariant} are fulfilled, 0 otherwise.

\section{Gaussian integration by parts}
\label{app_IPP}

This Appendix presents an alternative derivation of the expression (\ref{eq_hcR_Gaussian}) of $\hcR_p$ in the Gaussian noise case, avoiding the free probability combinatorial interpretation of the main text.

Before starting the derivation let us recall two well-known facts:
\begin{itemize}
\item if $G=(G_1,\dots,G_N)$ is a vector of independent centered Gaussian variables, an integration by parts reveals that
\beq
\E[G_i f(G)] = \E[G_i^2] \E\left[\frac{\partial f}{\partial G_i} \right] \ ,
\eeq
whenever the function $f$ is smooth enough and behaves at infinity in such a way that both members of this equality exist.

\item if $A(t)$ is a matrix dependent on a parameter $t$, the derivatives of its matrix powers read
  \beq
\frac{\dd}{\dd t} A(t)^p = \sum_{k=1}^p A(t)^{k-1} \frac{\dd A}{\dd t} A(t)^{p-k} \ .
  \eeq  
\end{itemize}

We have to compute the value of $\frac{1}{n}\E[ \Tr(Z Y^p) ]$, where $Y=S+Z$, with $S$ and $Z$ two independent symmetric matrices of size $n$, and specifically in the Gaussian noise case the matrix elements $\{Z_{i,j}\}_{i \le j}$ are independent centered Gaussian random variables, with $\E[(Z_{i,i})^2] = \frac{2 \Delta}{n}$, and $\E[(Z_{i,j})^2] = \frac{\Delta}{n}$ for $i<j$. The matrix elements $Z_{i,j}$ for $i>j$ are given as $Z_{i,j}=Z_{j,i}$, to ensure the symmetry. We thus write the above expression as
\begin{align}
\frac{1}{n}\E[ \Tr(Z Y^p) ] =  & \frac{1}{n} \sum_{i=1}^n \E[Z_{i,i} (Y^p)_{i,i} ] + \frac{2}{n} \sum_{1\le i<j \le n} \E[Z_{i,j} (Y^p)_{i,j} ] \\
  = & \frac{2 \Delta}{n^2} \sum_{i=1}^n \E\left[ \frac{\partial}{\partial Z_{i,i}} (Y^p)_{i,i} \right] + \frac{2 \Delta}{n^2} \sum_{1 \le i<j \le n} \E\left[\frac{\partial}{\partial Z_{i,j}} (Y^p)_{i,j} \right] \\
  = & \frac{2 \Delta}{n^2} \sum_{k=1}^p \left(
      \sum_{i=1}^n \E\left[ \left(Y^{k-1} \frac{\partial Y}{\partial Z_{i,i}}    Y^{p-k}\right)_{i,i}  \right]
+\sum_{1 \le i<j \le n} \E\left[ \left(Y^{k-1} \frac{\partial Y}{\partial Z_{i,j}}    Y^{p-k}\right)_{i,j}  \right]
      \right) \ ,
\end{align}
by applying the two facts recalled above. Since $\left(\frac{\partial Y}{\partial Z_{i,i}}\right)_{l,m} = \delta_{l,i}\delta_{m,i}$ and for $i<j$ $\left(\frac{\partial Y}{\partial Z_{i,j}}\right)_{l,m} = \delta_{l,i}\delta_{m,j} + \delta_{l,j}\delta_{m,i}$, because of the symmetry requirement, one has
\begin{align}
\frac{1}{n}\E[ \Tr(Z Y^p) ]  = & \frac{2 \Delta}{n^2} \sum_{k=1}^p \left(
      \sum_{i=1}^n \E\left[(Y^{k-1})_{i,i} (Y^{p-k})_{i,i}  \right]
      +\sum_{1 \le i<j \le n} \E\left[ (Y^{k-1})_{i,i} (Y^{p-k})_{j,j}  \right]
      +\sum_{1 \le i<j \le n} \E\left[ (Y^{k-1})_{i,j} (Y^{p-k})_{j,i}  \right]
      \right) \ , \nonumber \\
  = & \frac{\Delta}{n^2} \sum_{k=1}^p \left( \sum_{i,j=1}^n \E\left[ (Y^{k-1})_{i,i} (Y^{p-k})_{j,j}  \right] + \sum_{i,j=1}^n \E\left[ (Y^{k-1})_{i,j} (Y^{p-k})_{j,i}  \right]
      \right) \\
  = & \Delta \sum_{k=1}^p \E[\tr(Y^{k-1}) \tr(Y^{p-k}) ] + \frac{1}{n} \Delta p \E[\tr(Y^{p-1})] \ ,
\end{align}
with $\tr(\bullet) = \Tr(\bullet) /n$ the normalized trace. In the large $n$ limit the assumptions of convergence of the average moments and concentration expressed in Eqs.~(\ref{eq_convergence},\ref{eq_fluctuations}) imply that the second term above is of order $1/n$ and that the first one factorizes, yielding
\beq
\lim_{n \to \infty} \frac{1}{n}\E[ \Tr(Z Y^p) ] = \Delta \sum_{k=1}^p \mu_{Y,k-1} \mu_{Y,p-k} \ .
\eeq
This agrees indeed with (\ref{eq_hcR_Gaussian}), after the change of variable $j=k-1$.

\section{Degree 2 denoisers for arbitrary Wishart matrices perturbed by Wigner noise}
\label{app_order2}

We present in this Appendix the details on the computation whose results were stated in Sec.~\ref{sec_order2}.

\subsection{Off-diagonal terms}

We start with the determination of the matrix $\M^{(n,2)}$ defined in Eq.~(\ref{eq_order2_MR}), with the functions $b_1,\dots,b_4$ of Eq.~(\ref{eq_order2_b1b4}). A first step is to reduce its computation to those of joint moments of the matrix elements of $Y$, exploiting its invariance under permutation to group together equivalent terms:
\begin{align}
\M_{1,1}^{(n,2)}  = & (n-1)\E[Y_{1,2}^2] \ , \nonumber \\
\M_{1,2}^{(n,2)}  = & 2 (n-1)\E[Y_{1,1} Y_{1,2}^2] + (n-1) (n-2) \E[Y_{1,2} Y_{2,3} Y_{3,1}]  \ , \nonumber\\
\M_{1,3}^{(n,2)}  = & 2 (n-1)\E[Y_{1,1} Y_{1,2}^2] \ , \nonumber\\
\M_{1,4}^{(n,2)}  = & \sqrt{n} (n-1)\E[Y_{1,1} Y_{1,2}^2] \ , \nonumber\\
\M_{2,2}^{(n,2)}  = & 2 (n-1)\E[Y_{1,1}^2 Y_{1,2}^2] + 2 (n-1)\E[Y_{1,1} Y_{1,2}^2 Y_{2,2}] + 4 (n-1) (n-2) \E[Y_{1,1}Y_{1,2} Y_{2,3} Y_{3,1} ] \nonumber\\ & +(n-1)(n-2) \E[Y_{1,2}^2 Y_{2,3}^2] + (n-1)(n-2)(n-3) \E[Y_{1,2} Y_{2,3} Y_{3,4} Y_{4,1}] \ , \nonumber\\
\M_{2,3}^{(n,2)}  = & 2 (n-1)\E[Y_{1,1}^2 Y_{1,2}^2] + 2 (n-1)\E[Y_{1,1} Y_{1,2}^2 Y_{2,2}] + 3 (n-1) (n-2) \E[Y_{1,1}Y_{1,2} Y_{2,3} Y_{3,1} ] \ , \nonumber\\
\M_{2,4}^{(n,2)}  = & \sqrt{n} (n-1)\E[Y_{1,1}^2 Y_{1,2}^2] +\sqrt{n} (n-1) \E[Y_{1,1} Y_{1,2}^2 Y_{2,2}] + \sqrt{n} (n-1) (n-2) \E[Y_{1,1}Y_{1,2} Y_{2,3} Y_{3,1} ] \ , \nonumber\\
\M_{3,3}^{(n,2)}  = & 2 (n-1)\E[Y_{1,1}^2 Y_{1,2}^2] + 2 (n-1)\E[Y_{1,1} Y_{1,2}^2 Y_{2,2}] + (n-1) (n-2) \E[Y_{1,1}^2] \E[Y_{1,2}^2] \ , \nonumber\\
\M_{3,4}^{(n,2)}  = & \sqrt{n} (n-1)\E[Y_{1,1}^2 Y_{1,2}^2] + \sqrt{n} (n-1) \E[Y_{1,1} Y_{1,2}^2 Y_{2,2}] \ , \nonumber\\
\M_{4,4}^{(n,2)}  = & \frac{1}{2} n (n-1)\E[Y_{1,1}^2 Y_{1,2}^2] + \frac{1}{2} n (n-1) \E[Y_{1,1} Y_{1,2}^2 Y_{2,2}] \ .
\label{eq_order2_M_moments}
\end{align}
Let us explain the main ideas underlying the reasonings followed to obtain these expressions, taking for instance the term $\M_{1,2}^{(n,2)}$; starting from its definition,
\beq
\M_{1,2}^{(n,2)} = (n-1) \E[Y_{1,2} (Y^2)_{1,2}] = (n-1) \sum_{k=1}^n \E[Y_{1,2} Y_{1,k} Y_{k,2}] \ .
\eeq
Using the permutation invariance the two terms with $k \in \{1,2\}$ are both equal to $\E[Y_{1,1} Y_{1,2}^2]$, while the $n-2$ summands with $k\ge 3$ all yield $\E[Y_{1,2} Y_{2,3} Y_{3,1}]$. All the other lines in (\ref{eq_order2_M_moments}) are obtained with similar combinatorial reasonings, enumerating the number of contributions with equivalent patterns of indices under the permutation symmetry, and exploiting the following independence property: if $i_1,\dots,i_{2p}$ are indices in $I\subset [n]$, and $j_1,\dots,j_{2q}$ are in another subset $J$ disjoint from $I$, then $\E[Y_{i_1,i_2} \dots Y_{i_{2p-1},i_{2p}}Y_{j_1,j_2} \dots Y_{j_{2p-1},j_{2p}} ] = \E[Y_{i_1,i_2} \dots Y_{i_{2p-1},i_{2p}}] \E[Y_{j_1,j_2} \dots Y_{j_{2p-1},j_{2p}} ]$. Indeed the noise random variables $B_{k,l}$ are independent from one edge to the other, while the signal matrix elements $S_{k,l}$ are correlated but only through their dependency on $X_{k,\mu}$ and $X_{l,\mu}$. As a consequence the matrix elements $S_{i_a,i_b}$ only depend on $\{X_{i,\mu}\}_{i \in I}$ and the $S_{j_a,j_b}$ on the $\{X_{j,\mu}\}_{j \in J}$, hence the independence at the origin of this factorization.

Similar considerations yield the following expressions for the elements of the vector $\cR^{(n,2)}$ also defined in Eq.~(\ref{eq_order2_MR}):
\begin{align}
\cR_1^{(n,2)} = & (n-1) \E[S_{1,2} Y_{1,2} ] \ , \nonumber \\
\cR_2^{(n,2)} = & 2 (n-1) \E[S_{1,2} Y_{1,1} Y_{1,2}] + (n-1)(n-2) \E[S_{1,2} Y_{2,3} Y_{3,1}] \ , \nonumber \\
\cR_3^{(n,2)} = & 2 (n-1) \E[S_{1,2} Y_{1,1} Y_{1,2}] \ , \nonumber \\
\cR_4^{(n,2)} = & \sqrt{n} (n-1) \E[S_{1,2} Y_{1,1} Y_{1,2}] \ .
\label{eq_order2_R_moments}
\end{align}

To complete the computation it remains to determine more explicitly the joint moments of the matrix elements of $Y$. To do so one expands each $Y_{i,j}$ as $S_{i,j}+Z_{i,j}$; since the $Z$ are independent from the $S$, and independent from one edge to the other, one can easily perform the average on the former. To average over the $S$ it is necessary to express them in terms of the $X_{i,\mu}$, according to the definition (\ref{eq_def_Wishart_gene}), and then perform the averages exploiting the i.i.d. character of $X_{i,\mu}$, with again some combinatorial considerations to count the number of equivalent terms. The first order terms trivially vanish, $\E[Y_{1,1}]=\E[Y_{1,2}]=0$. The moments of order 2 that appear in (\ref{eq_order2_M_moments}) are
\beq
\E[Y_{1,1}^2] = \frac{1}{n}(\E[X_{1,1}^4] - 1 + \Delta \E[B_{1,1}^2]) \ , \qquad 
\E[Y_{1,2}^2] = \frac{1}{n}(1+\Delta) \ .
\eeq
Similarly one finds for the required third order terms:
\beq
\E[Y_{1,1} Y_{1,2}^2] = \frac{1}{n^2} \sqrt{\alpha} (\E[X_{1,1}^4] - 1) \ , \qquad
\E[Y_{1,2} Y_{2,3} Y_{3,1}] = \frac{1}{n^2} \sqrt{\alpha} \ ,
\eeq
and finally for the fourth order terms:
\begin{align}
\E[Y_{1,1}^2 Y_{1,2}^2] & = \frac{1}{n^2} (1+\Delta) (\E[X_{1,1}^4] - 1 + \Delta \E[B_{1,1}^2]) + \frac{1}{n^3} \alpha (\E[X_{1,1}^6] - 3 \E[X_{1,1}^4] +2 ) \ , \\
\E[Y_{1,1} Y_{1,2}^2 Y_{2,2}] & = \frac{1}{n^3} \alpha (\E[X_{1,1}^4] - 1)^2 \ , \\
\E[Y_{1,1}Y_{1,2} Y_{2,3} Y_{3,1} ] & = \frac{1}{n^3} \alpha (\E[X_{1,1}^4] - 1) \ , \\
\E[Y_{1,2}^2 Y_{2,3}^2] & = \frac{1}{n^2} (1+\Delta)^2 + \frac{1}{n^3} \alpha (\E[X_{1,1}^4] - 1)   \ , \\
\E[Y_{1,2} Y_{2,3} Y_{3,4} Y_{4,1}] & = \frac{1}{n^3} \alpha \ . 
\end{align}
The expressions for the joint moments between $S$ and $Y$ that appear in the expression (\ref{eq_order2_R_moments}) of $\cR^{(n,2)}$ are established through a similar expansion and read
\beq
\E[S_{1,2} Y_{1,2} ] = \frac{1}{n} \ , \qquad \E[S_{1,2} Y_{1,1} Y_{1,2}] = \frac{1}{n^2} \sqrt{\alpha} (\E[X_{1,1}^4] - 1) \ , \qquad
\E[S_{1,2} Y_{2,3} Y_{3,1}] = \frac{1}{n^2} \sqrt{\alpha} \ .
\eeq

Collecting all these relations gives explicit expressions for the matrix $\M^{(n,2)}$ and vector $\cR^{(n,2)}$ in terms of the parameters $\alpha$, $\Delta$, $n$, and the non-universal quantities $\E[X_{1,1}^4]$, $\E[X_{1,1}^6]$ and $\E[B_{1,1}^2]$. Taking their large $n$ limit then yields the expression (\ref{eq_order2_Minfty}) of the main text; with the help of a symbolic computing software we checked that the solution of $\M^{(n,2)} c = \cR^{(n,2)}$ has $c_4=0$ when plugging the values $\E[X_{1,1}^4]=3$, $\E[X_{1,1}^6]=15$ and $\E[B_{1,1}^2]=2$, restoring as it should the orthogonal invariance of the optimal estimator when the priors on the signal and noise are invariant under the orthogonal group. The off-diagonal contribution to the MMSE reads $\MMSEod^{(n,2)} = \left(1-\frac{1}{n}\right) - c^T \cR^{(n,2)}$, with $c$ the solution of $\M^{(n,2)} c = \cR^{(n,2)}$. With a symbolic expansion in the large $n$ limit we obtained the $1/n$ term stated in Eq.~(\ref{eq_order2_odcorrections}), as well as the $1/n$ corrections on the coefficients $c_1$ and $c_2$, and the dominant contribution for the coefficients of the optimal estimator that vanished in the large $n$ limit:
\begin{align}
c_3^{(n)} & = - \frac{1}{n} \frac{\alpha^\frac{3}{2} \Delta (\E[X_{1,1}^4]-1)}{(1+\Delta) ((1+\Delta)^3 + \alpha \Delta) (\E[X_{1,1}^4]-1+ \Delta \E[B_{1,1}^2])}+ O\left(\frac{1}{n^2}\right)  \ , \\
c_4^{(n)} & = \frac{1}{\sqrt{n}} \frac{2 \sqrt{\alpha} \Delta^2 (\E[X_{1,1}^4]-1-\E[B_{1,1}^2])}{ ((1+\Delta)^3 + \alpha \Delta) (\E[X_{1,1}^4]-1+ \Delta \E[B_{1,1}^2])}+ O\left(\frac{1}{n^{3/2}}\right)  \ .
\end{align}

\subsection{Diagonal terms}

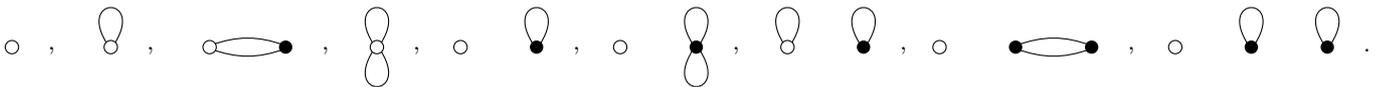
\begin{figure}
\begin{tikzpicture}
\fill[white] (0,0) circle (2.5pt);
\draw (0,0) circle (2.5pt);
\draw (0,0) node [right] {\phantom{=} ,};
\begin{scope}[xshift=1.3cm]
\draw (-.02,0) ..controls (-.5,.7) and (.5,.7).. (.02,0);
\fill[white] (0,0) circle (2.5pt);
\draw (0,0) circle (2.5pt);
\draw (0,0) node [right] {\phantom{=} ,};
\end{scope}
\begin{scope}[xshift=2.6cm]
\draw (0,0.02) to[in=160,out=20] (1,0.02);
\draw (0,-0.02) to[in=200,out=-20] (1,-0.02);
\fill[white] (0,0) circle (2.5pt);
\draw (0,0) circle (2.5pt);
\fill[black] (1,0) circle (2.5pt);
\draw (1,0) node [right] {\phantom{=} ,};
\end{scope}
\begin{scope}[xshift=4.8cm]
\draw (-.02,0) ..controls (-.5,.7) and (.5,.7).. (.02,0);
\draw (-.02,0) ..controls (-.5,-.7) and (.5,-.7).. (.02,0);
\fill[white] (0,0) circle (2.5pt);
\draw (0,0) circle (2.5pt);
\draw (0,0) node [right] {\phantom{=} ,};
\end{scope}

\begin{scope}[xshift=5.9cm]
\draw (0,0) circle (2.5pt);
\fill[black] (1,0) circle (2.5pt);
\draw (.98,0) ..controls (.5,.7) and (1.5,.7).. (1.02,0);
\draw (1,0) node [right] {\phantom{=} ,};
\end{scope}

\begin{scope}[xshift=8cm]
\draw (0,0) circle (2.5pt);
\fill[black] (1,0) circle (2.5pt);
\draw (.98,0) ..controls (.5,.7) and (1.5,.7).. (1.02,0);
\draw (.98,0) ..controls (.5,-.7) and (1.5,-.7).. (1.02,0);
\draw (1,0) node [right] {\phantom{=} ,};
\end{scope}

\begin{scope}[xshift=10.2cm]
\fill[black] (1,0) circle (2.5pt);
\draw (.98,0) ..controls (.5,.7) and (1.5,.7).. (1.02,0);
\draw (-.02,0) ..controls (-.5,.7) and (.5,.7).. (.02,0);
\fill[white] (0,0) circle (2.5pt);
\draw (0,0) circle (2.5pt);
\draw (1,0) node [right] {\phantom{=} ,};
\end{scope}

\begin{scope}[xshift=12.2cm]
\draw (1,0.02) to[in=160,out=20] (2,0.02);
\draw (1,-0.02) to[in=200,out=-20] (2,-0.02);
\draw (0,0) circle (2.5pt);
\fill[black] (1,0) circle (2.5pt);
\fill[black] (2,0) circle (2.5pt);
\draw (2,0) node [right] {\phantom{=} ,};
\end{scope}

\begin{scope}[xshift=15.3cm]
\draw (0,0) circle (2.5pt);
\fill[black] (1,0) circle (2.5pt);
\draw (.98,0) ..controls (.5,.7) and (1.5,.7).. (1.02,0);
\fill[black] (2,0) circle (2.5pt);
\draw (1.98,0) ..controls (1.5,.7) and (2.5,.7).. (2.02,0);
\draw (2,0) node [right] {\phantom{=} .};
\end{scope}

\end{tikzpicture}
\caption{The nine graphs in $\Ad^{(2)}$ that contribute to the diagonal terms of the estimators at order 2, in presence of the inversion symmetry. White circles represent the marked vertex, black ones the unmarked vertices. }
\label{fig_bG_diag}
\end{figure}

We shall now explain the main steps that brought us to the expansion (\ref{eq_order2_d}) for the contributions of the diagonal terms to the MMSE. According to the rules explained in Sec.~\ref{sec_universality_equivariant} there are 9 graphs, represented on Fig.~\ref{fig_bG_diag} that contribute: these are indeed the unlabelled graphs with one marked vertex (since we consider the diagonal terms of the estimator), at most two edges (since we only consider polynomials of degree at most two), no isolated unmarked vertex, and all vertices with even degrees (assuming the inversion symmetry).
These estimators $b_\beta(Y)$ have by definition $(b_\beta(Y))_{i,j}=0$ for $i \neq j$, and their respective values for $(b_\beta(Y))_{i,i}$ read:
\beq
1 \ , \quad Y_{i,i} \ , \quad \sum_{j \neq i} Y_{i,j}^2 \ , \quad Y_{i,i}^2 \ , \quad \sum_{j \neq i} Y_{j,j} \ , \quad 
\sum_{j \neq i} Y_{j,j}^2 \ , \quad Y_{i,i }\sum_{j \neq i} Y_{j,j} \ , \quad 
\sum_{\substack{j,k \neq i \\ j \neq k}} Y_{j,k}^2 \ , \quad 
\sum_{\substack{j,k \neq i \\ j \neq k}} Y_{j,j} Y_{k,k} \ . 
\eeq
One is thus led to consider a matrix $\M^{(n,2,{\rm d})}$ and a vector $\cR^{(n,2,{\rm d})}$, of dimension 9, defined as $\M_{\beta,\beta'}^{(n,2,{\rm d})} = \E[(b_\beta(Y))_{1,1} (b_{\beta'}(Y))_{1,1} ]$ and $\cR_\beta^{(n,2,{\rm d})} = \E[S_{1,1} (b_\beta(Y))_{1,1} ]$. Their matrix elements can be expressed in terms of joint moments of $Y$ with the same kind of computations as those detailed in the off-diagonal case. There are some additional moments that appear:
\begin{align}
\E[Y_{1,1}^3]  = & \frac{1}{n^2} \sqrt{\alpha} (\E[X_{1,1}^6] - 3 \, \E[X_{1,1}^4] +2)\ , \\
\E[Y_{1,1}^4]  = & \frac{1}{n^2} ( 3 (\E[X_{1,1}^4]-1)^2 + 6 \Delta (\E[X_{1,1}^4]-1) \E[B_{1,1}^2] + \Delta^2 \E[B_{1,1}^4] ) \\
& + \frac{1}{n^3} \alpha ( \E[X_{1,1}^8] - 3 \, \E[X_{1,1}^4]^2 - 4 \, \E[X_{1,1}^6] + 12 \, \E[X_{1,1}^4] - 6 ) \ , \\
\E[Y_{1,2}^4]  = & \frac{1}{n^2} (3+6\Delta + \Delta^2 \E[B_{1,1}^4]  )  + \frac{1}{n^3} \alpha ( \E[X_{1,1}^4]^2 -3 )\ , \\
\E[S_{1,1} Y_{1,1}] = & \frac{1}{n} (\E[X_{1,1}^4]-1) \ , \\
\E[S_{1,1} Y_{1,1}^2]  = &  \frac{1}{n^2} \sqrt{\alpha} (\E[X_{1,1}^6] - 3 \, \E[X_{1,1}^4] +2) \ , \\
\E[S_{1,1} Y_{1,2}^2]  = & \frac{1}{n^2} \sqrt{\alpha} (\E[X_{1,1}^4] - 1) \ . 
\end{align}
Following the steps encountered several times before one computes the optimal diagonal MSE as
\beq
\MMSEd^{(n,2)} = \E[S_{1,1}^2] - c^T \cR^{(n,2,{\rm d})} = \frac{1}{n}  (\E[X_{1,1}^4] - 1) - c^T \cR^{(n,2,{\rm d})} \ ,
\eeq
where $c$ is the 9-dimensional vector solution of $\M^{(n,2,{\rm d})} c = \cR^{(n,2,{\rm d})}$. With the help of a symbolic computing software we thus obtained the expansion stated in Eq.~(\ref{eq_order2_d}). We spare the reader the details of this nine-dimensional optimization problem since actually the leading $1/n$ order can be determined in a simpler way, working with the reduced set of diagonal basic functions $(b_1(Y))_{i,i}=1$, $(b_2(Y))_{i,i}=Y_{i,i}$, $(b_3(Y))_{i,i}=(Y^2)_{i,i}$ (the discarded six-dimensional variational space only contribute to the order $1/n^2$ in the expansion of $\MMSEd^{(n,2)}$). The elements of the corresponding matrix and vector read
\begin{align}
\M_{1,1} & = 1 \ , \\
\M_{1,2} & = 0 \ , \\
\M_{1,3} & = \E[Y_{1,1}^2] + (n-1) \E[Y_{1,2}^2] \ , \\
\M_{2,2} & = \E[Y_{1,1}^2] \ , \\
\M_{2,3} & = \E[Y_{1,1}^3] + (n-1) \E[Y_{1,1} Y_{1,2}^2] \ , \\
\M_{3,3} & = \E[Y_{1,1}^4] + 2 (n-1) \E[Y_{1,1}^2 Y_{1,2}^2] + (n-1) \E[Y_{1,2}^4] + (n-1)(n-2) \E[Y_{1,2}^2 Y_{2,3}^2] \ , \\
\cR_1 & = 0 \ , \\
\cR_2 & = \E[S_{1,1} Y_{1,1}] \ , \\
\cR_3 & = \E[S_{1,1} Y_{1,1}^2] + (n-1) \E[S_{1,1} Y_{1,2}^2] \ .
\end{align}
The expressions of the moments of $Y$ stated above allow to give a completely explicit form for the three-dimensional system of equations, with again an expansion in the large $n$ limit obtained symbolically that yields Eq.~(\ref{eq_order2_d}).

\section{Additional details on the universality conjecture for arbitrary $D$}
\label{app_universality}

We continue here the discussion of Sec.~\ref{sec_universality_allD} and explain more precisely the concrete implementation of the computation. One missing ingredient was the definition of the estimators $b_G(Y)$, a possible choice is constructed as follows. For $H=(V,E)$ an (unmarked) graph (with possibly multiple edges and self-loops), let us denote
\beq
T_H(Y) = \sum_{\phi \in \cF(V,[n]) } \prod_{e=\{a,b\} \in E} Y_{\phi(a),\phi(b)} \ ,
\eeq
where $\cF(E,F)$ is the set of functions from $E$ to $F$. This quantity is called the trace of the graph $H$, labelled by matrices $Y$, in~\cite{Ma20,CeDaMa16}. If $G=(V,E,v,w)$ is a marked graph we shall call ${\cal C}_{\rm u}$ the set of its connected components that do not contain any marked vertex, while $V_{\rm m}$ and $E_{\rm m}$ will denote the vertices and edges of the connected components that contain at least one marked vertex. We then define
\begin{align}
(b'_G(Y))_{i,j} & = n^{x_G} \left(\sum_{\substack{\phi \in \cF(V_{\rm m},[n]) \\ \phi(v)=i,\phi(w)=j}} \prod_{e=\{a,b\} \in E_{\rm m}} Y_{\phi(a),\phi(b)} \right) \prod_{H \in {\cal C}_{\rm u}} (T_H(Y) - \E[T_H(Y) ] ) \ , \\
(b_G(Y))_{i,j} & = \frac{1}{2} ( (b'_G(Y))_{i,j} + (b'_G(Y))_{j,i}) \ ,
\label{eq_b_perm}
\end{align}
with $x_G$ an exponent discussed later on. The linear combinations of such $b_G$, with $G \in \A^{(D)}$, span all the permutation equivariant polynomials of degree at most $D$. We call good graphs those which decompose in terms of their connected components as a path of $p$ edges with the two endvertices marked, and for $i=1,\dots,D$ $q_i$ simple cycles of $i$ edges, and set $x_G=0$ for these good graphs. This definition coincides precisely with the one of Eq.~(\ref{eq_b_rot}), thus the $\{b_G\}_{G \in \A^{(D,{\rm g})} }$ span indeed the equivariant functions under the orthogonal group; all other graphs are placed in the bad set $\A^{(D,{\rm b})} $. Note that the $b_G$ associated to the good graphs were seen in Sec.~\ref{sec_rot} to be properly normalized, in the sense that the block $\M^{(n,D,{\rm gg})}$ admits a finite invertible limit as $n\to\infty$.

Let us now discuss the choice of the exponent $x_G$ for the bad graphs, that should yield a finite limit for the matrix elements of $\M^{(n,D)}$ of $\cR^{(n,D)}$, and the invertibility of $\M^{(\infty,D)}$ (recall the objective set in Eq.~(\ref{eq_objective})); a necessary (but not sufficient) condition for the invertibility of a Gram matrix is that its diagonal elements are strictly positive. We will argue shortly that there exists indeed a choice of $x_G$ such that $\M^{(\infty,D)}_{G,G} > 0$ (for the only bad graph at $D=2$ one had $x_G=1/2$, see the definition of $b_4$ in (\ref{eq_order2_b1b4})); note meanwhile that the Cauchy-Schwartz inequality gives
\beq
|\M_{G,G'}^{(n,D)}| \le \sqrt{\M_{G,G}^{(n,D)} }\sqrt{\M_{G',G'}^{(n,D)} } \ \ , \qquad
|\cR_G^{(n,D)}| \le \sqrt{\E[\tr((S^{(n)})^2)]} \sqrt{\M_{G,G}^{(n,D)} } \ ,
\eeq
hence this choice of $x_G$ implies that the off-diagonal elements of $\M^{(n,D)}$ and those of $\cR^{(n,D)}$ are bounded (and the forthcoming argument will show that they admit a finite limit as $n \to \infty$).

A moment of thought reveals that $\M^{(n,D)}_{G,G'}$ is of the form
\beq
n^{x_G+x_{G'}-1} \sum_{i=1}^s C_i \prod_{j=1}^{t_i} \E[T_{H_{i,j}}(Y)] \ , 
\label{eq_M_expansion}
\eeq
where $s$, $\{C_i,t_i\}$ and the unmarked graphs $\{H_{i,j}\}$ depends on $(G,G')$. Indeed when computing $\E[\tr( b_G(Y) b_{G'}(Y) )]$ one has to join the two marked graphs $G$ and $G'$ by identifying their marked vertices, which yields an unmarked graph; because of the centering of the contribution of the unmarked components in (\ref{eq_b_perm}) the expansion of the product gives rise to the form (\ref{eq_M_expansion}) for some real coefficients $C_i$. We shall see that $\E[T_H(Y)] = n^{y_H}$ times a polynomial in $1/\sqrt{n}$, with $y_H$ a well-defined exponent, hence $\M^{(n,D)}_{G,G}$ has an algebraic large $n$ behavior with an exponent $2 x_G -1+\max_i \sum_j y_{H_{i,j}}$. One can thus set $x_G$ in such a way that the exponent of  $\M^{(n,D)}_{G,G}$ vanishes, which enforces the existence and the positivity of its limit, $\M^{(\infty,D)}_{G,G} > 0$.

In order to express $\E[T_H(Y)]$ in a more explicit form (see Appendix~\ref{app_order2} for some concrete examples of low orders) one should exploit the independence of the random variables forming the Wishart and GOE matrices; it is thus wise to introduce a variant of $T_H$ called the injective trace in~\cite{Ma20,CeDaMa16},
\beq
T_H^{\rm inj}(Y) = \sum_{\phi \in \cI(V,[n]) } \prod_{e=\{a,b\} \in E} Y_{\phi(a),\phi(b)} \ .
\eeq
By decomposing the sum over $\phi \in \cF(E,F)$ as a sum over the partitions $\pi$ of $E$ and a sum over the injections from the blocks of $\pi$ towards $F$, one can reduce the computation of $\E[T_H(Y)]$ to the ones of $\E[T_{H^\pi}^{\rm inj}(Y)]$, with $H^\pi$ the graphs obtained from $H$ by gluing together its vertices according to the various partitions $\pi$. One advantage of this injective formulation is the factorization over connected components it ensures, thanks to the independence structure of GOE and Wishart matrices. Since in addition each edge of $Y_{i,j}$ can be decomposed as $S_{i,j} + Z_{i,j}$, which are independent one of the other, one further reduces the computation to the one of $\E[T_H^{\rm inj}(S)]$ and $\E[T_H^{\rm inj}(Z)]$, where $H$ is an arbitrary connected graph. For the GOE part one can write a very explicit formula by exploiting the independence of the matrix elements:
\beq
\E[T_H^{\rm inj}(Z)] = \left(\frac{\Delta}{n} \right)^{|E|/2} n (n-1) \dots (n-|V|+1) \prod_{e \in E_{\neq}} \E[B_{1,2}^{d_e}] \prod_{e \in E_=} \E[B_{1,1}^{d_e}] \ ,
\label{eq_moments_Z}
\eeq
where $|E|$ is the number of edges of $H$, $|V|$ its number of vertices, and we called $E_{\neq}$ (resp. $E_=$) the set (without repetitions) of edges between distinct variables (resp. self-loops), $d_e$ being the multiplicity of the edge $e$ in $H$. By the centering of the $B_{i,j}$ this contribution vanishes if there are edges with a multiplicity of 1, and one can use the known values of the moments of Gaussian random variables to further simplify this expression. The algebraic dependency on $n$ claimed above is apparent on this expression. The Wishart counterpart, $\E[T_H^{\rm inj}(S)]$, requires an additional step: from the expression (\ref{eq_def_Wishart_gene}) of the matrix elements $S_{i,j}$ one has to introduce a summation over an index $\mu_e=1,\dots,r$ for each edge of $H$. This can be represented by a bipartite graph, with two types of vertices representing the indices $i$ and $\mu$ respectively, and edges corresponding to the random variables $X_{i,\mu}$ joining them. Introducing once again a sum over the partitions of the $\mu$-vertices, and identifying them along the blocks, one obtains a bipartite graph in which all the $i$ indices and all the $\mu$ indices are distinct. The average over the $X_{i,\mu}$ can then be performed very easily thanks to their independence, and one obtains a formula similar to (\ref{eq_moments_Z}), with products of moments of $X_{1,1}$ according to the degeneracy of each edge (with again a vanishing of the corresponding contribution if some edges are simple). The combinatorics of the sum over the $\mu$'s yields a polynomial factor in $r=n/\alpha$, which is thus polynomial in $n$.

The conclusion of this series of observations is that $\E[T_H(Y)]$ has indeed the claimed dependency in $n$ of the form $n^{y_H}$ times a polynomial in $1/\sqrt{n}$, with coefficients that depend on $\alpha,\Delta$, and on the moments of $X$. The exponent $y_H$ depends on the graph $H$ in a complicated way, through the various graphical transformations we just exposed, but in principle allows to fix the exponents $x_G$ of the bad graphs. Note that the computation of $\cR^{(n,D)}_G$ follows from the same graphical considerations, by adding to the graph $G$ one edge labelled with $S$ between its marked vertices.

We ensured a necessary condition for the invertibility of $\M^{(\infty,D)}$, it would remain to prove that this matrix is indeed invertible, as well as the other conditions in (\ref{eq_objective}), a seemingly challenging task in combinatorics and graph theory. Note indeed that most random matrix computations focus on the leading order behavior of $\E[\Tr(Y^p)]$, or covariances of the form $\E[\Tr(Y^p) \Tr(Y^q)] - \E[\Tr(Y^p)] \E[\Tr(Y^q)]$. In graphical terms this corresponds to $\E[T_H(Y)]$ for relatively simple graphs $H$ with one or two disjoint cycles, whose injective trace is dominated respectively by cacti graphs~\cite{CeDaMa16,MaFoLaKrMeZd19} (a cactus graph being defined as a graph in which each edge belongs exactly to one simple cycle), and annular non-crossing partitions~\cite{MaMiPeSp22}. On the contrary we require here a control of the trace of all graphs, at their leading order; the theory of traffic distributions~\cite{Ma20,CeDaMa16} deals precisely with those objects, but with a different normalization, that forces the limit of the injective trace of non-cacti graphs to vanish~\cite{CeDaMa16}. Upper bounds have been obtained for the traces of arbitrary graphs in~\cite{MiSp12}, but again we would need tighter results and matching lower bounds to achieve our goal.

Note finally that (\ref{eq_b_perm}) is by far not the only possible choice for the basic functions $b_G$; its advantage was to make obvious the distinction between good and bad graphs, yet replacing the summations there by their injective versions, as well as replacing the monomials on the edges by the Hermite polynomials, orthogonal with respect to the Gaussian measure (as done in~\cite{MoWe22} in the proof of their proposition 3.3) might instead simplify the derivation of the properties (\ref{eq_objective}).

\section*{Acknowledgments}

I warmly thank Guillaume Barraquand, Djalil Chafa\"\i, Laura Foini, Florent Krzakala, Antoine Maillard, Cris Moore, Soledad Villar and Lenka Zdeborov\'a for useful discussions, and in particular A. M. for providing some numerical data from~\cite{MaKrMeZd22} and for a careful reading of the manuscript, and C. M. for sharing the draft of~\cite{KuMoWe24} prior to publication and for pointing the reference~\cite{Pr76}. I also benefited from stimulating exchanges with Jean Barbier, Francesco Camilli and Marc M\'ezard, and from useful suggestions of two anonymous referees.

\section*{Conflict of interest statement}

The author has no competing interests to declare that are relevant to the content of this article.

\section*{Data availability statement }

The data presented in the plots has been produced by the simple numerical procedures explained in the text, and can be made available upon request to the author.

\bibliography{biblio}

\end{document}